\documentclass[11pt,a4paper]{article}
\usepackage[pdftex]{graphics}
\usepackage{amsmath,amssymb,amsfonts}
\usepackage{tcolorbox}
\usepackage{enumitem}
\usepackage{multirow}
\usepackage{array,booktabs}
\usepackage{slashed}
\usepackage[utf8]{inputenc}
\usepackage[english]{babel}
\usepackage{amsthm}
\usepackage{graphicx}
\usepackage{float}

\usepackage{jheppub}
\usepackage{amsmath,amssymb,amsfonts}

\usepackage{tcolorbox}
\usepackage{enumitem} 
\usepackage{multirow}
\usepackage{array,booktabs}
\usepackage{slashed}
\usepackage{wallpaper}
\usepackage[utf8]{inputenc}
\usepackage[english]{babel}
\usepackage{amsthm}

\usepackage{tikz-cd}
\usetikzlibrary{positioning}

\usepackage{graphicx}
\usepackage{float}
\usepackage{cancel}
\usepackage{tabularx}
\usepackage{physics}
\usepackage{hyperref}
\usepackage{bbm}

\usepackage{booktabs}   
\usepackage{ragged2e}   
\usepackage{caption}    
\usepackage{lmodern}    
\newcommand{\bn}{\begin{enumerate}}
\newcommand{\en}{\end{enumerate}}
\newcommand{\bl}{\begin{align}}
\newcommand{\el}{\end{align}}
\newcommand{\ie}{\begin{equation}\begin{aligned}}
\newcommand{\fe}{\end{aligned}\end{equation}}

\newcommand{\vecslashed}[1]{{\slashed{#1}}}

\newcommand{\bm}{\boldsymbol}

\newcommand{\intp}[1]{\int \frac{d^3 p_{#1}}{(2\pi)^3}}
\newcommand{\zero}{\partial}
\newcommand{\bchi}{\bm{\chi}}
\newcommand{\bbarchi}{\bm{\bar\chi}}
\newcommand{\bpsi}{\bm{\psi}}
\newcommand{\bbarpsi}{\bm{\bar\psi}}
\newcommand{\disc}{\mathop{\mathrm{Disc}}}

\newcommand{\beqas}{\begin{eqnarray*}}
\newcommand{\eeqas}{\end{eqnarray*}}
\newcommand{\beqa}{\begin{eqnarray}}
\newcommand{\eeqa}{\end{eqnarray}}			
\newcommand{\beq}{\begin{equation}}
\newcommand{\eeq}{\end{equation}}
\newcommand{\vk}{{{k}}}

\newcommand{\braketw}[2]{\langle #1  #2 \rangle}

\newcommand{\tablewidth}{0.90\textwidth}   
\newcommand{\tableheight}{0.17\textheight} 
\newcommand{\leftcolwidth}{0.24\linewidth} 

\newcommand{\betweenTableAndCaption}{0.5em} 
\newcommand{\explainfontsize}{\small}      
\newcommand{\explainwidth}{0.92\textwidth} 

\def\ck{\mathcal{K}}

\DeclareUnicodeCharacter{2212}{-}

\preprint{}

\title{Fermionic Boundary Correlators in (EA)dS space}
\author[1]{Wei-Ming Chen}
\author[2,3,4]{Yu-tin Huang}
\author[2]{Zi-Xun  Huang}
\author[2]{Yohan Liu}
\affiliation[1]{Department of Physics, National Sun Yat-sen University, Kaohsiung 80424, Taiwan}
\affiliation[2]{Department of Physics and Center for Theoretical Physics, National Taiwan University, Taipei 10617, Taiwan}
\affiliation[3]{Physics Division, National Center for Theoretical Sciences, Taipei 10617, Taiwan}
\affiliation[4]{Max Planck{-}IAS{-}NTU Center for Particle Physics, Cosmology and Geometry, Taipei 10617, Taiwan}

\emailAdd{tainist@gmail.com}
\emailAdd{yutinyt@gmail.com}
\emailAdd{d11222011@ntu.edu.tw}
\emailAdd{youan1997@icloud.com}
\abstract{In this paper we bootstrap de Sitter wavefunction coefficients (WFCs) involving fermionic operators. Starting with a fixed total-energy pole order, we systematically impose the conformal Ward identities (CWI) together with cutting-rule constraints. We derive the relevant cutting rules for fermionic exchange for the first time, enabling a complete determination of fermionic three- and four-point WFCs. We show that CWI fixes the leading total-energy-pole residue to the flat-space amplitude and subleading residues to curvature induced corrections to bulk vertices. The structure of the Ward-Takahashi identities are similarly fully determined.  As an application, we derive four massless spin-1/2 WFC due to graviton exchange. We also revisit the tension between conserved spin-3/2 operators and de Sitter geometry. We demonstrate that the reality conditions appropriate to dS and Euclidean AdS (EAdS) lead to distinct three-point WFCs for two spin-3/2 operators and the stress tensor. Consequently, the residue of the leading total-energy pole for the four-point WFC receives graviton- and photon-exchange contributions with opposite signs in dS, whereas they appear with the same sign in EAdS. This result is reminiscent of the classic analysis by Pilch, van Nieuwenhuizen, and Sohnius, though formulated in an on-shell framework.}

\begin{document}

\maketitle
\section{Introduction}
In a companion work~\cite{Flatspace}, we developed a systematic “four-particle test’’ for flat space boundary correlators. Beginning with a consistent flat-space S-matrix, we iteratively reconstructed the Wave Function Coefficients (WFC) by exploiting their analytic behavior in the energy variables. These analytic features arise from general properties of bulk-to-bulk and bulk-to-boundary propagators, thereby reflecting the structure of the bulk itself. In flat space the bootstrap system is exactly determined: starting with a consistent S-matrix, the constraints in energy variables are one to one tied to the unfixed terms in the WFC. Thus a consistent S-matrix is equivalent to a consistent WFC. Said in another way,  the presence of a boundary does not impose new dynamical constraints on the bulk theory, beyond the specification of suitable boundary conditions.

In this paper, we extend our analysis to WFCs in dS/AdS backgrounds. The key new ingredient is the constraint imposed by conformal invariance, which reflects the underlying isometries of dS/AdS space. The Conformal Ward Identities (CWI) leads to an descending tower of singularities in energy variables, with the residue of the leading total energy pole being the flat-space amplitude. There are already significant progress in recent years on bootstrapping dS WFCs. Starting with external scalars~\cite{Arkani-Hamed:2018kmz, Baumann:2019oyu}, conformal symmetry and the absence of unphysical singularities were shown to fix the 4-pt function with arbitrary mass or spin exchange. These results were generalized to external conserved currents in~\cite{Baumann:2020dch}. An alternative, “boostless’’ bootstrap was later introduced, in which the flat-space S-matrix serves as the seed~\cite{Pajer:2020wxk}. Building on the analytic structure encoded in the singularities of the energy variables~\cite{Baumann:2021fxj}—as implied by the Cosmological Optical Theorem (COT)~\cite{Goodhew:2020hob}—BCFW-like recursion relations were formulated in~\cite{Jazayeri:2021fvk}. These determine the WFCs up to boundary terms, which were fixed using the constraint that the WFC stems from bulk physics, i.e. the Manifestly Local Test (MLT). This framework was subsequently used to obtain the four–stress-tensor WFC in~\cite{Bonifacio:2022vwa}.

Here, we aim to streamline the application of constraints, as to sharpen their tension when applied to consistent theories in flat space that do not extend to dS backgrounds. An important example is that of the conserved spin-3/2 operator. The Ward identity associated with the conserved current is related to bulk supersymmetry which does not foster a unitary representation consistent with dS isometry~\cite{Lukierski:1984it}. In particular, requiring that the fermionic charges to be embedded in SO(1,4), i.e. the existence of real killing spinors in dS$_4$, reality conditions on the fermionic charges leads to the algebra $\{Q,Q^\dagger\}=0$, which implies an indefinite metric for the Hilbert space. Indeed the explicit construction of an action with dS$_4$ local supersymmetry leads to either an imaginary mass for the gravitino or a wrong sign for the gravi-photon kinetic term~\cite{Pilch:1984aw}.  Alternatively, it was proposed that supersymmetric higher spin theories might accommodate stable (no ghosts) dS solutions. This is due to the anti-involution property of Hermitian conjugation in higher spin algebra \cite{Sezgin:2012ag}. Explicit computation of the partition function for vector models indicates that the dual wave function in the bulk is peaked at the undeformed dS space \cite{Hertog:2017ymy} (see  \cite{Anninos:2025mje} for recent discussion).

Note that these difficulties with spin-3/2 currents do not forbid supersymmetry in dS. Since dS is conformally equivalent to flat-space, super-conformal theories can be consistently put in dS background by invoking conformal killing spinors~\cite{Anous:2014lia}. However the resulting spin-3/2 current would couple to the gravitino of conformal instead of Poincaré supergravity. Such a setup was recently explored in the self-dual sector in ~\cite{Higuchi:2025pbc}.

We begin our bootstrap with the spin-1/2 and 3/2 two-point functions. For the spin-3/2 case, after solving the CWI, we then further apply the current conservation to determine the conformal dimension of the conserved case. These two-point functions then serve as one of the building blocks of the three-point bootstrap. For the three-point bootstrap, we start with the flat-space amplitude on the leading total energy pole as the initial condition. Specifically, we only use the derivative order of the amplitude to determine the leading total energy pole order. We will then build a general ansatz with terms containing \emph{all orders} of the total energy pole, which are SO(3) covariant tensors. Note that we do not utilize spinor-helicity formalism when solving CWI, since in general the helicity operator do not commute with conformal boost generator.  While for conserved currents they commute up to derivative terms~\cite{Caron-Huot:2021kjy}, this implies that the action of conformal boosts on a helicity eigenstate of a field carrying vector indices will lead to mixing with longitudinal components. Indeed the action of the boost generator in helicity variables, on helicity eigenstates yields~\cite{Baumann:2020dch}
\ie
    \tilde K^i \chi^{\pm} &= -\xi^{\chi}_{\pm,\alpha} K^i \chi^\alpha,  \quad\quad \tilde{K}^i
    =2(\sigma^i)_\alpha^\beta\frac{\partial^2}{\partial\lambda_\alpha\partial\bar\lambda^\beta} \nonumber\\
    \tilde{K}^i J^{\pm} &=
    \left(
    -\,\epsilon^{j}_{\pm}\,K^{i}
    \;+\;
    2\,\epsilon^{i}_{\pm}\,\frac{p^{j}}{E^{2}}
    \right) J_{j},
    \\
    \tilde{K}^i\left(\frac{T^\pm}{E}\right)&=\left(-\frac{1}{E}\epsilon^{j}_{\pm}\epsilon^{k}_{\pm}K^i+12\epsilon^i_{\pm}\frac{\epsilon_{\pm}^{j}p^{k}}{E^3}\right)T_{jk},\label{eq:KtildeonT}  
\fe
where $\chi^+ \equiv \bar{\lambda}_{\alpha} \chi^\alpha, \chi^- \equiv \lambda_{\alpha} \chi^\alpha, \xi^{\chi}_{+,\alpha} \equiv \bar{\lambda}_\alpha, \xi^{\chi}_{-,\alpha} \equiv \lambda_\alpha, J^\pm \equiv \epsilon_{\pm}^i J_{i}, T^{\pm}\equiv\epsilon_{\pm}^i\epsilon^j_{\pm}T_{ij}$ and $K^i$ is the momentum-space boost generator. Since the the WFC should vanish under the action of $K^i$, we see that for conserved spinning operators, the action of $\tilde{K}^i$ will generate longitudinal terms, and CWI becomes non-homogeneous constraint. Nevertheless, once the bootstrap result is obtained, it is often convenient to cast the final result in the helicity basis.

In our analysis, we will be agnostic to the explicit form of the Ward–Takahashi (WT) identity, other than that the longitudinal projection must be proportional to two-point functions. We find that already at 3-pts, CWI leads to new features compared to its ``boostless" flat-space counterpart:
\begin{itemize}
    \item With the leading total energy pole fixed by the flat-space amplitude, consistency between WT identities and CWI fixes the WT identity to be that of diffeomorphism for spin-2 and supersymmetry for spin-3/2. In comparison, the precise form of the WT identities must be an input in the flat space bootstrap.
    \item For the spin-1 two spin-3/2 WFC, $\langle J\bar\psi \psi\rangle$, we find that the  longitudinal projection of $J$  indicates that spin-3/2 current must be charged.
    \item In the helicity basis, while those compatible with flat space S-matrix will appear at leading in the total energy pole, other helicity structures emerge at subleading order. This is contrary to flat space where incompatible helicity structures are regular in the total energy. Importantly, we demonstrate that the subleading total energy singularities encode local $\frac{1}{H}$ corrections to the interactions. 
    \item When projected into the helicity basis, we see that there must be minimal coupling between spin-3/2 and the gravi-photon, appearing on the leading total energy pole for $\langle J^{+}\bar\psi^{-}\psi^{+}\rangle$ configuration. Albeit it is subleading when all helicity configurations are considered. 
\end{itemize}
Conformal invariance also requires that any condition imposed on the flat-space amplitude must be embedded in SO(1,4). In particular, the Majorana condition of Lorentzian/Euclidean flat-space amplitude that serves as the residue on the total energy pole for dS/EAdS must be related to reality conditions of the boundary profile, which in turn must compatible with the unique Majorana representation in SO(1,4), the Symplectic Majorana condition. As a consequence, the flat-space amplitude of 2 gravitinos coupled to a graviton, $M_3(h,\bar{\psi}^{\mathbb{I}},\psi^{\mathbb{J}})$ where $\mathbb{I}=1,2$ are the SO(2) R-symmetry indices, will have distinct structures. In particular, 
\begin{equation}
M(T\bar\psi\psi)= M(JJJ)\left(l_{\mathbb{I}\mathbb{J}}\bar{u}^\mathbb{I}\slashed{\epsilon}_1 u^\mathbb{J}\right)\,,
\end{equation}
where $M(JJJ)$ is the Yang-Mills 3-pts amplitude and $l_{\mathbb{I}\mathbb{J}}=(\epsilon_{\mathbb{I}\mathbb{J}}, \delta_{\mathbb{I}\mathbb{J}})$ for Euclidean and Lorentzian signature respectively. This difference is the source of the inconsistency at 4-pts. 

To extend the bootstrap beyond three-point WFCs, we make systematic use of partial-energy pole constraints: whenever a factorization channel is available, higher-point WFCs necessarily develop singularities in the corresponding partial-energy variable. Both the existence of these poles and the form of their residues can be inferred from cutting rules, which state that the WFC exhibits universal behavior when the internal energy flowing through a diagram is flipped. This universality was first identified as a consequence of unitary time evolution—the so-called Cosmological Optical Theorem (COT)~\cite{Goodhew:2020hob}—and can be derived more directly from the analytic structure of bulk-to-bulk propagators~\cite{Meltzer:2021zin, Baumann:2021fxj, Melville:2021lst}.

Armed with these cutting rules and the bootstrapped three-point WFCs, we show that all terms with partial-energy pole are straightforwardly determined by the cutting rules, which has already been observed in \cite{Baumann:2021fxj,Jazayeri:2021fvk,Bonifacio:2022vwa}. The unfixed terms are then confined to pure total energy pole singularities with polynomial numerators. The leading of course is determined by matching to flat-space amplitude, and the subleading will be determined by CWI. This allows us to setup a bootstrap procedure for the 4-pt WFC that mirrors our construction in flat space~\cite{Flatspace}. In particular, the transverse-traceless WFC takes the form, 
\ie
c^{\hat T}_4=\left\{\sum_{e \in s,t,u}\left(\frac{A_R^e}{(E_R^e)^{p_{R}}}+\frac{B_L^e}{(E_L^e)^{p_{L}}}\right)+\frac{C}{E_T^{p}}+\frac{ D_{odd}}{E_T^{p-1}}\right\}+\frac{D_{even}}{E_T^{p-1}}
\fe
where $(p, p_R, p_L)$ are the leading degrees in total energy and partial energy poles, $D_{even}$ are polynomials of even degrees in energy variables, while $D_{odd}$ are odd in at least one energy variable. The terms  in the curly brackets are completely determined by our input. Thus, the non-trivial task reduces to finding $D_{even}$ that solves CWI. We illustrate the procedure for conformally coupled scalar and, for the first time, massless spin $1/2$ due to graviton exchange.

When considering conserved spin-3/2, since the three-point WFC is completely determined, the 4-pt WFC is also determined up to total energy pole singularities. Here we see where the tension arises. As we have already seen at 3-pts the $\langle T \bar{\psi}^{\mathbb{I}}\psi^{\mathbb{J}}\rangle$ have distinct R-symmetry structure. When considering the leading total energy pole constraint at 4-pts, where the residue is the flat-space amplitude that factorizes into 3-pts, we find that the graviton channel will have a relative sign vs the gravi-photon channel for dS$_4$, This is precisely what was found in~\cite{Pilch:1984aw}, albeit stated in an on-shell manner. This obstruction also points to potential avenue out: one can introduce modifications of the total energy pole degree so that the leading term no-longer leads to the supergravity amplitude. We discuss various scenarios, with the presence of infinite higher spins as the likely possibility. 

This paper is organized as follows. In section \ref{sec: Bootstrap consistency conditions}, we review the definition of dS WFCs and discuss the their pole structures as well as constraints arising from symmetry requirements such as the CWIs and WT identities. We also include discussions on some general properties of the WT identities and a consistency condition whenever there are multiple currents (the WT consistency). In section \ref{sec3}, starting from general discussions on the ansatz structure, we systematically apply the consistency conditions to bootstrap the WFCs from lower to higher points. In section \ref{sec:dsinconsistency}, upon the previous results, we conduct an on-shell test on the validity of (EA)dS Poincaré supergravity via inspecting the $4\psi$ amplitudes in the leading $E_T$ residues and find that unitarity is violated in the dS. We then focus on the leading $E_T$ pole residue and discuss possible resolutions. Appendix \ref{sec: Conventions and notation} sets up the notation and convention in this paper. In the appendix \ref{sec:residualsym}, we discuss the boundary WT identities viewed as the bulk residual gauge symmetry. Specifically, comparison of the difference between the diffeomorphism and supersymmetry in flat and curved space reveals that additional redundancies arise in the curved space, thus resulting additional WT identities than the flat-space. In the appendix \ref{app:LeadingTotalEnergyPoleStructure}, we review and discuss how the flat-space amplitude arises in the leading $E_T$ pole residue from carefully examining the bulk time-integral. In the appendix \ref{app: Feynman Rule from boundary action}, we derive the massless spin-1/2 and 3/2 cutting rules in the (EA)dS. In the appendix \ref{sec:SO(1,4)details}, we discuss the Clifford algebra of $SO(1,4)$ in detail and the different realization of its SM condition in the 4D Lorentzian / Euclidean space which serves as the foundation of section \ref{sec:dsinconsistency}.

\section{Bootstrap consistency conditions}
\label{sec: Bootstrap consistency conditions}

This section reviews some basics of the dS wavefunction coefficients and the constraints they satisfy. For constraints also present in flat-space theories \cite{Flatspace}, we will pay special attention to the essential differences between the flat-space and dS/EAdS. The corresponding results in Euclidean Anti-de Sitter (EAdS) are generally related to those in de Sitter (dS) space by analytic continuation (or Wick rotation). For a detailed discussion of this crucial technique, see, e.g., \cite{Maldacena:2002vr}. Consequently, we will focus our primary setup and calculations within the dS spacetime.


\subsection{Wavefunction Coefficient}

Let us start from the Poincaré patch of the 4-dimensional dS spacetime, 
\begin{equation}
    ds^2=\frac{-d\eta^2+d\vec{x}^2}{H^2\eta^2},\quad-\infty<\eta<0, \label{eq:dsmetric}
\end{equation}
where the Hubble constant $H$ defines the radius $1/H$, $\eta$ is the conformal time coordinate and $\vec{x}$ denotes the 3-dimensional space at $\eta=0$. The observable is the equal (late) time correlator in the in-in formalism $
\bra{\Omega} \hat \varphi(  x_1) \hat \varphi(  x_2) \dots \hat \varphi(  x_n) \ket{\Omega}$, 
where $\hat \varphi$ represents the field operator on the equal-time boundary and $\ket{\Omega}$ is the Bunch-Davies vacuum. The observable can be computed via the wavefunction, 
\begin{equation}
    \Psi[\varphi_\zero] = \braket{\varphi_\zero}{\Omega},
\end{equation}
where $\varphi_\partial$ is the boundary profile of the field, i.e. $\varphi(t = 0) =\varphi_\partial$, understood as a Dirichlet boundary condition. The wavefunction is then the overlap between a state $\ket{\varphi_\zero}$ at $t = 0$ and the vacuum state $\ket{\Omega}$. The observable is then given as:
\begin{equation}
\label{inindef}
\bra{\Omega} \hat \varphi(  x_1) \hat \varphi(  x_2) \dots \hat \varphi(  x_n) \ket{\Omega}
=
\int \mathcal{D}\varphi_\zero \ \varphi_\zero(  x_1) \varphi_\zero(  x_2) \dots \varphi_\zero(  x_n) |\Psi[\varphi_\zero]|^2\,.
\end{equation}
It will be convenient to expand the wavefunction  in three-dimensional momentum-space eigenstates:
\begin{equation}
\label{eqn: expand wave function in wave function coefficient}
\log \Psi[\varphi_\zero]= \sum_{n=2}^\infty \int \frac{d^3\mathbf{k}_1 \dots d^3\mathbf{k}_n }{(2\pi)^{3n}} \, \delta^{(3)}\left(\sum_{i=1}^n \mathbf{k}_i\right)\, \varphi_{\zero,\mathbf{k}_1} \dots \varphi_{\zero,\mathbf{k}_n} \, c_n(\mathbf{k}_1, \dots, \mathbf{k}_n),
\end{equation}
where the functions $c_n(\mathbf{k}_1, \dots, \mathbf{k}_n)$ are referred to as \emph{wavefunction coefficients} (WFCs). For spinning fields, the WFCs carry explicit indices to be contracted with the spinning boundary profiles.\footnote{
    We use $i, j, k, \ldots$ for vector indices, $A, B, C, \ldots$ for four-component spinors, and $\alpha, \beta, \dot{\alpha}, \dot{\beta}, \ldots$ for two-component spinors. A generic WFC is denoted by $c_{n,AB}^{i_1 \dots i_n}$, while bracket notation refers to specific boundary operators:
    \begin{equation}
        c_{n,AB}^{i_1 \dots i_n}(p_1, p_2, p_3, \ldots) \rightarrow \langle J^{i_1}_1\, \bar\psi^{i_2}_{2,A}\, \psi^{i_3}_{3,B}\, \phi_4 \cdots \rangle,
    \end{equation}
    where subscripts label momenta. 
} For a comprehensive review on the relation between in-in correlator and WFCs, see \cite{Goodhew:2020hob}. We will focus on the "tree-level" approximation of $c_n$, which is given by substituting the classical solution $\varphi_{cl}$ into the action.\footnote{
For fermionic WFCs,  due to the action being first-derivative, one can only impose Dirichelet on \emph{half} of the fermion, since the other half is its conjugate. This also requires the inclusion of  boundary actions. See~\cite{Flatspace} for a recent review.} The WFC coefficients can be computed diagrammatically using Feynman rules. For examples, see \cite{Anninos:2014lwa, Baumann:2020dch,Salcedo:2022aal,Bonifacio:2022vwa,Chowdhury:2024snc}

\paragraph{Spinning WFC Decomposition}
As noted earlier, for spinning boundary profiles the uncontracted WFCs carry explicit spacetime (or spinor) indices. It is therefore natural to decompose both the operator and the corresponding WFCs into irreducible representations, allowing one to construct the different ansatz independently.

For a spin-1 operator, one simply decomposes into transverse and longitudinal pieces:
\begin{equation}
    \mathcal{O}^i(p)=\left(\delta^i_j-\hat{p}^i\hat{p}_j\right)\mathcal{O}^j+\hat{p}^i\hat{p}_j\mathcal{O}^j\equiv\pi^{ii'}\mathcal{O}_{i'}+\hat{p}^i\hat{p}_j\mathcal{O}^j\equiv\mathcal{O}_T^i+\mathcal{O}_L^i.\label{eq: decomposition}
\end{equation}
For the symmetric spin-2 current $T^{ij}$, one similarly decomposes into $T^{ij}_{TT}$, $T^{ij}_{TL}$ and $T^{ij}_{LL}$. It is useful to further decompose the transverse part into the following,
\ie
T^{ij}_{TT}=\pi^{i'i}\pi^{jj'}T_{i'j'}=&\left(\pi^{ii'}\pi^{jj'}-\frac{1}{2}\pi^{ij}\pi^{i'j'}\right)T_{i'j'}+\frac{1}{2}\pi^{ij}\pi^{i'j'}T_{i'j'}
\\
\equiv &\hat{\Pi}^{iji'j'}T_{i'j'}+\underline{\Pi}^{iji'j'}T_{i'j'}
\equiv T_{\hat{T}\hat{T}}^i+T_{\underline{TT}}^i,
\label{eq:Tdecompose},
\fe
where $\widehat{\Pi}^{iji'j'}$ projects onto the \emph{transverse-traceless} part $\delta_{ij}\widehat{\Pi}^{iji'j'}=\delta_{i'j'}\widehat{\Pi}^{iji'j'}=0$ and $\underline{\Pi}^{iji'j'}$ satisfies $\delta_{ij}\underline{\Pi}^{iji'j'}=\pi^{i'j'},\delta_{ji'}\underline{\Pi}^{iji'j'}=\frac{1}{2}\pi^{ij'}$.
For WFCs involving spin-3/2 currents $\psi^i$, it is also useful to further decompose the transverse projector into components that are orthogonal (parallel) to the gamma matrices,
\begin{equation}
\psi_{T}^i=\left(\pi^{ij}-\frac{1}{2}   \pi^{ii'}\pi^{jj'} \sigma_{i'}\sigma_{j'}\right)\psi_{j}+\left(\frac{1}{2}   \pi^{ii'}\pi^{jj'} \sigma_{i'}\sigma_{j'}\right)\psi_{j}\equiv\hat{\Pi}^{ij}\psi_j+\underline{\Pi}^{ij}\psi_j\equiv\psi_{\hat{T}}^i+\psi_{\underline{T}}^i,\label{eq:gammadecompose}
\end{equation}
where the $\hat{\Pi}^{ij}$ is orthogonal to the gamma matrices ($\sigma_i\hat{\Pi}^{ij}=\hat{\Pi}^{ij}\sigma_j=0$), and we have called such component as \emph{transverse-gamma-traceless}  whereas $\underline{\Pi}^{ij}$ satisfies $\sigma_i\underline{\Pi}^{ij}=\pi^{jj'}\sigma_{j'},\;\underline{\Pi}^{ij}\sigma_{j}=\pi^{ii'}\sigma_{i'}$.
In summary, a spinning WFC is decomposed as \eqref{eq: decomposition}, \eqref{eq:Tdecompose}, \eqref{eq:gammadecompose}. The general expression takes the form,
\begin{equation}
    \langle\mathcal{O}^{\{i_s\}}\dots\rangle=\left(\sum_g\mathbb{P}_g\mathbb{A}^g\right)^{\{i_s\}}\,,
    \label{eq:generalPdecompose}
\end{equation}
where $\{i_s\}$ densely labels the spin-$s$ Lorentz indices. The $g$ runs over various combinations of transverse ($T$), longitudinal ($L$) and trace projectors. Thus the non-trivial information of the WFC is encoded in $\mathbb{A}^g$ which is the main subject of this paper. The explicit index in $g$ reflects the nature of its representation as summarized in the table below. Note that when all indices of a WFC are of transverse (gamma)traceless we simply denote as $\hat{\mathbb{A}}$.
\begin{center}
\label{IndexTable}
  \fbox{
    \begin{minipage}[t][\tableheight]{\tablewidth}
      \vspace{3pt} 

      \begin{tabularx}{\linewidth}{@{}>{\RaggedRight\arraybackslash}p{\leftcolwidth}
    >{\RaggedRight\arraybackslash}X@{}}
        \textbf{Symbol} & \textbf{Meaning} \\
        \midrule
       $\mathbb{A}^{\mathcal{O}_L}$ & Longitudinal component of $\mathcal{O}$\\[8pt]
        $\Hat{\mathbb{A}}$ & Pure Transverse Traceless component\\[8pt]
        $\mathbb{A}^{\underline{T}}$ & Trace of Transverse component \\[8pt]
    $\mathbb{A}^{\Hat{T}}$ & Transverse Traceless component \\[8pt]
      \end{tabularx}

    \end{minipage}%
  } 
\end{center}
 \vspace{\betweenTableAndCaption}

\noindent{\explainfontsize\ }
\begin{minipage}{\explainwidth}
Table 1: Examples of $\mathbb{A}^g$, where all Lorentz indices are suppressed. Note that for $g$ carrying various $\Hat{T},\underline{T}$, the displayed order follows the order of the corresponding operators. For instance, $\mathbb{A}^{\Hat{T}\underline{T}\Hat{T}}$ is the component of $\langle\mathcal{O}_{\Hat{T}}\mathcal{O}_{\underline{T}}\mathcal{O}_{\Hat{T}} \rangle$.
\end{minipage}
\subsection{Ward Identities}
\label{sec:cwi}
We now review the constraints on the WFCs that can be inferred from symmetry principles. 

\paragraph{CWIs}



Bulk fields in de Sitter space are constrained by the $SO(1,4)$ isometry group. To determine how these isometries restrict the WFCs, one can examine the late-time behavior of the equations of motion for a spin-$\ell$ bulk field. At the future boundary, the spatial profile of the dominant solution transforms as a conformal primary operator with scaling dimension $\Delta'$. Furthermore, because the WFCs are defined by operators conjugate to the bulk fields, an integration by parts of the action demonstrates that the WFCs inherit the same isometries.  For a bulk field with scaling dimension $\Delta'$, the associated boundary operator is characterized by the conformal dimension $\Delta = 3 - \Delta'$. A detailed discussion of bulk fields and WFCs can be found in \cite{Baumann:2020dch}.

Specifically, $\Delta$ is related to the bulk mass $m$ and spin $\ell$:
\begin{itemize}
\item integer spin is dS
\ie
&\text{Scalar}:\quad\Delta=\frac{3}{2}+\sqrt{\frac{9}{4}-\frac{m^2}{H^2}}.
\\
&\text{Spin}-l:\quad\Delta=\frac{3}{2}+\sqrt{\left(l-\frac{1}{2}\right)^2-\frac{m^2}{H^2}}\,.\label{eq:(EA)dSDelta}\\
\fe
\item dS half-integer spin \cite{Anguelova:2003kf} (also see EAdS half-integer spin \cite{Henningson:1998cd,Mueck:1998iz,Volovich:1998tj,Corley:1998qg})
\ie
&\hspace{-4.5mm}\text{Spin-1/2 and 3/2}:\quad \Delta=\frac{3}{2}+i\frac{m}{H}\,.
\fe
\end{itemize}
When performing the 3-point bootstrap, we choose specific $\Delta$ that correspond to the specific masses of bulk fields, according to the equations here.

Moving forward, we focus on the isometries satisfied by the WFCs, which manifest as Conformal Ward Identities (CWIs). Note that, since there is an overall momentum-conserving delta function support, operations on $c_n$ will be on the $n-1$ independent momentum. Let us start from the dilatation, which gives
\begin{equation}
\left\{-3+\sum_{m=1}^{n}\left(3-\Delta_m\right)+\sum_{m=1}^{n-1}p_m^i\partial_{p_m^i}\right\}c_n=0,\label{eq:dilatationcwi}
\end{equation}
which controls the momentum dimension ($\sum_{m=1}^{n-1}p_m^i \partial_{p_m^i}$) of the WFCs. The action of the conformal boost depends directly on the indices of the underlying operators and, consequently, on the structure of the coefficient $c_n$	
 \cite{Bzowski:2013sza}. We must therefore make these indices explicit, which yields, 
\begin{equation}
  \sum_{m=1}^{n-1}\left\{ \left[2\left(3-\Delta_m\right)\partial_{p_m^i}+2p_m^j\partial_{p_m^j}\partial_{p_{m}^i}-p^i_m\partial_{p_m^j}\partial_{p_{m}^j}\right]c_n^{\dots a\dots }-2\left(S_m^{ij}\right)^a_{b_m}\partial_{p_m^j}c_n^{\dots b_m\dots}\right\}=0\label{eq:boostcwi}
\end{equation}
where $\left(S_m^{ij}\right)^a_{b_m}$ denotes Lorentz spin generator for the $m$th leg. $a,b_
m$ represent the Lorentz indices of a spinning field and $b_m$ runs over all the indices carries by the $m$-th operator. For example, it acts on spin-1 and spin-1/2 operators in the following way \cite{Bzowski:2013sza,Iliesiu:2015qra},
\begin{equation}
    \text{Spin-1}: (S^{ij})^k_l=\delta^{ik}\delta^j_l-\delta^{i}_l\delta^{jk},\quad \text{Spin-1/2}: (S^{ij})^\alpha_\beta= \frac{1}{2}(\sigma^{[ij]})^\alpha_\beta
\end{equation}
all operations on other spinning fields can be obtained accordingly.

Note that in general CWI are anomalous in momentum space. A straightforward way to see this is to consider position space correlators, which satisfy CWI as long as the operators are at distinct points. However, upon  Fourier transforming divergences might appear where regularization is needed. In such a case, the right-hand sides of \eqref{eq:dilatationcwi} and \eqref{eq:boostcwi} are non zero, but acquire anomalous terms dressed by the delta function \cite{Cespedes:2020xqq,Gillioz:2022yze, Nakayama:2019mpz}. This situation occurs when the power counting in time becomes negative, see for instance \cite{Bzowski:2015pba,Goodhew:2022ayb}. In this work, we proceed under the assumption there are no anomalies. This can be argued for any theory with at least two derivatives, which will be our focus~\cite{Goodhew:2022ayb}. 

\paragraph{General Structure of the WT Identity}
\label{sec: WTgeneral}

We will focus on conserved operators, satisfying $\partial_i \mathcal{J}^i=0$. In correlation functions this is reflected in WT identities: which relate the divergence to lower point WFCs, 
\begin{equation}
    \frac{\partial}{\partial x^i}\langle\mathcal{J}_x^{i\{j_{s{-}1}\}}\mathcal{O}_1\mathcal{O}_2\rangle=\delta^3(x-x_1)\langle\delta^{\{j_{s{-}1}\}}\mathcal{O}_1\mathcal{O}_2\rangle+\delta^3(x-x_2)\langle\mathcal{O}_1\delta^{\{j_{s{-}1}\}}\mathcal{O}_2\rangle,\label{eq:wtposition}
\end{equation}
where $\delta^{\{j_{s{-}1}\}}\mathcal{O}_m$ represents the transformation of the operator with respect to the underlying symmetry. If the transformation is local,
\begin{equation}
 \langle\delta^{\{j_{s{-}1}\}}\mathcal{O}_1\mathcal{O}_2\rangle= \mathbb{M}_1^{\{j_{s{-}1}\}}\langle\mathcal{O}_1\mathcal{O}_2\rangle \label{eq:WTlocal}  
\end{equation}
where the spin-$s-1$ tensor $\mathbb{M}_m^{\{j_{s{-}1}\}}$ is constructed out of $(x_m, \partial_m)$.\footnote{We are slightly abusing the notation here, as the transformation of an operator can be proportional to another operator. Eq.(\ref{eq:WTlocal}) is to be understood as allowing such possibility.  } Upon Fourier transforming to momentum space,
\begin{equation}
p_{i}\langle\mathcal{J}_p^{i\{j_{s{-}1}\}}\mathcal{O}_1\mathcal{O}_2\rangle=\mathbb{M}_1^{\{j_{s{-}1}\}}\langle\mathcal{O}_{1+p}\mathcal{O}_2\rangle+\langle\mathcal{O}_1\mathcal{O}_{2+p}\rangle\mathbb{M}_2^{\{j_{s{-}1}\}}\label{eq:wtmomentum},
\end{equation}
where the subscripts now label the momentum dependence.

 For the stress tensor $T^{ij}$ and spin-3/2 currents $\bar\psi^i,\psi^j$, we have additional constraints that arise from taking the (gamma)trace. For example, the gamma-trace identities will take the form, 
\begin{equation}
\sigma_{i}\langle\bar\psi_p^{i}\mathcal{O}_1\mathcal{O}_2\rangle=\mathbb{M}_1\langle\mathcal{O}_{1+p}\mathcal{O}_2\rangle+\langle\mathcal{O}_1\mathcal{O}_{2+p}\rangle\mathbb{M}_2\label{eq:gammatransatz}.  
\end{equation}
These are associated with (super)Weyl symmetry, and the identity takes the form
\ie
&\text{Weyl:}\quad\delta_{ij}\langle T_p^{ij}\mathcal{O}_1\dots\mathcal{O}_n\rangle=-\sum_{m=1}^n\Delta_m\langle\dots\mathcal{O}_{m+p}\dots\rangle,\label{eq:weyltransf}
\\
&\text{SuperWeyl:}\quad\sigma_j \langle J^i\bar\psi^j\psi^k\rangle=-\langle J_{1}^iJ^k\rangle,\quad \sigma_k\langle T^{ij} \bar\psi^k\psi^l\rangle&=\frac{1}{2}\sigma^{(i}\langle\bar\psi_{-3}^{j)}\psi^l\rangle\,
\fe
where $\Delta_{m}$ is the conformal dimension of the $m$th primary operator $\mathcal{O}_{m+p}$ and for superWeyl we have listed identities for $\bar\psi$ \cite{Papadimitriou:2017kzw}. 

In our bootstrap setup, we will not assume the explicit form of $\mathbb{M}_m$. Rather, they will be fixed by CWI as well as consistency conditions which occur when multiple currents are present. For example, for $\langle T\bar\psi \psi\rangle$, the longitudinal (trace) ansatz for each current should satisfy,
\begin{equation}
    p_{1i}p_{2k}\langle T^{ij}\bar\psi^k\psi^l\rangle=p_{1i}\mathbb{A}^{ijl}_{\bar\psi_L}=p_{2k}\mathbb{A}^{jkl}_{T_L},
\end{equation}
where $\mathbb{A}^{ijl}_{\bar\psi_L}$ is the ansatz for the longitudinal part of $\bar\psi$ while $\mathbb{A}^{jkl}_{T_L}$ is the ansatz for the longitudinal part of $T$. 

Beyond their role in relating correlation functions, the WT identities can also be viewed as reflecting the residual gauge symmetries of the bulk gauge field. For details, see appendix \ref{sec:residualsym}.

\subsection{Total energy pole singularities}
\label{subsec:totalenergypole}
As the bulk geometry does not have time-like killing vectors, energy is not conserved. However, if one were to analytic continue the energy variables such that they sum to zero, one restores the full Poincaré invariance. Given how massless flatspace amplitudes in many theories are uniquely fixed by symmetry and locality considerations, one expects that the former should appear as the leading term in the total energy expansion ~\cite{Raju:2012zr, Raju:2012zs, Arkani-Hamed:2015bza}.  
 
In App.~\ref{app:LeadingTotalEnergyPoleStructure}, although the derivation is more involved than in flat space, we can show that the residue and order of the leading total-energy pole can be obtained by replacing the propagators in the Feynman rules with their asymptotic form in the far past limit and then performing the time integral. In this limit, the propagators are essentially the exponential factor $e^{iE\eta}$ dressed with a certain power of time. The relevant integral is approximately of the form
\ie
\label{rational exp integral}
\lim_{E_T \rightarrow 0} \int_0^{\infty} dz\; e^{-E_T z} z^L  \propto \begin{cases}
	\frac{1}{E_T^{L+1}},\quad L\geq 0,  \\
	\frac{1}{E_T^{L+1}}\log(E_T),\quad L<0,
\end{cases}
\fe
where $L$ is the total power of time counted from the contributions of the metric contractions at the vertices and of the propagators. The resulting singularity structure takes the form 
\ie
\label{eq: total energy pole}
\lim_{E_T\to0}c_n = \frac{P_T}{E_T^p} \times M, \quad p = [M] + (n - 3)
\fe
for $p \geq 1$. For $p <1$, the total energy pole will become a branch point and the RHS of \eqref{eq: total energy pole} should be multiplied by a logarithmic factor $\log E_T$. Here, $M$ is always the \emph{massless} amplitude, regardless of whether the field is massive or massless in (EA)dS, $[M]$ denotes its kinematical (momentum/energy/polarization) mass dimension and $n$ is the number of external legs. We define $P_T \equiv \left(\prod_{a}E_a^{k_a}\right)$, where $k_a = \Delta_a -2$ for bosonic external legs and $k_a = \Delta_a -3/2$ for fermionic external legs, which represents the polynomial prefactor. The details of this calculation can be found in App.~\ref{app:LeadingTotalEnergyPoleStructure}.

Note that, as stressed in \cite{Flatspace}, the amplitude $M$ could be rewritten by the transverse-traceless polarization and the spinor boundary condition,
\ie
\label{eq:amp transform to boundary}
\epsilon_{\mu_1} \epsilon_{\mu_2} \dots M^{\mu_1 \mu_2 \dots}
=
\epsilon^T_{i_1} \epsilon^T_{i_2} \dots M^{i_1 i_2 \dots}\,, \quad
u =
(1 - i\vecslashed{\hat{p}})\bm{\chi}_{\zero}\,, \quad
\bar u =
\bm{\bar\chi}_{\zero}(1 + i\vecslashed{\hat{p}}).
\fe
Then under our decomposition,  if we require the leading total energy pole to be the amplitude as \eqref{eq: total energy pole}, it thus should only appear in the transverse-traceless part. That is the reason why we write ansatz of this part with the total energy pole order beginning at $p$ in Sec. \ref{sec3}.

\subsection{Cutting Rules and Partial Energy Pole Singularities}
\label{subsec:cuttingrules}
The cutting rules describe the universal properties of bulk-to-bulk propagator under the flipping of internal energies. In particular, we have \cite{Meltzer:2021zin,Baumann:2020dch,Goodhew:2020hob}:
\ie
\label{eq:disc of boson G}
\disc_{E_s}  G(E_s, t, t') = \disc_{E_s} K_\phi(E_s, t) \cdot \left(-\frac{i\,C_2(\vec p_s)}{2 E_s^{2\Delta_s-3}}\right) \cdot \disc_{E_s} K_\phi(E_s, t').
\fe
where we use the definition 
\ie
\disc_{E_s} f(E_s)
	&:=  f(E_s) - f(-E_s)\,. \label{discofcorr}
\fe
Here $\Delta_s$ is the conformal dimension of the exchanged scalar field. For general spins, including half-integers, they take the form
\ie
 	\label{discCOT}
	\disc_{E_s} c_{4, s}
	=
	\disc_{E_s}c_{3,L,i_1 \dots }({p}_{1} ,{p}_{2} ,{p}_{s}) \cdot \frac{C_{2}^{ i_1 j_1 \dots}({p}_{s})}{2 E_s^{2\Delta_s-3}}\cdot\disc_{E_s}c_{3,R, j_1 \dots}(-{p}_{s} ,{p}_{3} ,{p}_{4}),
\fe
for the scalars and conserved operators with the individual gluing factors,
\ie
\label{gluing factor of fermion}
C_{2,\phi}({p}_{s}) = 1, \;
C_{2,J}^{i_1 j_1}(\vec p_s) = \pi_{s}^{i_1j_1}, \;
C_{2,T}^{i_1 i_2 j_1 j_2}(\vec p_s) = \Pi_{s,(2,2)}^{i_1 i_2 j_1 j_2}\,.
\fe
The fermion cutting rules take the same form as the above with extra projectors and
\begin{equation}
C_{2,\chi}(\vec p_s) = i\vecslashed{\hat p}_{s}, \;
C_{2,\psi}^{ij}(\vec p_s) = i\hat{\Pi}^{ij}_{s} \vecslashed{\hat p}_{s}\,.
\end{equation}
For example for spin-3/2 we have
\begin{equation}
	\label{ds fermion discCOT}
	\begin{aligned}
	&\disc_{E_s} c_{4,s}
	=
	\disc_{E_s} c_{3, A, i} (\vec p_1, \vec p_2, E_s) 
	\cdot \left[\frac{1+i\gamma_0}{2} \cdot \frac{i\hat{\Pi}^{ij}_s \vecslashed{\hat p}_{s}}{2 E_s^2} \cdot \frac{1-i\gamma_0}{2}\right]^{AB}
	\disc_{E_s} c_{3, B, j} (\vec p_3, \vec p_4, E_s).
	\end{aligned}
\end{equation}
We present the derivation for the above in App. \ref{app: Feynman Rule from boundary action}, where we use the equation of motion (EOM) that the propagator satisfies. This approaches streamlines the derivation (EA)dS space without the need for the explicit solution. Moreover, given the fact that the in-in correlators can be built up by the WFCs, the discontinuity of the WFCs then controls the discontinuity of the in-in correlators \cite{Goodhew:2020hob, Das:2025qsh} (see the appendix C of \cite{Das:2025qsh} for a detailed discussion).

The cosmological cutting rule~\eqref{discCOT} enables one to \emph{fully determine} both the leading and subleading residues of the partial energy poles of $c_4$. This is because, on the left-hand side, the $c_4|_{E_s\to -E_s}$ does not contain $E_{L/R}^e$ poles due to the energy flipping in the internal energy $E_s$. Thus, taking the residue at a partial energy pole on both sides gives, for the $s$-channel as an example,
\begin{eqnarray}
	\label{eq: bosonic partial energy pole}
    \lim_{E_{L}^s\to0} c_4
    &=& \left(  \lim_{E_{L}^s\to0}c_{3L} \right)\cdot \frac{C_2}{2 E_s^{2\Delta_s-3}}\cdot \disc_{E_s} c_{3R}\,.
\end{eqnarray}
Note that since the CWI requires the 3-pt WFC to have subleading total energy poles, this translates to a series of subleading partial energy poles $E_L^s$ above.

Note that from the singularity structure of eqs.~(\ref{discCOT}) and 
(\ref{eq: bosonic partial energy pole}) one can deduce several non-trivial
consistency conditions. In particular, since the left-hand side of 
eq.~(\ref{eq: bosonic partial energy pole}) is the four-point WFC, which must 
vanish upon imposing the CWI, the equation becomes a constraint on the 
right-hand side that is solved order by order in the partial-energy pole 
\(E_L\). For eq.~(\ref{discCOT}), the key observation is that \(E_s\) does not correspond 
to a pole on the left-hand side. This imposes non-trivial constraints on the 
expansion of the three-point WFC in powers of \(E_s\) that appear on the 
right-hand side. This diagnostic is known as the \emph{manifest locality test} 
(MLT)~\cite{Jazayeri:2021fvk, Bonifacio:2022vwa}.

Since the discontinuities of WFCs are odd functions of \(E_s\), the first 
non-vanishing constraint from eq.~(\ref{discCOT}) arises at \(\Delta = 3\) for 
scalars, while for fermions the first non-trivial constraint appears at 
\(\Delta = 5/2\). To see this, consider the 3-pt WFC involving a spin-3/2, we have
\begin{equation}
 \disc_{E_s} c_3^i = \left[
    \left(c_{3,\mathcal{I}_s,j}(E_s) - c_{3,\mathcal{I}_s,j}(-E_s)\right) + \left(c_{3,\mathcal{P}_s,j}(E_s) + c_{3,\mathcal{P}_s,j}(-E_s)\right) \vecslashed{\hat p}_s  
 \right]
 \hat{\Pi}^{ji}_s  
\end{equation}
where we have decomposed
\begin{equation}
    c_{3}^i = \left( c_{3,\mathcal{I}_s,j} + c_{3,\mathcal{P}_s,j} \vecslashed{\hat p}_s \right) \hat{\Pi}^{ji}_s.
\end{equation}
Here, the second term is proportional to $\vecslashed{\hat p}_s$. Note that in this procedure, only the energy $E_s$ is flipped, while the spatial vector $\vec{p}_s$ remains fixed, so $\hat{p}_s = \vec{p}_s / E_s$ is likewise flipped.  Consequently, only the function multiplying $\vecslashed{\hat p}_s$ is \emph{even} in $E_s$, and the cancellation of the $1/E_s^2$ pole imposes:
\begin{equation}
    \left. c_{3,\mathcal{P}_{5/2},j} \right|_{E_{5/2}=0} = 0,
    \label{eq:gravitinomlt}
\end{equation}
where $E_{5/2}=E_s$ is the energy of the $\Delta = 5/2$ state. 

In our bootstrap procedure, the 3-pt WFC is fixed before implementing the above constraints, and thus the latter will serve as a consistency check of the result.

\section{Bootstrapping the WFCs}
\label{sec3}

Our bootstrap procedure can be summarized by the following schematic:
\begin{tcolorbox}
    \ie
\text{2pt}:&\quad\text{Ansatz (General $\Delta$)}\rightarrow\text{CWI}\rightarrow\text{WT : $\Delta$ for Conserved Currents}
\\
\text{3pt}:&\quad\text{Ansatz (Fixed $\Delta$)}\rightarrow\text{WT Consistency (Multiple Currents)}\rightarrow\text{CWI}\\
\text{4pt}:&\quad\text{Match $E_{L/R}$ and leading $E_T$ residues}\\
&\rightarrow\text{Ansatz for Subleading term in $E_T$}\rightarrow\text{MLT/soft limit}\rightarrow\text{CWI}
\fe
\end{tcolorbox}
\paragraph{2-point WFCs}
For 2-point functions with general conformal dimension, we simply put the most general ansatz compatible with the momentum dimension given by \eqref{eq:dilatationcwi}. Not surprisingly, CWI fixes the two-point function up to position space delta functions, which can be removed by contact terms. 

\paragraph{3-point WFCs}
At 3-pts, since we will be working with conserved currents $\mathcal{J}$, we require that the longitudinal and trace components are proportional to 2-pt functions as \eqref{eq:wtmomentum}. For the pure transverse piece, the leading order of the total energy pole, $p$, is given by \eqref{eq:contactleadingresidue}. In general, takes the form,
\begin{equation}
    \mathbb{A}^{T}= \frac{\mathbb{M}\otimes\text{Poly}\left(E_T,E_2,E_3\right)}{E_T^p}, \label{eq:3pttransverseansatz}
\end{equation}
where $\mathbb{M}$ represents different tensor structures each dressed with a polynomial Poly$\left(E_T,E_2,E_3\right)$, where $p{+}3$ minus the combined momentum dimension of $\mathbb{M}$ and Poly must be equal to the sum of momentum dimensions given by \eqref{eq:dilatationcwi}. 

Now, after constructing the ansatz, we can apply constraints. For WFCs with a single current ($\langle J\bar\chi\chi\rangle, \langle T\bar\chi\chi\rangle$), the only consistency condition we apply is the CWI. For WFCs involving multiple currents ($\langle J\bar\psi\psi\rangle, \langle T\bar\psi\psi\rangle$), we will first reduce the ansatz by WT self-consistency outlined in section \ref{sec: WTgeneral}. We will find that CWIs are sufficient to determine the ansatz up to contact terms. Importantly, \textit{the flat-space amplitude is not needed as a seed}, and the result automatically reproduces the flat-space amplitude on the leading total energy pole. Furthermore, the precise form of WT identity as well as the passing of MLT is fully determined.

\paragraph{4-point  WFCs}
We begin our analysis of the 4-pt WFCs by systematically constructing the transverse, traceless component from the residues of the partial-energy poles and the leading total-energy pole, using the flat-space amplitude and the previously bootstrapped three-point WFCs as input data. This procedure completely determines \emph{all} contributions carrying partial-energy poles. The remaining part consists solely of subleading total-energy poles, for which we introduce an ansatz. The corresponding coefficients are then fixed by imposing the CWI on the full WFC, including its longitudinal and trace components. Schematically the transverse-traceless WFC takes the form, 
\ie
\label{eqn: 4-point WFC ansatz}
c^{\hat T}_4=\left\{\sum_{e \in s,t,u}\left(\frac{A_R^e}{(E_R^e)^{p_{R}}}+\frac{B_L^e}{(E_L^e)^{p_{L}}}\right)+\frac{C}{E_T^{p}}+\frac{ D_{odd}}{E_T^{p-1}}\right\}+\frac{D_{even}}{E_T^{p-1}}
\fe
where $D_{even}$ are polynomials of even degree in energy variables, while $D_{odd}$ are odd in at least one energy variable.  Terms in the curly brackets are completely determined by the inputs as mentioned earlier and $D_{even}$ is the only undetermined part.
Here, $p_{R},p_{L},p$ are the leading total energy pole order of the left, right 3-pt  and the 4 pt WFC, respectively, 
\ie
p_{R} = [ M_R], \quad
p_{L} = [ M_L], \quad
p = [ M_4] + 1= p_{R} + p_{L} -1
\fe
and we focus on the case where $p \geq p_{R}, p_{L} \geq 1$. Let us now see how each structure is related to the constraints. 

Note that our procedure is reminiscent of the recursion relations developed in \cite{Jazayeri:2021fvk,Bonifacio:2022vwa}.  There, complex deformations for the partial energies were introduced such that only partial- and not total-energy poles are exposed. The WFC is then built from the residues of the partial energy poles as well as terms arising from the pole at infinity in the deformation parameter. The latter is determined by a variety of constraints, including the cancellation of spurious poles, MLT as well as soft-constraints. In comparison, we construct our WFC with manifest physical singularities.

We now give in detail the precise form of each term in the curly brackets of eq.(\ref{eqn: 4-point WFC ansatz}).

\begin{enumerate}

    \item \emph{Matching Partial Energy Poles} ($A^e_{R},B_L^e$)

	We begin by matching the term singular in the partial energy pole $E_R^e$ in each channel. The resulting leading residue 
    will by the product of the  amplitude and the discontinuity of WFCs on the other side. Importantly on the support of $E_R^e \rightarrow 0$, 
    \begin{equation}
(E_L^e)^{-m}
 \Bigg|_{E_e\rightarrow -E_e}
=
(E_T-E_R^e)^{-m}\,
= (E_T)^{-m} \left[\sum_{n=0}^{\infty} \mathbb{C}^{m+n-1}_n \, \left(\frac{E_R^e}{E_T}\right)^n\right] ,\label{eq:ELtoET}
    \end{equation}
where $\mathbb{C}^m_n$ is the binomial coefficient. Thus the discontinuity of the WFC will introduce total energy pole. Let us firstly focus on the $E_R^e\to0$ limit. After using the result in Sec. \ref{subsec:cuttingrules}, we find

	\ie
    \label{eq:partialenergypole_4pt_L}
   \lim_{E_R^e \rightarrow 0}  c_4^{\hat T}
         &=
         \left(\frac{P_{T,L} M_{L}}{(E_{L}^e)^{p_{L}}} + \sum_{q=1}^{p_{L}-1}\frac{N_{L}^{(q)}}{(E_{L}^e)^{q}}\right)\cdot\frac{C_2^e}{2E_e^{2\Delta_e -3}}
         \cdot
         \left(\frac{P_{T,R} M_{R}}{(E_{R}^e)^{p_{R}}} + \sum_{q=1}^{p_{R}-1}\frac{N_{R}^{(q)}}{(E_{R}^e)^{q}}\right)\\
         &-
         \frac{P'_{T,L} M_{L}}{(E_T)^{p_{L}}}
         \left[
             \sum_{n=0}^{p_R-1} \mathbb{C}^{p_{L}+n-1}_n \left(\frac{E_R^e}{E_T}\right)^n\right]\cdot \frac{C_2^e}{2E_e^{2\Delta_e -3}}
             \cdot
             \frac{P_{T,R}M_R}{(E_{R}^e)^{p_R}}
         \\
         &-
         \sum_{q=1}^{p_{R}-1} \frac{P'_{T,L} M_{L}}{(E_T)^{p_{L}}}
         \left[
             \sum_{n=0}^{q-1} \mathbb{C}^{p_{L}+n-1}_n \left(\frac{E_R^e}{E_T}\right)^n\right]\cdot \frac{C_2^e}{2E_e^{2\Delta_e -3}}
             \cdot
             \frac{N^{(q)}_R}{(E_{R}^e)^{q}}
         \\
         &-
         \sum_{q=1}^{p_{L}-1}
            \frac{N_{L}^{(q)}}{(E_T)^{q}}
            \left[
                \sum_{n=0}^{p_R-1} \mathbb{C}^{q+n-1}_n \, \left(\frac{E_R^e}{E_T}\right)^n\right]\cdot
                \frac{C_2^e}{2E_e^{2\Delta_e -3}}
            \cdot
            \frac{P_{T,R}M_R}{(E_{R}^e)^{p_R}}
         \\
         &-
         \sum_{q_L=1}^{p_{L}-1} \sum_{q_R=1}^{p_{R}-1}
            \frac{N_{L}^{(q_L)}}{(E_T)^{q_L}}
            \left[
                \sum_{n=0}^{q_R-1} \mathbb{C}^{q_L+n-1}_n \, \left(\frac{E_R^e}{E_T}\right)^n\right]\cdot
                \frac{C_2^e}{2E_e^{2\Delta_e -3}}
                \cdot\frac{N_{R}^{(q_R)}}{(E_{R}^e)^{q_R}}\\
            &+P_L\cdot \frac{C_2^e}{2E_e^{2\Delta_e -3}}
            \cdot
            \left(
               \frac{P_{T,R} M_{R}}{(E_{R}^e)^{p_{R}}} 
               + 
               \sum_{q=1}^{p_{R}-1}\frac{N_{R}^{(q)}}{(E_{R}^e)^{q}}
            \right)\\
	\fe
where we have reorganized product of discontinuities of 3-pt WFCs based on the singularity structure and the presence of the total energy poles is due to \eqref{eq:ELtoET}. Specifically, the $P_{T,L}M_L,\;P_{T,R}M_R$ correspond to the leading residue of the left/ right 3-pt WFCs as denoted in \eqref{eq: total energy pole}, $N^{(q)}_{L}$ which is the $q$-th order subleading residue of the left depends only on $(E_1,E_2)$ while $N^{(q)}_{R}$ depends only on $(E_3,E_4)$. Moreover, we define $
P_{T,L}^{'}\equiv P_{T,L}\big|_{E_e \to -E_e}$ and $P_{L/R}:=\sum_{q'=0}^{\infty} N_{L/R}^{(-q')} E_{L/R}^{q'}-N_{L/R}^{(-q')} (E_{L/R}|_{E_e \to -E_e})^{q'}$ which is the polynomial terms in the left/right part of the cutting without any singularity.\footnote{
    The Left/Right 3-pt WFCs now are written by $P_T,M,N$ as an expansion of $E_L/E_R$. For example,
    \ie
    c_{3,L}=\frac{P_{T,L} M_{3,L}}{E_L^{p_L}}+\sum_{q=1}^{p_L-1}\frac{N^{(q)}_L}{E_L^q} +\sum_{q'=0}^{\infty} N^{(-q')}_{L} E_L^{q'},
    \fe
} Importantly, these terms are completely determined in the bootstrap of 3-pt WFC.

$c_4^{\hat{T}}$ under $E_R^e\to 0$ has a similar expansion and we provide its explicit form below for completeness,
\ie
\lim_{E_L^e \rightarrow 0} c_4^{\hat T}
            &=
            \left(\frac{P_{T,L} M_{L}}{(E_{L}^e)^{p_{L}}} + \sum_{q=1}^{p_{L}-1}\frac{N_{L}^{(q)}}{(E_{L}^e)^{q}}\right)\cdot\frac{C_2^e}{2E_e^{2\Delta_e -3}}
            \cdot
            \left(\frac{P_{T,R} M_{R}}{(E_{R}^e)^{p_{R}}} + \sum_{q=1}^{p_{R}-1}\frac{N_{R}^{(q)}}{(E_{R}^e)^{q}}\right)\\
            &-
            \frac{P_{T,L}M_L}{(E_{L}^e)^{p_L}}
            \cdot
            \frac{C_2^e}{2E_e^{2\Delta_e -3}}
            \cdot
            \frac{P'_{T,R} M_{R}}{(E_T)^{p_{R}}}
            \left[
                \sum_{n=0}^{p_L-1} \mathbb{C}^{p_{R}+n-1}_n \left(\frac{E_L^e}{E_T}\right)^n\right]
            \\
            &-
            \sum_{q=1}^{p_{R}-1}
            \frac{N^{(q)}_L}{(E_{L}^e)^{q}}
            \cdot \frac{C_2^e}{2E_e^{2\Delta_e -3}}
            \cdot
            \frac{P'_{T,R} M_{R}}{(E_T)^{p_{R}}}
            \left[
                    \sum_{n=0}^{q-1} \mathbb{C}^{p_{R}+n-1}_n \left(\frac{E_L^e}{E_T}\right)^n
            \right]
            \\
            &-
            \sum_{q=1}^{p_{R}-1}
                \frac{P_{T,L}M_L}{(E_{L}^e)^{p_L}}
               \cdot
                   \frac{C_2^e}{2E_e^{2\Delta_e -3}}
               \cdot
               \frac{N_{R}^{(q)}}{(E_T)^{q}}
               \left[
                   \sum_{n=0}^{p_L-1} \mathbb{C}^{q+n-1}_n \, \left(\frac{E_L^e}{E_T}\right)^n\right]
            \\
            &-
            \sum_{q_L=1}^{p_{L}-1} \sum_{q_R=0}^{p_{R}-1}
               \frac{N_{L}^{(q_L)}}{(E_L^e)^{q_L}}
               \cdot
                   \frac{C_2^e}{2E_e^{2\Delta_e -3}}
                   \cdot\frac{N_{R}^{(q_R)}}{(E_{T})^{q_R}}
                \left[
                    \sum_{n=0}^{q_L-1} \mathbb{C}^{q_R+n-1}_n \, \left(\frac{E_L^e}{E_T}\right)^n\right]\\
               &+\left(
                \frac{P_{T,L} M_{L}}{(E_{L}^e)^{p_{L}}} 
                + 
                \sum_{q=1}^{p_{L}-1}\frac{N_{L}^{(q)}}{(E_{L}^e)^{q}}
             \right)
             \cdot 
             \frac{C_2^e}{2E_e^{2\Delta_e -3}}
               \cdot
            P_R.
\fe

We can readily determine $A_R^e$ by matching \eqref{eq:partialenergypole_4pt_L}:
	\ie
		A_{R}^e
        &=
        \left(\frac{P_{T,L} M_{L}}{(E_{L}^e)^{p_{L}}} + \sum_{q=1}^{p_{L}-1}\frac{N_{L}^{(q)}}{(E_{L}^e)^{q}}\right)\cdot\frac{C_2^e}{2E_e^{2\Delta_e -3}}
         \cdot
         \left(P_{T,R} M_{R} + \sum_{q=1}^{p_{R}-1}N_{R}^{(q)}(E_R^e)^{p_R-q}\right)\\
         &-
         \frac{P'_{T,L} M_{L}}{(E_T)^{p_{L}}}
         \left[
             \sum_{n=0}^{p_R-1} \mathbb{C}^{p_{L}+n-1}_n \left(\frac{E_R^e}{E_T}\right)^n\right]\cdot \frac{C_2^e}{2E_e^{2\Delta_e -3}}
             \cdot
             P_{T,R}M_R
         \\
         &-
         \sum_{q=1}^{p_{R}-1} \frac{P'_{T,L} M_{L}}{(E_T)^{p_{L}}}
         \left[
             \sum_{n=0}^{q-1} \mathbb{C}^{p_{L}+n-1}_n \left(\frac{E_R^e}{E_T}\right)^n\right]\cdot \frac{C_2^e}{2E_e^{2\Delta_e -3}}
             \cdot
             N^{(q)}_R
             (E_R^e)^{p_R-q}
         \\
         &-
         \sum_{q=1}^{p_{L}-1}
            \frac{N_{L}^{(q)}}{(E_T)^{q}}
            \left[
                \sum_{n=0}^{p_R-1} \mathbb{C}^{q+n-1}_n \, \left(\frac{E_R^e}{E_T}\right)^n\right]\cdot
                \frac{C_2^e}{2E_e^{2\Delta_e -3}}
            \cdot
            P_{T,R}M_R
         \\
         &-
         \sum_{q_L=1}^{p_{L}-1} \sum_{q_R=1}^{p_{R}-1}
            \frac{N_{L}^{(q_L)}}{(E_T)^{q_L}}
            \left[
                \sum_{n=0}^{q_R-1} \mathbb{C}^{q_L+n-1}_n \, \left(\frac{E_R^e}{E_T}\right)^n\right]\cdot
                \frac{C_2^e}{2E_e^{2\Delta_e -3}}
                \cdot N_{R}^{(q_R)} (E_R^e)^{p_R-q_R}\\
            &+P_L\cdot \frac{C_2^e}{2E_e^{2\Delta_e -3}}
            \cdot
            \left(
               P_{T,R} M_{R} 
               + 
               \sum_{q=1}^{p_{R}-1}N_{R}^{(q)}(E_R^e)^{p_R-q}
            \right)
	\fe
	Then, it is straightforward to write $B_L^e$ based on $A_R^e$ to match the expected singular term in the other partial energy pole, $E_L^e$:
	\ie
	B_{L}^e
        &=
        -P_{T,L}M_L
            \cdot
            \frac{C_2^e}{2E_e^{2\Delta_e -3}}
            \cdot
            \frac{P'_{T,R} M_{R}}{(E_T)^{p_{R}}}
            \left[
                \sum_{n=0}^{p_L-1} \mathbb{C}^{p_{R}+n-1}_n \left(\frac{E_L^e}{E_T}\right)^n\right]
            \\
        &-\sum_{q=1}^{p_{R}-1}
            N^{(q)}_L (E_{L}^e)^{p_L-q}
            \cdot 
            \frac{C_2^e}{2E_e^{2\Delta_e -3}}
            \cdot
            \frac{P'_{T,R} M_{R}}{(E_T)^{p_{R}}}
            \left[
                    \sum_{n=0}^{q-1} \mathbb{C}^{p_{R}+n-1}_n \left(\frac{E_L^e}{E_T}\right)^n
            \right]
            \\
            &-
            \sum_{q=1}^{p_{R}-1}
                P_{T,L}M_L
               \cdot
                   \frac{C_2^e}{2E_e^{2\Delta_e -3}}
               \cdot
               \frac{N_{R}^{(q)}}{(E_T)^{q}}
               \left[
                   \sum_{n=0}^{p_L-1} \mathbb{C}^{q+n-1}_n \, \left(\frac{E_L^e}{E_T}\right)^n\right]
            \\
            &-
            \sum_{q_L=1}^{p_{L}-1} \sum_{q_R=1}^{p_{R}-1}
               N_{L}^{(q_L)}E_L^{p_L-q_L}
               \cdot
                   \frac{C_2^e}{2E_e^{2\Delta_e -3}}
                   \cdot \frac{N_{R}^{(q_R)}}{(E_T)^{q_R}}
                \left[
                    \sum_{n=0}^{q_L-1} \mathbb{C}^{q_R+n-1}_n \, \left(\frac{E_L^e}{E_T}\right)^n\right]\\
            &+\left(
                P_{T,L} M_{L}
                + 
                \sum_{q=1}^{p_{L}-1}N_{L}^{(q)}(E_L^e)^{p_L-q}
                \right)
                \cdot 
                \frac{C_2^e}{2E_e^{2\Delta_e -3}}
                \cdot
                P_R,
	\fe

    Again $A_{R}^e$ and $B_{L}^e$ has poles in $E_e$ and their cancellation corresponds to the MLT which puts constraints on the 3-pt WFCs. However the since the latter is fully determined, these poles are guaranteed to cancel.  
    
    \item \emph{Matching Amplitude-Limit} ($C$)
    
	Next, we examine the leading $E_T$ behavior of the WFCs constructed from $A_R^e$ and $B_L^e$ as $E_T \rightarrow 0$:
	\ie
	\label{eqn: total pole of A B}
	&\lim_{E_T \rightarrow 0} 
	\sum_{e \in s,t,u}\left(\frac{A_R^e}{(E_R^e)^{p_{R}}}+\frac{B_L^e}{(E_L^e)^{p_{L}}}\right)\\ 
	&=
	{-}\sum_{e \in s,t,u}
    \frac{\mathbb{C}^{p_{L}+p_{R}-2}_{p_{L}-1}}{E_T^{p}}
    \cdot
    \left(
        P'_{T,L} M_{L} \cdot
        \frac{C_2^e}{2E_e^{2\Delta_e -3}}
        \cdot
        \frac{P_{T,R} M_{R}}{E_R^e}
        {+}
        \frac{P_{T,L} M_{L}}{E_L^e}
        \cdot
        \frac{C_2^e}{2E_e^{2\Delta_e -3}}
        \cdot
        P'_{T,R} M_{R}
    \right)\\
	&=
	{-}\sum_{e \in s,t,u}
    \frac{(-1)^{k_e}\mathbb{C}^{p_{L}+p_{R}-2}_{p_{L}-1} P_{T}}{E_T^{p}}
    \cdot
    \left(
        M_{L} \cdot
        \frac{C_2^{(flat),e}}{2E_e}
        \cdot
        \frac{M_{R}}{E_R^e}
        {+}
        \frac{M_{L}}{E_L^e}
        \cdot
        \frac{C_2^{(flat),e}}{2E_e}
        \cdot
         M_{R}
    \right)\\
    &\equiv
    \frac{P_T}{E_T^p} \cdot M_{\text{fact}}
	\fe
    in which we already use $p=p_{L}+p_{R}-1$ and insert the explicit form of the $P_T$. Here, the $C_2^{(flat),e}$ is the gluing factor in the flat space cutting rule \cite{Flatspace} for \emph{massless} particles. As shown in \cite{Flatspace}, the expression $M_{\text{fact}}$ already aligns with the amplitude at the factorization pole. Then the discrepancy between the leading total energy pole residue and the amplitude is the contact term, which we address by including $\frac{C}{E_T^{p}}$ in the ansatz \eqref{eqn: 4-point WFC ansatz} and reads
    \ie
	C =
	P_T\cdot\left[
		M_4
	- 
		\lim_{E_T \rightarrow 0} M_{\text{fact}}
	\right].
	\fe
Let us pause here to emphasize that, the amplitude encoded in the leading $E_T$ pole residue are \emph{massless}. 
    \item \emph{Back to the Cutting Rules} ($D_{odd}$)

    In the previous steps, we already match the singular parts of the cutting rules  \eqref{discCOT} (i.e. the partial energy poles), now we return to the polynomial part of the cutting rules in $s,t,u$ one by one to verify their validity. If they are not satisfied, we add the subleading terms one after another without partial energy poles to restore the consistency. In the end, we could collect all the subleading terms into $D^e_{odd}$ and 
    \ie
    &\frac{D_{odd}}{E_T^{p-1}} 
    = \frac{1}{2}\left( 
        \disc_{E_s} c_L \cdot \frac{C_2^s}{2E_s^{2\Delta_s-3}} \cdot \disc_{E_s} c_R 
        -
        \disc_{E_s} \left( \frac{A_R^s}{(E_R^s)^{p_{R}}}+\frac{B_L^s}{(E_L^s)^{p_{L}}}+\frac{C}{E_T^{p}}\right)
        \right)\\
    &+ \frac{1}{2}\left( 
            \disc_{E_t} c_L \cdot \frac{C_2^t}{2E_t^{2\Delta_t-3}} \cdot \disc_{E_t} c_R 
            -
            \disc_{E_t} \left( \frac{A_R^t}{(E_R^t)^{p_{R}}}+\frac{B_L^t}{(E_L^t)^{p_{L}}}+\frac{C}{E_T^{p}}
            +\frac{D^s_{odd}}{E_T^{p-1}}
            \right)
            \right)\\
    &+ \frac{1}{2}\left( 
            \disc_{E_u} c_L \cdot \frac{C_2^u}{2E_u^{2\Delta_u-3}} \cdot \disc_{E_u} c_R 
            -
            \disc_{E_u} \left( \frac{A_R^u}{(E_R^u)^{p_{R}}}+\frac{B_L^u}{(E_L^u)^{p_{L}}}+\frac{C}{E_T^{p}}
            +\frac{D^s_{odd}+D^t_{odd}}{E_T^{p-1}}
            \right)
            \right)
    \fe

    \item \emph{CWI, MLT, and Soft limits} ($D_{even}$ and longitudinal-trace sector)
    
So far, all factors have been written entirely in terms of the input data. The remaining factor, $D_{\text{even}}$, together with the longitudinal–trace sector, must instead be obtained from an ansatz. This ansatz is then fixed by imposing the CWI. In favorable cases, the size of the ansatz for $D_{\text{even}}$ can first be reduced by enforcing the MLT or soft limits.\footnote{For theories containing fields with a shift symmetry (for example, a massless scalar minimally coupled to gravity), the WFCs must vanish when the momentum of any such field is taken to zero. For external conserved spins $3/2$ and $2$, one has manifest locality, as discussed at the end of Sec.~\ref{subsec:cuttingrules}.} For instance, in the case of $\langle TTTT \rangle$ generated by bulk Einstein gravity studied in~\cite{Bonifacio:2022vwa}, the MLT reduces the undetermined parameters in $D_{\text{even}}$ to four, which can be identified with (non-local) field redefinitions. The ansatz for the longitudinal–trace sector is in turn partially inherited from the three-point bootstrap, from which the linear part of the symmetry transformation can be extracted. Consequently, only terms proportional to the two-point function are unknown, capturing the non-linear part of the symmetry transformation. This completes the construction of the ansatz for $D_{\text{even}}$, together with the longitudinal–trace terms. If no solution to the CWI can be found within these ansatz, then the theory is inconsistent.

\end{enumerate}

\subsection{2-pt and 3-pt WFCs}

As we will be interested in fermion WFCs, the function carries space time vector as well as spinor indices. The indices can be carried by (unit) momentum, Pauli matrices $\sigma$, Kronecker delta $\delta$ and the 3D Levi-Civita symbol $\epsilon$. 
Since products of more than one $\sigma$ can always be reduced,  the WFC with two fermions can be decomposed into 4 independent basis depending on whether it contains $\sigma$ or $\epsilon$ tensors or both, thus each ansatz $\mathbb{A}^g$ in \eqref{eq:generalPdecompose} can be written as (suppressing all indices),
\begin{equation}
    \mathbb{A}^g= \mathbb{A}^g_{\mathbb{I}}+\mathbb{A}^g_{\sigma}+\mathbb{A}^g_{\epsilon\cdot\sigma}+\mathbb{A}^g_{\epsilon\mathbb{I}}, \label{eq:4indep}
\end{equation}
where $\mathbb{I}$ is the 2-dimensional identity matrix and $\epsilon\cdot\sigma\equiv\epsilon_{ijk}\sigma^k$.\footnote{Since $\epsilon^{ijk}\sigma^l=\left(\delta^{il}\epsilon^{jkh}+\delta^{kl}\epsilon^{ijh}+\delta^{jl}\epsilon^{kih}\right)\sigma_h$, for product of $\epsilon$ and $\sigma$ it is sufficient to consider $\epsilon\cdot\sigma$.}

\subsubsection{2-pt}
\label{sec: 2ptfunction}
Let us firstly begin by bootstrapping the spin-1/2 and 3/2 2-point WFCs. Since in position space, the two-point function (at separate points) will be diagonal in conformal dimension as space-time representation \cite{Osborn:1993cr}, here we will only consider such cases.  

\paragraph{$\langle\bar{\chi}\chi\rangle$} 

The most general ansatz with the right mass weight is
\begin{equation}
\langle\bar{\chi}_p\chi_{-p}\rangle=E^{2\Delta-3}\left( a_0+a_1 \slashed{\hat{p}}\right),
\end{equation}
where $a_0,a_1$ are the coefficients and the CWI fixes it to,
\begin{equation}
\langle\bar{\chi}_p\chi_{-p}\rangle=E^{2\Delta-3}\left( a_0\delta_{\Delta,\frac{3}{2}}+a_1 \slashed{\hat{p}}\right), 
\end{equation}
Note that for $\Delta=\frac{3}{2}$, the term proportional to $a_0$ corresponds to a delta function $\delta^3(x)$ in position space. For two-point functions satisfying $\Delta_1+\Delta_2=d$, these are independent conformal invariants~\cite{Nakayama:2019mpz, Sun:2021thf} and can be removed by local contact terms at the boundary~\cite{Maldacena:2002vr}. Since we focus on the structure of the WFCs, all results are reported using an overall normalization of 1:
\begin{tcolorbox}
    \begin{equation}
\langle\bar{\chi}_p\chi_{-p}\rangle\equiv E^{2\Delta-3}\slashed{\hat{p}}.  \label{eq:chichi} 
\end{equation}
\end{tcolorbox}

\paragraph{$\langle \bar{\psi}\psi\rangle$}

For the spin-3/2 2-point function, with arbitrary conformal dimensions, since the WFC is decomposed into the gamma-trace less and gamma trace part, with the latter just being a spin-1/2, we will consider the gamma trace-less two-point function,
\begin{equation}
\sigma_i \langle\bar{\psi}^i\psi^j\rangle=\langle\bar{\psi}^i\psi^j\rangle\sigma_j=0.   
\end{equation}

\paragraph{CWI}
The most general gamma-traceless ansatz for all $\Delta$ is,
\begin{equation}
\langle\bar{\psi}_p^i\psi^j_{-p}\rangle=E^{2\Delta-3}\Pi_\sigma^{ii'}\left(a_1\delta_{i'j'}+a_2\hat{p}_{i'}\hat{p}_{j'}+a_3\delta_{i'j'}\slashed{\hat{p}}+a_4\hat{p}_{i'}\hat{p}_{j'}\slashed{\hat{p}}\right)\Pi_\sigma^{j'j}, 
\end{equation}
where $\Pi_\sigma^{ij}\equiv\delta^{ij}-\frac{1}{3}\sigma^i\sigma^j$ is the gamma-traceless projector.\footnote{One might wonder why don't we add the two further terms $i\epsilon_{i'j'k}\sigma^k, i\epsilon_{i'j'k}\hat{p}^k$. Because when contracted with the projectors, they become minus the terms labeled by $a_1,a_3$, respectively.} CWI fixes the ansatz into
\ie
&\langle\bar{\psi}_p^i\psi^j_{-p}\rangle
 \\
 &=E^{2\Delta-3}\Pi_\sigma^{ii'}\left\{a_3\left(\delta_{i'j'}\slashed{\hat{p}}-\frac{4(\Delta-2)}{2\Delta-1}\hat{p}_{i'}\hat{p}_{j'}\slashed{\hat{p}}\right)+a_1\left(\delta_{\Delta,\frac{3}{2}}\delta_{i'j'}+\delta_{\Delta,\frac{5}{2}}\left(\delta_{i'j'}-\frac{3}{2}\hat{p}_{i'}\hat{p}_{j'}\right)\right)\right\}\Pi_\sigma^{j'j}, \label{eq:psipsicwi}
\fe
where the result shows that when the conformal dimension equals to $\frac{3}{2}$ or $\frac{5}{2}$ there are two additional terms allowed. Note that the term labeled by $a_3$ in \eqref{eq:psipsicwi} coincides with the previous bulk calculation via the AdS/ CFT correspondence \cite{Corley:1998qg,Volovich:1998tj}. 
The additional terms when $\Delta=\frac{3}{2},\frac{5}{2}$ are proportional to derivative operators acting on the delta function in the position space and hence removable. Thus we find,
 \begin{tcolorbox}
    \begin{equation}
\langle\bar{\psi}_p^i\psi^j_{-p}\rangle=E^{2\Delta-3}\Pi_\sigma^{ii'}\left(\delta_{i'j'}\slashed{\hat{p}}-\frac{4(\Delta-2)}{2\Delta-1}\hat{p}_{i'}\hat{p}_{j'}\slashed{\hat{p}}\right)\Pi_\sigma^{j'j}.\label{eq:psipsicwiunique}
    \end{equation}
\end{tcolorbox}
If we further demand \eqref{eq:psipsicwiunique} to be transverse $p_i\langle\bar{\psi}_p^i\psi^j_{-p}\rangle=p_j\langle\bar{\psi}_p^i\psi^j_{-p}\rangle=0$ we will obtain $\Delta=5/2$, which is the protected dimension for conserved spin-3/2 current.

\subsubsection{3-pt}
In this sub-section, we  bootstrap the WFC functions of two spin-1/2, 3/2 with conserved spin-1 and 2.

\paragraph{$\underline{\langle J \bar{\chi}\chi\rangle}$}~\\

The conformal dimensions are $\Delta_J=2,\quad\Delta_{\bar{\chi}}=\Delta_{\chi}=\frac{3}{2}$, therefore the 3-point function has mass weight -1 by the dilatation symmetry \eqref{eq:dilatationcwi}. The amplitude-limit is minimally coupled QED. The counting formula \eqref{eq:contactleadingresidue} then tells us that the order of the leading total energy pole is 1. We can now begin our general bootstrap.
\paragraph{General Ansatz}
We begin by decomposing the WFC into the transverse part and the longitudinal part, writing down the ansatz separately as  

\begin{equation}
    \langle J^i\bar{\chi}\chi\rangle=\pi_1^{ii'}\mathbb{A}^T_{i'}+\hat{p}_1^i\mathbb{A}^L.
\end{equation}
Following the same spirit with \eqref{eq:3pttransverseansatz}, $\mathbb{A}^T_{i}$ can be written as
\begin{equation}
\mathbb{A}^T_{i}=\frac{\mathbb{M}_{i}}{E_T},    
\end{equation}
where Poly in this case can only be $1$, since the leading pole order is 1 while the overall momentum dimension is -1 and $\mathbb{M}$ can not have negative dimension. In other words, this restricts $\mathbb{M}_{i}$ to be dimensionless. The goal now for writing down the general $\mathbb{M}$ is to arrange the single free index into the dimension-0 blocks,
\ie
\mathbb{M}_i=&\left(a_1\hat{p}_{2i}+a_2\hat{p}_{3i}\right)_{\mathbb{A}^g_{\mathbb{I}}}+\left(a_3\sigma_{i}+a_4\sigma_{i}(\hat{p}_2\cdot\hat{p}_3)+a_5\hat{p}_{2i}\slashed{\hat{p}_3}+a_6\hat{p}_{3i}\slashed{\hat{p}_2}\right)_{\mathbb{A}^g_{\sigma}}
\\
+&\left(a_7i\epsilon_{ijk}\hat{p}_{2}^j\sigma^k+a_8i\epsilon_{ijk}\hat{p}_{3}^j\sigma^k\right)_{\mathbb{A}^g_{\epsilon\cdot\sigma}}+\left(a_9i\epsilon_{ijk}\hat{p}_{2}^j\hat{p}_{3}^k\right)_{\mathbb{A}^g_{\epsilon\mathbb{I}}},
\label{eq:jchichiansatz}
\fe
where subscript of the parentheses indicates the four sectors in \eqref{eq:4indep}, the number of unit vectors on the fermion legs follows from the propagator structure \eqref{eq:dSbulkbdypropgs}. For the longitudinal part, the most general ansatz reads,
\begin{equation}
\mathbb{A}^L=\frac{1}{E_1}p_{1i}\langle J^i\bar{\chi}\chi\rangle=\frac{1}{E_1}\left(a\langle\bar{\chi}_{1+2}\chi\rangle+b\langle\bar{\chi}\chi_{1+3}\rangle\right),
\end{equation}
where since the dimension of $p_{1i}\langle J^i\bar{\chi}\chi\rangle$ and $\langle\bar
\chi\chi\rangle$ are both 0, thus the only transformation patterns are simply constants $a,b$. 

\paragraph{CWI}

After acting with the boost operators, we found that the CWI uniquely determines the WFC,
\begin{tcolorbox}
\begin{equation}
\mathbb{A}^T_{i}=\frac{1}{E_T}\left(\sigma_{i}+\slashed{\hat{p}}_2\sigma_{i}\slashed{\hat{p}}_3\right),\quad \mathbb{A}^L=\frac{1}{E_1}\left(\langle\bar{\chi}_{1+2}\chi\rangle-\langle\bar{\chi}\chi_{1+3}\rangle\right).\label{eq:Jchichicwi}
\end{equation}    
\end{tcolorbox}
\paragraph{Spinor-Helicity form}
Below we project the transverse component onto various helicity configurations. We use the conventions of appendix~\ref{sec: Conventions and notation}  as well as C.35 of \cite{Baumann:2020dch}, 
\begin{tcolorbox}
 \begin{equation}
     \langle J^+\bar\chi^+\chi^-\rangle=\frac{\langle \bar{1}\bar{2} \rangle^2}{\langle\bar{2}\bar{3}\rangle}\left(\frac{E_T-2E_1}{E_TE_1}\right),\quad  \langle J^+\bar\chi^-\chi^-\rangle= \langle J^+\bar\chi^+\chi^+\rangle=0.
 \end{equation} 
\end{tcolorbox}

\paragraph{Discussions}
 Note that using the
relation between spin-1/2 boundary profiles and the polarizations in eq.(\ref{eq:amp transform to boundary}), one can see that the residue of the total energy pole is precisely the flat space amplitude even when the only input is the momentum dimension of the residue. This result has also been computed in \cite{Chowdhury:2024snc} using Witten diagrams and  is identical to the flat-space \cite{Flatspace}. From the longitudinal part, we readoff the WT identity is that of U(1) transformation.  

\paragraph{$\underline{\langle T \bar{\chi}\chi\rangle}$}~\\

Replacing the current with stress-tensor, dilatation symmetry \eqref{eq:dilatationcwi} fixes the momentum dimension to be 0. Here we also work with bulk minimally-coupled theory, meaning that the corresponding amplitude has 1 derivative and the leading total energy pole order is 2.

\paragraph{General Ansatz}
We now have,
\begin{equation}
    \langle T^{ij}\bar{\chi}\chi\rangle=\Hat{\Pi}_1^{iji'j'}\Hat{\mathbb{A}}_{i'j'}+\underline{\Pi}_1^{iji'j'}\mathbb{A}^{\underline{T}}_{i'j'}+\left(\pi_1^{ii'}\hat{p}_1^j+\pi_1^{ji'}\hat{p}_1^i+\hat{p}_1^i\hat{p}_1^j\hat{p}_1^{i'}\right)\mathbb{A}^L_{i'},
\end{equation}
where $\Hat{\mathbb{A}}_{ij}$ denotes the transverse traceless, $\mathbb{A}^L_i$  the longitudinal ansatz and $\mathbb{A}^{\underline{T}}_{ij}$ encodes the trace. 
First, $\Hat{\mathbb{A}}_{ij}$ can be written as a product form of a tensor and a scalar as in \eqref{eq:3pttransverseansatz}. The trace ansatz relates $\mathbb{A}^{\underline{T}}$ and $\mathbb{A}^L$ as the following way,
\begin{equation}
    \mathbb{N}\equiv\langle T_i^i\bar\chi\chi\rangle=\pi^{ij}\mathbb{A}^{\underline{T}}_{ij}+\hat{p}_1^i\mathbb{A}_i^L=\sum_{p}\underline{\mathbb{M}}_p\langle\bar\chi_p\chi\rangle.
\end{equation}
where by dimension counting $\underline{\mathbb{M}}_p$ can only be constant. Note that the above equation allows us to write the transverse-trace component in terms of the trace and longitudinal ansatz,
\begin{equation}
\underline{\Pi}_1^{iji'j'}\mathbb{A}^{\underline{T}}_{i'j'}=\frac{1}{2}\pi_1^{ij}\left(\mathbb{N}-\hat{p}_1^i\mathbb{A}_i^L\right).\label{eq:tchichitransatz}    
\end{equation}
\paragraph{CWI}
CWI fixes the WFC into the following form,
\begin{tcolorbox}
\ie
\label{eq:Tchichi}
&\Hat{\mathbb{A}}_{ij}=\left(\frac{E_1+E_T}{E_T^2}\right)(p_{2i}-p_{3i})\left(\sigma_{j}+\hat{\slashed{p}}_2\sigma_{j}\hat{\slashed{p}}_3\right)
\\
&\mathbb{A}_{i}^{L}=\frac{1}{E_1}\left\{p_{3i}\langle\bar\chi_2\chi\rangle-p_{2i}\langle\bar\chi_3\chi\rangle+\frac{1}{8}\left[\left(\slashed{\vec{p}}_1\sigma_i-\sigma_i\slashed{\vec{p}}_1\right)\langle\bar\chi_3\chi\rangle+\langle\bar\chi_2\chi\rangle\left(\slashed{\vec{p}}_1\sigma_i-\sigma_i\slashed{\vec{p}}_1\right)\right]\right\}
\fe
The trace identity
\begin{equation}
   \mathbb{N}=\langle T^i_i\bar\chi\chi\rangle=\frac{3}{2}\left(\langle\bar\chi_{-2}\chi\rangle+\langle\bar\chi_{3}\chi\rangle\right)\label{eq:tchichitr}
\end{equation}
\end{tcolorbox}

\paragraph{Spinor-Helicity form}
Below we project the transverse traceless component onto various helicity configurations, the results read,

\begin{tcolorbox}
\begin{equation}
    \langle T^+\bar\chi^+\chi^-\rangle=-\frac{\langle \bar{1}\bar{2} \rangle^3\langle \bar{3}\bar{1} \rangle}{\langle \bar{2}\bar{3} \rangle^2}\left(\frac{\left(2E_1+E_{23}\right)\left(E_1-E_{23}\right)^2}{2E_T^2E_1^2}\right),\quad\langle T^+\bar\chi^-\chi^-\rangle=\langle T^+\bar\chi^+\chi^+\rangle=0
\end{equation}   
\end{tcolorbox}

\paragraph{Discussions}

The leading pole residue reads, 
\begin{equation} \lim_{E_T\to0}\left(E_T\Hat{\mathbb{A}}_{ij}\right)=E_1(p_{2i}-p_{3i})\left(\sigma_{j}+\hat{\slashed{p}}_2\sigma_{j}\hat{\slashed{p}}_3\right)\label{eq:tchichileading}.
\end{equation}
which again is precisely the $P_T=E_1$ times 3-point amplitude for two spin-1/2 coupled to a graviton as we shown in \eqref{eq: total energy pole}.  For the longitudinal component $\mathbb{A}^L_i$, it matches the expected diffeomorphism transformation on the spin-1/2 particles \cite{Flatspace} while the trace component \eqref{eq:tchichitr} matches the Weyl transformation \eqref{eq:weyltransf}.

\paragraph{$\underline{\langle J \bar{\psi}\psi\rangle}$}~\\

Here we encounter our first case of multiple currents WFC, whose conformal dimension read $\Delta_J=2,\Delta_{\bar\psi}=\Delta_\psi=\frac{5}{2}$. Dilatation symmetry \eqref{eq:dilatationcwi} then requires the overall momentum dimension to be 1. We also assume the corresponding bulk interaction consists of one derivative, this then fixes the leading total energy pole order to be 2 by \eqref{eq: total energy pole}.

\paragraph{General Ansatz}
We begin by decomposing the WFC into independent components,
\ie
 &\langle J^i\bar{\psi}^j\psi^k\rangle=\langle J_T^i\bar{\psi}_{\Hat{T}}^j\psi_{\Hat{T}}^k\rangle+\left(\langle J_T^i\bar{\psi}_{\Hat{T}}^j\psi_{\underline{T}}^k\rangle+\langle J_T^i\bar{\psi}_{\underline{T}}^j\psi_{\Hat{T}}^k\rangle+\langle J_T^i\bar{\psi}_{\underline{T}}^j\psi_{\underline{T}}^k\rangle\right)+\langle J_L^i\bar{\psi}_T^j\psi_T^k\rangle
    \\
    &\quad\quad\quad\quad+\left( \langle J^i\bar{\psi}_T^j\psi_L^k\rangle+\langle J^i\bar{\psi}_L^j\psi_T^k\rangle+\langle J^i\bar{\psi}_L^j\psi_L^k\rangle\right)
 \\
=&\pi_1^{ii'}\Hat{\Pi}_2^{jj'}\Hat{\mathbb{A}}_{i'j'k'}\Hat{\Pi}_3^{k'k}+\pi_1^{ii'}\left(\Hat{\Pi}_2^{jj'}\mathbb{A}_{i'j'k'}^{\Hat{T}\underline{T}}\underline{\Pi}_3^{k'k}+\underline{\Pi}_2^{jj'}\mathbb{A}_{i'j'k'}^{\underline{T}\Hat{T}}\Hat{\Pi}_3^{k'k}+\underline{\Pi}_2^{jj'}\mathbb{A}_{i'j'k'}^{\underline{T}\underline{T}}\underline{\Pi}_3^{k'k}\right)+\hat{p}_1^i\pi_2^{jj'}\pi_3^{kk'}\mathbb{A}_{j'k'}^{J_L}
\\
&+\left(\pi_2^{jj'}\hat{p}_3^k\mathbb{A}_{j'}^{i,\psi_L}+\pi_3^{kk'}\hat{p}_2^j\mathbb{A}_{k'}^{i,\bar\psi_L}+\hat{p}_2^j\hat{p}_3^k\hat{p}_3^{k'}\mathbb{A}_{k'}^{i,\bar\psi_L}\right)
\label{eq:jpsipsidecomp}
\fe
where we organize the components based on the transversality of $\bar{\psi}$ and $\psi$ and all the projectors are defined in \eqref{eq:gammadecompose}. Let us now quickly go over the ansatz for each component.

First is the transverse gamma-traceless component $\Hat{\mathbb{A}}$, which can be written in the form similar to \eqref{eq:3pttransverseansatz}.
Next we move on to longitudinal components. For the $J$'s longitudinal part $\mathbb{A}^{J_L}_{jk}$ we assumed they are written as transformations of $\langle\bar\psi_p\psi\rangle$ ($p=2,3$) while for the $\bar\psi,\psi$'s longitudinal components $\mathbb{A}^{\bar\psi_L}_i,\mathbb{A}^{\psi_L}_i$, we assume they are composed of transformation on $\langle JJ\rangle,\langle\bar\psi\psi\rangle$\footnote{The conserved $\langle J^i J^j\rangle=E\pi^{ij}$.}.
At last, the $\mathbb{A}^{\Hat{T}\underline{T}},\mathbb{A}^{\underline{T}\Hat{T}},\mathbb{A}^{\underline{T}\underline{T}}$ components can be related to the gamma trace ansatz and the longitudinal components as,
\ie
&\pi_1^{ii'}\left(\Hat{\Pi}_2^{jj'}\mathbb{A}_{i'j'k'}^{\Hat{T}\underline{T}}\underline{\Pi}_3^{k'k}+\underline{\Pi}_2^{jj'}\mathbb{A}_{i'j'k'}^{\underline{T}\Hat{T}}\Hat{\Pi}_3^{k'k}+\underline{\Pi}_2^{jj'}\mathbb{A}_{i'j'k'}^{\underline{T}\underline{T}}\underline{\Pi}_3^{k'k}\right)
    \\
&=\slashed{\pi}_2^j\left(\mathbb{N}_{\bar\psi}^{ik'}-(\mathbb{L}_{\bar\psi})^{ik'}\right)\Hat{\Pi}_3^{k'k}+\left(\mathbb{N}_{\psi}^{ij}-(\mathbb{L}_{\psi})^{ij}\right)\slashed{\pi}_3^k,\label{eq:jpsipsitranstransatz}
\fe
where  $\mathbb{N}_{\bar\psi}^{ik}=\sigma_j\langle J^i\bar\psi^j\psi^k\rangle=\sum_{p}\mathbb{S}_{p,i'k'}^{ik}\langle J_p^{i'}J^{k'}\rangle$ ($\mathbb{S}$ denotes the transformation ansatz) and $\mathbb{L}_{\bar\psi}$ also abbreviates all the longitudinal $\mathbb{A}^g$ entering the RHS of \eqref{eq:jpsipsidecomp} after contracting with $\sigma_j$ (similarly for $\psi$'s trace identity),
\begin{equation}
\mathbb{L}_{\bar\psi}^{ik}=\sigma_j\left\{\hat{p}_1^i\pi_2^{jj'}\pi_3^{kk'}\mathbb{A}_{j'k'}^{J_L}+\left(\pi_2^{jj'}\hat{p}_3^k\mathbb{A}_{j'}^{i,\psi_L}+\pi_3^{kk'}\hat{p}_2^j\mathbb{A}_{k'}^{i,\bar\psi_L}+\hat{p}_2^j\hat{p}_3^k\hat{p}_3^{k'}\mathbb{A}_{k'}^{i,\bar\psi_L}\right)\right\}.
\end{equation}
\paragraph{WT Consistency}
The linearly independent ansatz can be further reduced by imposing the consistency of multi longitudinal and trace projections, including
\begin{align}
    &\langle J_L\bar{\psi}_L\psi\rangle,\quad \langle J_L\bar{\psi}\psi_L\rangle,\quad \langle J\bar{\psi}_L\psi_L\rangle,\nonumber\\
    \quad \sigma_o\langle J_L\bar{\psi}^o\psi\rangle,\quad& \langle J_L\bar{\psi}\psi^q\rangle\sigma_q,\quad\sigma_o\langle J\bar{\psi}^o\psi_L\rangle,\quad\langle J\bar{\psi}_L\psi^q\rangle\sigma_q,\quad\sigma_o\langle J\bar{\psi}^o\psi^q\rangle\sigma_q.
\end{align}

\paragraph{CWI}
We find that 3-pt WFC is fully fixed by CWI up to 4 separate polynomial terms, where the latter are separately invariant by themselves. The part that contains the total energy pole is given by
\begin{tcolorbox}
\ie    
&\Hat{\mathbb{A}}_{ijk}
    =
    \frac{E_2E_3}{E_T^2}
    \left(
        \slashed{\hat{p}}_2-\slashed{\hat{p}}_3
    \right)
    \left(
        \delta_{ij}p_{1k}-\delta_{ik}p_{1j}
    \right)
    -
    \left(
        \frac{E_1 E_2 E_3}{E_T^2}
        -
        \frac{E_2 E_3}{E_T}
    \right)
    \left(
        -1+ \slashed{\hat{p}}_2 \slashed{\hat{p}}_3
    \right)
    \epsilon_{ijk}
        \\
    &\quad\quad\quad+
    \frac{1}{E_T}
    \left(
        \slashed{p}_2-\slashed{p}_3
    \right)
    \left(
        \delta_{ij}p_{1k}-\delta_{ik}p_{1j}
    \right)
    -\left(
        \frac{E_3^2+E_2^2}{E_T}
    \right)\epsilon_{ijk}
    \label{eq:Jpsipsitranssymmetrized},
    \\
   &\mathbb{A}^{J_L}_{jk}=\frac{1}{E_1}\left(-\langle\bar{\psi}_{j,2}\psi_{k}\rangle+\langle\bar{\psi}_{j,-3}\psi_{k}\rangle\right), 
   \\
&\mathbb{A}^{\bar{\psi}_L}_{ik}=\frac{1}{E_2}\frac{1}{2}\left(\langle\bar{\psi}_{i,-3}\psi_{k}\rangle-\slashed{\vec{p}}_3\langle J_{i,1}J_{k}\rangle-\epsilon_{hlk}p_3^l\langle J_{i,1}J^{h}\rangle+\sigma_{k}p_{3h}\langle J_{i,1}J^{h}\rangle\right),
\\
&\mathbb{A}^{\psi_L}_{ik}=\frac{1}{E_3}\frac{1}{2}\left(-\langle\bar{\psi}_{i,2}\psi_{k}\rangle-\slashed{\vec{p}}_2\langle J_{i,1}J_{k}\rangle-\epsilon_{hlk}p_2^l\langle J_{i,1}J^{h}\rangle+\sigma_{k}p_{2h}\langle J_{i,1}J^{h}\rangle\right).
\fe
$\bar{\psi}$ and $\psi$'s trace identity,
\begin{equation}
\mathbb{N}_{\bar\psi}^{ik}=\sigma_j \langle J^i\bar\psi^j\psi^k\rangle=-\langle J_{1}^iJ^k\rangle,\quad \mathbb{N}_{\psi}^{ij}=\langle J^i\bar\psi^j\psi^k\rangle\sigma_k =-\langle J_{1}^iJ^j\rangle\label{eq:jpsipsitrwt}.
\end{equation}
\end{tcolorbox}
\noindent while the unfixed polynomial terms are:
\begin{equation}
\sigma^j\langle J_1^iJ^k\rangle,\quad  \sigma^k\langle J_1^iJ^j\rangle,\quad \delta^{jk}\sigma_{i'}\langle J_1^iJ^{i'}\rangle,\quad i\epsilon^{jki'}\langle J_1^iJ^{i'}\rangle.\label{eq:jpsipsiunfixed}
\end{equation}
Given that their forms are all proportional to 2-pt functions, they correspond to field redefinitions. For example, the first term can be generated from the following field redefinition, $A_{i,\partial}(x) \to A_{i,\partial}(x)+ \bar\psi_{j,\partial}(x) \sigma^j \psi_{i,\partial}(x)$, since,
\ie
\log \Psi' = \log \Psi + 2 \int d^3 x d^3y \langle J^i(x) J^k(y) \rangle A_{\partial,i}(x) (\bar\psi_{j,\partial}(y) \sigma^j \psi_{k,\partial}(y)) + \dots
\label{eq:JpsipsiFR}
\fe
Note that a field redefinition would also modify the action of symmetry transformations, and hence the WT identities. Indeed the unfixed polynomial terms in eq.(\ref{eq:jpsipsiunfixed}) have non-trivial longitudinal projections.

\paragraph{Spinor-Helicity form}
Below we project the transverse gamma-traceless component onto various helicity configurations, 
\begin{tcolorbox}
    \ie
&\langle J^+\bar\psi^-\psi^- \rangle=\frac{\langle 23 \rangle^4}{\langle 12 \rangle\langle 31 \rangle}\left(\frac{(E_{12}-E_3)^2(E_{13}-E_2)^2E_{23}}{8E_T^2E_1E_2E_3}\right) 
\\
&\langle J^+\bar\psi^+\psi^+ \rangle=\langle\bar{1}\bar{2}\rangle\langle\bar{2}\bar{3}\rangle^2\langle\bar{3}\bar{1}\rangle\frac{E_{23}}{8E_1E_2E_3}
\\
&\langle J^+\bar\psi^+\psi^- \rangle=-\langle\bar{1}\bar{2}\rangle\langle\bar{2}3\rangle^2\langle\bar{1}3\rangle\left(\frac{(E_2^2+E_3^2)+E_1(E_2+E_3)}{8E_TE_1E_2E_3}\right)\,.
\label{eq:Jpsipsihelicity}
    \fe
\end{tcolorbox}

\paragraph{Discussions}
Firstly, as expected, the leading total energy pole residue is proportional to the amplitude. One finds, 
\begin{equation}
  \lim_{E_T\to 0} \left( E_T^2\Hat{\mathbb{A}}\right)=2E_2E_3\left(\slashed{\hat{p}}_2-\slashed{\hat{p}}_3\right)\left(\delta_{ij}p_{1k}+\frac{1}{2}\delta_{jk}\left(p_{2i}-p_{3i}\right)-\delta_{ik}p_{1j}\right).
\end{equation}
which matches the amplitude up to the factor $P_T=E_2 E_3$ defined in \eqref{eq: total energy pole}. This is even more transparent in the spinor-helicity form for $\langle J^+\bar\psi^-\psi^- \rangle$ in eq.(\ref{eq:Jpsipsihelicity}). Note that there is an $1/E_T$ pole in the non-amplitude helicity- configuration  $\langle J^+\bar\psi^+\psi^- \rangle$. Importantly, viewed from the helicity-basis, the $1/E_T$ would be ``leading" in the $(++-)$ configuration. This contribution can in fact be traced to the present of an $\bar\psi A\psi$ (suppressing gamma matrices) vertex in $\mathcal{N}=2$ gauged supergravity in curved space \cite{Ortin:2015hya}. Thus we see that the helicity-basis, poles in $E_T$ reflect bulk interactions. \textit{The poles for non-amplitude helicity-configurations reflect higher order $H$ corrections of the interactions}. We will see similar phenomenon in the $\langle T\bar\psi\psi \rangle$ WFC. Finally, while the $(+++)$ configuration does not carry any $E_T$ pole, they cannot be removed via linear combinations of the unfixed terms in \eqref{eq:jpsipsiunfixed}, i.e. they can not be removed by field redefinition.\footnote{We thank Harry Goodhew for raising this question.} 

Let us now turn our attention to the longitudinal and trace components. First, we find that $J$'s longitudinal part $\mathbb{A}^{J_L}$ follows the U(1) transformation pattern \cite{Flatspace}, which shows that the CWI requires the $\bar\psi,\psi$ to be  U(1)-charged Dirac fermions. Furthermore, $(\mathbb{A}^{\bar\psi_L}, \mathbb{A}^{\psi_L})$ reflects the 3-dimensional $\mathcal{N}=2$ supersymmetry while the gamma-trace identities reflect the superWeyl symmetry \cite{Papadimitriou:2017kzw}.

\paragraph{$\underline{\langle T \bar{\psi}\psi\rangle}$}~\\

Our last 3-point WFC contains a  $\Delta=3$ stress tensor and two $\Delta=5/2$ conserved spin-3/2 currents, assuming the bulk-coupling contains a single derivative. All together these fix the leading total energy pole is 2 \eqref{eq: total energy pole}.
\paragraph{General Ansatz}
We begin again by decomposing the WFC similar to the pattern \eqref{eq:jpsipsidecomp} and comment on each of them,
\ie
&\langle T^{ij}\bar{\psi}^{k}\psi^l\rangle=\langle \Hat{T}^{ij}\bar{\psi}^k_{\Hat{T}}\psi_{\Hat{T}}^l\rangle+\left(\langle \Hat{T}^{ij}\bar{\psi}^k_{\Hat{T}}\psi^l_{\underline{T}}\rangle+\langle \Hat{T}^{ij}\bar{\psi}^k_{\underline{T}}\psi^l_{\Hat{T}}\rangle+\langle \Hat{T}^{ij}\bar{\psi}^k_{\underline{T}}\psi^l_{\underline{T}}\rangle\right)+\langle \underline{T}^{ij}\bar{\psi}^k_T\psi^l_T\rangle+\langle T^{ij}_L\bar{\psi}^k_T\psi^l_T\rangle
    \\
    &\quad\quad\quad\quad+\left( \langle T^{ij}\bar{\psi}^k_T\psi^l_L\rangle+\langle T^{ij}\bar{\psi}^k_L\psi^l_T\rangle+\langle T^{ij}\bar{\psi}^k_L\psi^l_L\rangle\right),
    \\
&=\Hat{\Pi}^{ii'jj'}_1\Hat{\Pi}^{kk'}_2\Hat{\mathbb{A}}_{i'j'k'l'}\Hat{\Pi}^{l'l}_3+\Hat{\Pi}^{ii'jj'}_1\left(\Hat{\Pi}_2^{kk'}\mathbb{A}^{\Hat{T}\Hat{T}\underline{T}}_{i'j'k'l'}\underline{\Pi}_3^{l'l}+\underline{\Pi}_2^{kk'}\mathbb{A}^{\Hat{T}\underline{T}\Hat{T}}_{i'j'k'l'}\Hat{\Pi}^{l'l}_3+\underline{\Pi}_2^{kk'}\mathbb{A}_{i'j'k'l'}^{\Hat{T}\underline{T}\underline{T}}\underline{\Pi}^{l'l}_3\right)
    \\
    &+\underline{\Pi}_1^{iji'j'}\mathbb{A}^{\underline{T}}_{i'j'k'l'}\pi^{kk'}_2\pi^{ll'}_3+\pi_2^{kk'}\pi_3^{ll'}\left(\pi_1^{ii'}\hat{p}_1^j+\pi_1^{ji'}\hat{p}_1^i+\hat{p}_1^i\hat{p}_1^j\hat{p}_1^{i'}\right)\mathbb{A}^{T_L}_{i'k'l'}
    \\
&+\left(\pi_2^{kk'}\hat{p}_3^l\mathbb{A}^{ij,\psi_L}_{k'}+\pi_3^{ll'}\hat{p}_2^k\mathbb{A}^{ij,\bar\psi_L}_{l'}+\hat{p}_2^k\hat{p}_3^l\hat{p}_3^{l'}\mathbb{A}^{ij,\bar\psi_L}_{l'}\right)
    \label{eq:tpsipsidecomp}
\fe
For the transverse-traceless component $\Hat{\mathbb{A}}$, the ansatz will again take a form \eqref{eq:3pttransverseansatz}, while the longitudinal part of the stress tensor $\mathbb{A}^{T_L}_{ijk}$ will be proportional to the two point function $\langle\bar\psi\psi\rangle$ and that of the spin-3/2 currents are proportional to $\langle TT\rangle$ and $\langle\bar\psi\psi\rangle$.

 Since each of the three operators have their trace components, the trace structure will be slightly more complicated. Let us first focus on the stress tensor. Taking the trace of \eqref{eq:tpsipsidecomp}, one can then express the $\underline{\Pi}_1\mathbb{A}^{\underline{T}}\pi_2\pi_3$ component in terms of the trace ansatz and the longitudinal ansatz similar to \eqref{eq:tchichitransatz},
\begin{equation}
\underline{\Pi}_1^{iji'j'}\mathbb{A}^{\underline{T}}_{i'j'k'l'}\pi^{kk'}_2\pi^{ll'}_3=\frac{1}{2}\pi_1^{ij}\left(\mathbb{N}_{\underline{T}}^{kl}-\mathbb{L}_{\underline{T}}^{kl}\right),
\end{equation}
where following the abbreviation in \eqref{eq:jpsipsitranstransatz}, we have densely written the trace ansatz as $\mathbb{N}_{\underline{T}}^{kl}\equiv\delta_{ij}\langle T^{ij}\bar{\psi}^{k}\psi^l\rangle=\sum_p \mathbb{M}_{p,k'l'}^{kl}\langle\bar\psi_p^{k'}\psi^{l'}\rangle$ and collected all terms involving longitudinal components as $\mathbb{L}_{\underline{T}}^{kl}$,
\begin{equation}
\mathbb{L}_{\underline{T}}^{kl}=\pi_2^{kk'}\pi_3^{ll'}\hat{p}_1^{i'}\mathbb{A}^{T_L}_{i'k'l'}+\delta_{ij}\left(\pi_2^{kk'}\hat{p}_3^l\mathbb{A}^{ij,\psi_L}_{k'}+\pi_3^{ll'}\hat{p}_2^k\mathbb{A}^{ij,\bar\psi_L}_{l'}+\hat{p}_2^k\hat{p}_3^l\hat{p}_3^{l'}\mathbb{A}^{ij,\bar\psi_L}_{l'}\right).
\end{equation}
Next, similar to \eqref{eq:jpsipsitranstransatz}, terms involving $\mathbb{A}^{\Hat{T}\Hat{T}\underline{T}},\mathbb{A}^{\Hat{T}\underline{T}\Hat{T}},\mathbb{A}^{\Hat{T}\underline{T}\underline{T}}$ can also be expressed via the gamma-trace ansatz as well as the longitudinal components,
\ie
&\Hat{\Pi}^{ii'jj'}_1\left(\Hat{\Pi}_2^{kk'}\mathbb{A}^{\Hat{T}\Hat{T}\underline{T}}_{i'j'k'l'}\underline{\Pi}_3^{l'l}+\underline{\Pi}_2^{kk'}\mathbb{A}^{\Hat{T}\underline{T}\Hat{T}}_{i'j'k'l'}\Hat{\Pi}^{l'l}_3+\underline{\Pi}_2^{kk'}\mathbb{A}_{i'j'k'l'}^{\Hat{T}\underline{T}\underline{T}}\underline{\Pi}^{l'l}_3\right)
\\
&=\Hat{\Pi}_{1,i'j'}^{ij}\left[\slashed{\pi}_2^k\left(\mathbb{N}^{i'j'l'}_{\bar\psi_{\sigma}}-\mathbb{L}_{\bar\psi_{\sigma}}^{i'j'l'}\right)\Hat{\Pi}_3^{l'l}+\left(\mathbb{N}^{i'j'k}_{\psi_{\sigma}}-\mathbb{L}_{\psi_{\sigma}}^{i'j'k}\right)\slashed{\pi}_3^l\right],\label{eq:Tpsipsitransgammatransatz}
\fe
where the overall $\Hat{\Pi}_1$ eliminates the $\mathbb{A}^{\underline{T}}$ terms in the gamma trace expressions and $\mathbb{N}_{\bar\psi},\mathbb{L}_{\bar\psi}$ serve the same abbreviation as in \eqref{eq:jpsipsitranstransatz} $\mathbb{N}_{\bar\psi}^{ijl}=\sigma_k\langle T^{ij}\bar{\psi}^{k}\psi^l\rangle=\sum_{p}\mathbb{M}_{p,i'j'}^{ijl}\langle \bar\psi_p^{i'}\psi^{j'}\rangle$, 
\ie
\mathbb{L}_{\bar\psi}^{ijl}=\sigma_k&\Big\{\pi_2^{kk'}\pi_3^{ll'}\left(\pi_1^{ii'}\hat{p}_1^j+\pi_1^{ji'}\hat{p}_1^i+\hat{p}_1^i\hat{p}_1^j\hat{p}_1^{i'}\right)\mathbb{A}^{T_L}_{i'k'l'}
\\
&+\left(\pi_2^{kk'}\hat{p}_3^l\mathbb{A}^{ij,\psi_L}_{k'}+\pi_3^{ll'}\hat{p}_2^k\mathbb{A}^{ij,\bar\psi_L}_{l'}+\hat{p}_2^k\hat{p}_3^l\hat{p}_3^{l'}\mathbb{A}^{ij,\bar\psi_L}_{l'}\right)\Big\}
\fe
Note that the (gamma)trace ansatz here contains only transformations on $\langle\bar\psi\psi\rangle$, since by dimension counting transformation matrices acting on $\langle TT \rangle$ would be $-1$ dimension and thus are forbidden by our locality assumption.

\paragraph{WT Consistency}
The linearly independent ansatz can be further reduced by imposing the consistency of multi longitudinal and trace projections, including
\ie
&\langle T_L\bar{\psi}_L\psi\rangle,\quad\langle T_L\bar{\psi}\psi_L\rangle,\quad \langle T\bar{\psi}_L\psi_L\rangle,
\\
&\sigma_k\langle T_L\bar{\psi}^k\psi\rangle,\quad \langle T_L\bar{\psi}\psi^l\rangle\sigma_l,\quad\sigma_k\langle T\bar{\psi}^k\psi_L\rangle,\quad\langle T\bar{\psi}_L\psi^l\rangle\sigma_l,
\\
&\sigma_k\langle T\bar{\psi}^k\psi^l\rangle\sigma_l,\quad\sigma_k\langle T_i^i\bar{\psi}^k\psi\rangle,\quad \langle T_i^i\bar{\psi}\psi^l\rangle\sigma_l.
\fe
\paragraph{CWI}
Similar to $\langle J\bar\psi\psi\rangle_{E_T}$, CWI drastically fixes the ansatz up to two pure polynomial terms. The pole containing solution is given as:
\begin{tcolorbox}
\ie
\label{eq:Tpsipsiresult}
&\Hat{\mathbb{A}}_{ijkl}=-\left(\frac{2E_TE_2E_3-E_2E_3^2-E_3E_2^2}{E_T^2} \right)\left(\sigma_{i}+\hat{\slashed{p}}_2\sigma_{i}\hat{\slashed{p}}_3\right)\left(\delta_{jk}p_{2l}+\text{cyclic}\right)  
\\
&\quad\quad\quad+\left(\frac{E_T^2-E_1(E_2+E_3)}{E_T} \right)\sigma_i\left(\delta_{jk}p_{2l}+\text{cyclic}\right)+\left(E_2^2\hat{\slashed{p}}_2-E_3^2\hat{\slashed{p}}_3\right)\delta_{ik}\delta_{jl}
\\
&\mathbb{A}_{T_L}^{jkl}=\frac{1}{E_1}\left\{p_2^j\langle\bar\psi_3^k\psi^l\rangle-p_3^j\langle\bar\psi_2^k\psi^l\rangle+\frac{1}{8}\left[\left(\slashed{\vec{p}}_1\sigma^j-\sigma^j\slashed{\vec{p}}_1\right)\langle\bar\psi_3^k\psi^l\rangle+\langle\bar\psi_2^k\psi^l\rangle\left(\slashed{\vec{p}}_1\sigma^j-\sigma^j\slashed{\vec{p}}_1\right)\right]\right\}
\\
&\mathbb{A}_{\bar\psi_L}^{ijl}=\frac 
{1}{E_2}\left(-\frac{1}{4}\sigma_k\langle 
T^{ij}_1T^{kl} \rangle+\frac{1}{16}p_{1k}\sigma^{[kj]}\langle\bar\psi_3^i\psi^l\rangle-\frac{1}{8}p_{1k}\sigma^{[ki]}\langle\bar\psi_3^j\psi^l\rangle\right)
\\
&\mathbb{A}_{\psi_L}^{ijk}=\frac 
{1}{E_3}\left(-\frac{1}{4}\sigma_l\langle 
T^{ij}_1T^{kl} \rangle+\frac{1}{16}p_{1l}\langle\bar\psi_2^i\psi^k\rangle\sigma^{[lj]}-\frac{1}{8}p_{1l}\langle\bar\psi_2^j\psi^k\rangle\sigma^{[li]}\right)
\fe    
The $T,\bar\psi,\psi$'s trace identities
\ie
\label{eq:tpsipsitraceidentity}
\mathbb{N}_{\underline{T}}^{kl}=\langle T^i_i \bar\psi^k\psi^l\rangle&=\frac{1}{2}\left(\langle\bar\psi_{-2}^k\psi^l\rangle+\langle\bar\psi_3^k\psi^l\rangle\right),
\\
\mathbb{N}_{\bar\psi}^{ijl}=\sigma_k\langle T^{ij} \bar\psi^k\psi^l\rangle&=\sigma^{(i}\langle\bar\psi_{-3}^{j)}\psi^l\rangle,\quad \mathbb{N}_{\psi}^{ijk}=\langle T^{ij} \bar\psi^k\psi^l\rangle\sigma_l=\langle\bar\psi_{2}^{(j}\psi^k\rangle\sigma^{i)}. 
\fe
\end{tcolorbox}
\noindent The unfixed polynomial terms are of the form, 
\begin{equation}
 \delta^{k(j}\langle \bar\psi_3^{i)}\psi^l\rangle,\quad \delta^{l(j}\langle \bar\psi_{2}^{i)}\psi^k\rangle\,.
\end{equation}
Given that the polynomial terms are all proportional to spin-3/2 2-pt functions, one identifies this as field redefinition ambiguities similar to \eqref{eq:JpsipsiFR}.

\paragraph{Spinor-Helicity form}
Below we project the pure transverse traceless component onto various helicity configurations, 
\begin{tcolorbox}
    \ie
&\langle T^+\bar\psi^+\psi^- \rangle=-\frac{\langle \bar{1}\bar{2} \rangle^5}{\langle \bar{2}\bar{3} \rangle^2\langle \bar{3}\bar{1} \rangle}\left(\frac{\left(E_{T}-2E_2\right)\left(E_{T}-2E_{23}\right)^2}{16 E_T^2E_1^2E_2E_3}\right)\times
\\
&\left(4E_2E_3^2E_{23}+2E_3^2(E_3-3E_2)E_T-(3E_2^2+E_3^2)E_T^2+(E_2+3E_3)E_T^3-E_T^4\right)
\\
&\langle T^+\bar\psi^+\psi^+ \rangle=-\frac{1}{16}\langle\bar{1}\bar{2}\rangle^2\langle\bar{2}\bar{3}\rangle\langle\bar{3}\bar{1}\rangle^2\left[\frac{\left(E_T^2-E_1(E_2+E_3)\right)+\left(E_2^2+E_3^2\right)}{E_1^2 E_2E_3}\right]
\\
&\langle T^+\bar\psi^-\psi^- \rangle=\frac{1}{16}\langle\bar{1}2\rangle^2\langle 23\rangle\langle3\bar{1}\rangle^2\left[\frac{2E_{23}^3-\left(5E_2^2+6E_2E_3+5E_3^2\right)E_T+3E_{23}E_T^2-E_T^3}{E_TE_1^2E_2E_3}\right]
\label{eq:Tpsipsihelicity}
    \fe
\end{tcolorbox}

\paragraph{Discussions}

Once again the transverse (gamma) traceless component contains the leading $E_T$ pole, whose residue reads,
\begin{equation}    \lim_{E_T\to0}\left(E_T^2\Hat{\mathbb{A}}_{ijkl}\right)=-E_1E_2 E_3 \left(\sigma_{i}+\hat{\slashed{p}}_2\sigma_{i}\hat{\slashed{p}}_3\right)\left(\delta_{jk}p_{2l}+\text{cyclic}\right),  \label{eq:tpsipsileading}
\end{equation}
where the tensor structure matches the $P_T=E_1 E_2 E_3$ times $T\bar\psi\psi$ amplitude with one derivative as we shown \eqref{eq: total energy pole}. Similar to the previous $J\bar\psi\psi$ case, in the helicity-basis, we see that for the non-amplitude $(+--)$ configuration there is an $E_T$ pole, which can be traced to a $R\bar\psi \psi$ term in the bulk. This is consistent with the general description that the gravitino requires a mass term from the cosmological constant \cite{Freedman:2012zz}.  At last, one can also show that the $+++$ case cannot be removed by field redefinition.

 Next we turn our attention to the longitudinal components and (gamma)trace identities. First, the trace identity of the stress tensor shows that $\bar\psi,\psi$ are both weight-5/2 under the Weyl transformation and $\mathbb{A}_{T_L}$ exhibits the standard diffeomorphism transformation of spin-3/2 fields. The $\mathbb{A}_{\bar\psi_L},\mathbb{A}_{\psi_L}$ give the $\mathcal{N}=1$ supersymmetry transformation and the gamma trace identity gives the superWeyl transformation, as can all be verified via the variation form of the boundary profiles in the appendix \ref{sec:residualsym}. Note that by studying the position space 3-pt function, it was shown that conformal invariance does not imply that conserved spin-3/2 currents lead to supersymmetry~\cite{Buchbinder:2022mys}. This does not contradict our result as we have one additional input, our analysis assumes a rational function in momentum and energy variables. Thus under  the above assumption, \textit{ CWI and dimensional analysis fixes the algebra of the symmetry transformation responsible for the WT identities.}  

 \subsection{4-pt fermion WFCs}
We now consider the bootstrap of the four-point massless spinor WFC $\langle \bar\chi\,\chi\,\bar\chi\,\chi \rangle$ in QED and gravity. 

We begin with the exchange of a conserved vector in QED. Using the total-energy pole formula for the 3-point function $\langle J\,\bar\chi\,\chi \rangle$ (see \eqref{eq: total energy pole}), the total-energy pole order is $p=1$ with an energy prefactor $P_T=1$, which coincides with the flat-space WFC total energy pole limit. Therefore, we can directly insert this result into the cutting rule for $\langle \bar\chi\,\chi\,\bar\chi\,\chi \rangle$, which yields exactly the same cutting rule as in flat space \cite{Flatspace}. Moreover, applying the same total-energy pole formula \eqref{eq: total energy pole} to $\langle \bar\chi\,\chi\,\bar\chi\,\chi \rangle$ itself shows that the total-energy pole constraint remains identical to the flat-space WFC, again with total pole order $p=1$. Consequently, all bootstrap constraints obtained from the pole conditions and cutting rule (i.e., steps 1--3 of the 4-point bootstrap) match those of flat space, and thus the bootstrap solution agrees with the flat-space result:
\ie
\label{eq: chi4}
\langle \bar\chi \chi \bar\chi \chi \rangle
&=
\frac{\left(\bar u_1 \gamma^i u_2\right)\pi_{ij}^s \left(\bar u_3 \gamma^ju_4\right)}{E_T E_{L}^s E_{R}^s} -\frac{\left(\bar u_1 \gamma_0 u_2\right) \left(\bar u_3 \gamma_0 u_4\right)}{E_T E_s^2} + (t),
\fe
where the spinor indices in each square bracket are fully contracted. Here, $(t)$ denotes the replacements $E_s\to E_t$ in the first two terms, as well as the exchange of kinematic labels 1 and 3. In step 4 of our bootstrap, the CWI does not generate any new structures because there are no subleading $D_{\text{even}}$ (or $D_{\text{odd}}$) terms that can be added. As mentioned in the previous subsection, the CWI fixes the total-energy pole to coincide with the flat-space amplitude, which is already supplied by steps~1--3 above (see \eqref{eq: chi4}). 
\footnote{
    The bootstrap result is identical to the flat-space case. This is also expected from the Feynman-rule perspective: for massless spinors and conserved vectors, all propagators agree with their flat-space counterparts up to overall powers of $\eta$. However, these powers cancel against the metric factors appearing in the Feynman rules of $\langle \bar\chi\,\chi\,\bar\chi\,\chi \rangle$, reducing the result to the flat-space expression.}

Having established this, we now turn to an example where the total-energy pole formula \eqref{eq: total energy pole} yields a higher total-energy pole order $p>1$. It is instructive to first apply our procedure for a simpler example: the 4-point function of conformally coupled scalars ($\Delta=2$), denoted $\langle O O O O \rangle$, exchanging a graviton. This serves to verify our method by recovering the result established in \cite{Baumann:2020dch}. We start from the transverse-traceless part\footnote{The transverse-traceless part is the only part relevant to the cutting rule and the residues of the partial energy poles of $\langle O O O O \rangle$.} of the 3-point WFC, which can be easily determined by our 3-pt bootstrap:
\ie
\langle T^{\hat{T}\hat{T},i j}  O O \rangle =\left(\frac{E_1}{E_T^2} + \frac{1}{E_T}\right) p_{2,k} p_{2,l} \hat\Pi^{kl i j}_{1}.
\fe
This matches the result obtained from weight shifting operator \cite{Baumann:2020dch}.
Next, we determine $A_R^s$ for $\langle O_1O_2O_3 O_4 \rangle$:
\ie
    \frac{A_{R}^s}{(E_R^s)^2}
    &=
    \frac{3}{4} \frac{\alpha^s_{i_s}\alpha^s_{j_s}\hat{\Pi}_s^{i_s j_s l_s m_s}\beta^s_{l_s}\beta^s_{m_s}}{(E_R^s)^2 E_s^3}
     \left( 
    \begin{aligned}
        \frac{(E_s+ E^s_{L})(E_s+ E^s_{R})}{(E^s_{L})^2}
        +2\frac{E_s^2 E^s_{R}}
        {E_T^3}
        +\frac{E_s^2}{E_T^2}
        -\frac{E_s+E^s_{R}}{E_T}
    \end{aligned}
    \right)
\fe
where $\alpha^s_i=p_{1i}-p_{2i},\;\beta^s_i=p_{3i}-p_{4i}$, and $\hat{\Pi}^{ijkl}$ is defined in \eqref{eq:Tdecompose}. It is then straightforward to derive $B_L^s$ from $A_R^s$ to match the expected singular term in the other partial energy pole, $E_L^s$:
\ie
\frac{B_{L}^s}{(E_L^s)^2}
&=
\frac{3}{4} \frac{\alpha^s_{i_s}\alpha^s_{j_s}\hat{\Pi}_s^{i_s j_s l_s m_s}\beta^s_{l_s}\beta^s_{m_s}}{(E_L^s)^2 E_s^3}
\left(
    2\frac{E_s^2 E^s_{L}}
        {E_T^3}
        +\frac{E_s^2}{E_T^2}
        -\frac{E_s+E^s_{L}}{E_T}
\right)
\fe
Similarly, one can obtain the terms associated with the partial energy poles in the $t$ and $u$ channels. Next, we examine the leading $E_T^{-3}$ behavior of the WFCs constructed from $A_R^e$ and $B_L^e$ as $E_T \to 0$, and determine the contact term $C$ by matching it with the flat-space amplitude:
\ie
M_{OOOO}=3\frac{\left[(p_1-p_2)^\mu (p_3-p_4)_\mu\right]^2}{S} + (t) + (u)
\fe
in which $(t)$ and $(u)$ denote the replacements $E_s\to E_t, E_u$ in the first term, as well as the exchange of kinematic labels $1 \leftrightarrow 3$ and $2 \leftrightarrow 3$, respectively. We then utilize the identities listed in our companion paper \cite{Flatspace} to obtain $C$:
\ie
\frac{C}{E_T^3} &= -\frac{1}{E_T^3}\left(
    \begin{aligned}
    &6 \hat\alpha^s \hat\beta^s (\alpha^s \cdot \pi_s \cdot \beta^s)-
    3 E_{12s}E_{34s} (\hat\alpha^s)^2 (\hat\beta^s)^2\\ 
    &+\frac{3}{2} \Pi^C_{1,OOOO}
    +
    \frac{3}{2} E_{12s}E_{34s} \Pi^C_{2,OOOO}
    \end{aligned}
     \right)+ (t) + (u)
\fe
where we define $\hat{\alpha}^s=(E_1-E_2)/E_s,\;\hat{\beta}^s=(E_3-E_4)/E_s$, and 
\ie
\Pi^C_{1,OOOO} &= -(\hat \alpha^s)^2 (-E_s^2+E_{12}^2) - (\hat \beta^s)^2 (-E_s^2+E_{34}^2)\\
\Pi^C_{2,OOOO} &= - \left(1+ (\hat \alpha^s)^2 (\hat \beta^s)^2 \right).
\fe
We fix $D_{odd}$ by comparing the difference of the $A, B, C$ insertions on the LHS of the cutting rule with the RHS given by $\langle TOO \rangle$ in the $s, t, u$ channels individually:
\ie
\frac{D_{odd}}{E_T^2} &= -\frac{E_s}{E_T^2}\left(\frac{9}{2}(\hat\alpha^s)^2(\hat\beta^s)^2+\frac{3}{2}\right)+(t)+(u)
\fe
Finally, we propose an ansatz for $D_{even}$, which is completely fixed by the CWI:
\ie
\frac{D_{even}}{E_T^2} &=-\frac{1}{E_T^2}\left[\frac{1}{3}E_1+E_2+ \frac{3}{2}(\hat\alpha^s)^2 (E_1+E_2)+\frac{3}{2}(\hat\beta^s)^2 (E_3+E_4)+(t)+(u)\right]
\fe 
Combining all these terms in \eqref{eqn: 4-point WFC ansatz}, one can show that the 4-point WFC can be reorganized as
\begin{tcolorbox}
\ie
\langle O O O O \rangle&=
\frac{E_s\mathcal{Q}_{O,2s}}{E_T^2(E_{L}^s)^2(E_{R}^s)^2}+\frac{\mathcal{P}_{O,2s}}{E_T^3E_L^sE_R^s} +\mathcal{R}_{\mathcal{O},2s} + (t) + (u)
\fe
where we define
\ie
\mathcal{Q}_{O,2s} &=
    3\alpha^s_{i_s}\alpha^s_{j_s}\hat{\Pi}_s^{i_s j_s l_s m_s}\beta^s_{l_s}\beta^s_{m_s}-\frac{1}{2} (E^{s}_L)^2 (E_{R}^s)^2 (1-3(\hat\alpha^s)^2)(1-3(\hat\beta^s)^2)\\
\mathcal{P}_{O,2s} &=
    3\alpha^s_{i_s}\alpha^s_{j_s}\hat{\Pi}_s^{i_s j_s l_s m_s}\beta^s_{l_s}\beta^s_{m_s}
    -6
    E^{s}_L E_{R}^s
    (\alpha^s \cdot \pi_s \cdot \beta^s)\hat \alpha^s \hat \beta^s\\
    &+
    \frac{1}{2} (E^{s}_L)^2 (E_{R}^s)^2 (1-3(\hat\alpha^s)^2)(1-3(\hat\beta^s)^2)\\
\mathcal{R}_{\mathcal{O},2s} &=-\frac{1}{E_T^2}\left(\frac{E_1}{3}+E_2+E_s\right)+\frac{E_{R}^s E_{L}^s}{E_T^3}.
\fe
\end{tcolorbox}
If we examine our $s$-channel result, we find a mismatch of $R_{O,s}$ compared to the result presented in \cite{Baumann:2020dch}, which is itself a single-channel CWI solution. This mismatch term also possesses a leading total energy pole residue, which is simply the Mandelstam variable $S=E_{12s}E_{34s}$. At the leading total energy pole, this term is canceled under the $s, t, u$ channel combination. More generally, it can be shown that
\ie
\mathcal{R}_{O,2s} + (t) + (u) = 0.
\fe
After summing over the $s, t,$ and $u$ channels, our result matches the solution provided in \cite{Baumann:2020dch}. This agreement arises because only the combined $s, t, u$ channel CWI solution ensures that the residue of the leading total energy pole correctly reproduces the amplitude which is our input. Consequently, while standard approaches often start from conformal symmetry and focus on a single channel, our bootstrap procedures base on the full amplitude. This yields a combined-channel WFC that is more natural from a bulk perspective, as it shows a clear amplitude structure at the leading total energy pole.

Finally, we turn to $\langle \bar{\chi}\chi\bar{\chi}\chi \rangle$, four massless spin-1/2 fermions exchanging a graviton. Using the 3-point result \eqref{eq:Tchichi}, we determine $A_R^s$ for $\langle \bar{\chi}\chi\bar{\chi}\chi \rangle$:
\ie
\frac{A_R^s}{(E_R^s)^2} &=\frac{3}{4} \frac{\alpha^s_{i_s}\rho^s_{j_s}\hat{\Pi}_s^{i_s j_s l_s m_s}\beta^s_{l_s}\tau^s_{m_s}}{(E_R^s)^2 E_s^3}
         \left( 
        \begin{aligned}
            \frac{(E_s+ E^s_{L})(E_s+ E^s_{R})}{(E^s_{L})^2}
            +2\frac{E_s^2 E^s_{R}}
            {E_T^3}
            +\frac{E_s^2}{E_T^2}
            -\frac{E_s+E^s_{R}}{E_T}
        \end{aligned}
        \right)
\fe
where $\rho^s_i \equiv \bar u_1 \gamma_i u_2$ and $\tau^s_i \equiv \bar u_3 \gamma_i u_4$ are fully contracted spinor bilinears, and $\alpha^s, \beta^s$ are defined as in the previous example. It is also straightforward to write $B_L^s$ based on $A_R^s$ to match the expected singular term in the other partial energy pole, $E_L^s$:
\ie
\frac{B_{L}^s}{(E_L^s)^2}
&=
\frac{3}{4} \frac{\alpha^s_{i_s}\rho^s_{j_s}\hat{\Pi}_s^{i_s j_s l_s m_s}\beta^s_{l_s}\tau^s_{m_s}}{(E_L^s)^2 E_s^3}
\cdot
\left(
    2\frac{E_s^2 E^s_{L}}
        {E_T^3}
        +\frac{E_s^2}{E_T^2}
        -\frac{E_s+E^s_{L}}{E_T}
\right)
\fe
Similarly, one can obtain the terms associated with the partial energy poles in the $t$ and $u$ channels. Next, we examine the leading $E_T^{-3}$ behavior of the WFCs constructed from $A_R^e$ and $B_L^e$ as $E_T \to 0$, and determine the contact term $C$ by matching it with the flat-space amplitude:
\ie
M_{\bar\chi \chi \bar\chi \chi}=3\frac{(p_1-p_2)^{\mu_1} \bar u_1 \gamma^{\mu_2}  u_2  \cdot
\eta_{(\mu_1 (\nu_1} \eta_{\mu_2) \nu_2)} \cdot
(p_3-p_4)^{\nu_1} \bar u_3 \gamma^{\nu_2} u_4}{S} + (t) 
\fe
We then utilize the identities listed in our companion paper \cite{Flatspace} to obtain $C$:
\ie
\frac{C}{E_T^3} &=\frac{3}{2E_T^3} \left(
    \begin{aligned}
    &-\hat \alpha^s \hat \beta^s (\rho^s \cdot \pi_s \cdot \tau^s)
    -
    \hat \rho^s \hat \tau^s (\alpha^s \cdot \pi_s \cdot \beta^s)\\
    &-
    \hat \alpha^s \hat \tau^s (\rho^s \cdot \pi_s \cdot \beta^s)
    -
    \hat \rho^s \hat \beta^s (\alpha^s \cdot \pi_s \cdot \tau^s)
    +
    E_{12s}E_{34s}\hat \alpha^s \hat \beta^s \hat \rho^s \hat \tau^s
    \end{aligned} 
\right)+(t)
\fe
where $\hat\rho^s \equiv \bar u_1 \gamma_0 u_2 /E_s$ and $\hat\tau^s \equiv \bar u_3 \gamma_0 u_4 /E_s$ are additional fully contracted spinor bilinears, and $\hat\alpha^s, \hat\beta^s$ are defined as in the previous example. We fix $D_{odd}$ by comparing the difference of the $A, B, C$ insertions on the LHS of the cutting rule with the RHS given by $\langle T\bar\chi \chi \rangle$ in the $s$ and $t$ channels individually:
\ie
\frac{D_{odd}}{E_T^2} &= -\frac{3E_s}{2E_T^2}\left(
    \hat \alpha^s \hat \beta^s \hat \rho^s \hat \tau^s
    \right)+(t)
\fe
Finally, we propose an ansatz for $D_{even}$, which is completely fixed by the CWI:
\ie
\frac{D_{even}}{E_T^2} &= \frac{3}{2E_T}\left(
    \hat \alpha^s \hat \beta^s \hat \rho^s \hat \tau^s +(t)
    \right)
\fe 
Combining all these terms in \eqref{eqn: 4-point WFC ansatz}, one can show that the 4-point WFC can be reorganized into the following compact form:
\begin{tcolorbox}
\begin{equation}
    \langle \bar{\chi}\chi\bar{\chi}\chi \rangle=\frac{E_s\mathcal{Q}_{2s,\chi}}{E_T^2(E_{L}^s)^2(E_{R}^s)^2}+\frac{\mathcal{P}_{2s,\chi}}{E_T^3E_L^sE_R^s}+(t),
\end{equation}
where we adopt notation similar to \cite{Baumann:2020dch}:
\ie
\mathcal{Q}_{2s,\chi}=&3\Hat{\Pi}_s^{ijkl}\alpha^s_i\rho^s_j\beta^s_k\tau^s_l-\frac{9}{2}(E_L^s)^2(E_R^s)^2\hat{\alpha}^s\hat{\beta}^s\hat{\rho}^s\hat{\tau}^s,
\\
\mathcal{P}_{2s,\chi}=&3\Hat{\Pi}_s^{ijkl}\alpha^s_i\rho^s_j\beta^s_k\tau^s_l\\
-&
\frac{3}{2}\left(\hat{\rho}^s\hat{\tau}^s\alpha^s_i\pi_s^{ij}\beta^s_j+\hat{\alpha}^s\hat{\beta}^s\rho^s_i\pi_s^{ij}\tau^s_j+\hat{\alpha}^s\hat{\tau}^s\rho^s_i\pi_s^{ij}\beta^s_j+\hat{\beta}^s\hat{\rho}^s\tau_i^s\pi_s^{ij}\alpha^s_j\right).
\fe
\end{tcolorbox}
Note that all terms above contain products of two spinor bilinear forms. Given these products, by using the explicit forms of the gamma matrices \eqref{eq:gammamatrices} and the spin-1/2 polarizations \eqref{eq:spin1/2polarization}, one finds that the non-amplitude helicity configurations $(++++)$ and $(+++-)$ vanish.

\section{Consistency Between the Amplitude-Limit and Bulk Unitarity}
\label{sec:dsinconsistency}
We now turn to the tension between conserved spin-3/2  operator and dS isometry. In particular, we will pin-point the contradiction from our on-shell view point, which begins with the consistent flat-space amplitude. It is well known that consistent four spin-3/2 amplitude requires the presence of graviton exchange as well as reality conditions on the fermion (see appendix D of ~\cite{Flatspace}). This reality condition will be reflected on the boundary profile, which in turn, must be compatible with reality condition in the bulk which has $SO(1,4)$ isometry. There is a unique reality condition, which is  \emph{symplectic Majorana} (SM) condition:
\ie
\bar\chi^{\mathbb{I}}_5=\chi^{\mathbb{J},T}_5\epsilon^{\mathbb{J}\mathbb{I}}C_5,
\label{eq:5D SM}
\fe
where $\mathbb{I},\mathbb{J}$ are SO(2) indices, $C_5$ is the charge conjugate matrix and $\chi_5,\bar\chi_5$ are the 5D spinors. While $\text{EAdS}_4/\text{dS}_4$ share the \emph{same} isometry, the reality condition realized \emph{differently} in the four-dimensional Euclidean/Lorentzian flat-space amplitude-limit. As a consequence, the flat-space amplitude will have distinct couplings in SO(2) for the   $A\bar\psi\psi$ and $h\bar\psi\psi$ amplitude. This leads to a sign difference between the graviton and photon exchange in the $4\psi$ amplitude in Lorentzian but not in Euclidean signature. This unitarity-violation echos the wrong sign for gravi-photon kinetic term in \cite{Pilch:1984aw}. 

For concreteness,
we streamline the strategy in the following graph,

\begin{tikzpicture}[
  node distance=0.5cm, 
  >=Stealth,
  every node/.style={align=center}
]
\node (sm) {SO(1,4) SM on \(\bar u^\mathbb{I}\)\(\Big\{\begin{array}{rcl}&\text{Lorentzian}\\
&\text{Euclidean}\;
\end{array}\)};
\node[right=of sm] (lij) {\(l_\mathbb{IJ}\;\text{in}\; T\bar\psi\psi \;\Big\{\begin{array}{rcl}&\delta_\mathbb{IJ}\\
&\epsilon_\mathbb{IJ}\;
\end{array}\)}
;
\node[right=of lij] (amp) {Unitarity \(\Big\{\begin{array}{rcl}&\text{\;Inconsistent}\\
&\text{\;Consistent}.
\end{array}\)};

\draw[->] (sm) -- (lij);
\draw[->] (lij) -- (amp);
\end{tikzpicture}
\subsection{Reality conditions in the Amplitude-Limit}
Staring with the SO(1,4) SM condition in eq.(\ref{eq:5D SM}), we can identify the four-dimensional embedding by choosing the SO(4) or SO(1,3) sub-group for Euclidean and Lorentzian space. The details of the representation for the different Clifford algebras are given in the appendix \ref{sec:SO(1,4)details}. Thus the four-dimensional SM condition on the fermionic wave factors $\bar{u}$ becomes,  
\ie
   \text{Lorentzian: } \bar u^{\mathbb{I}}=(u^{\mathbb{J}})^TC_{+,L}\epsilon^{\mathbb{J}\mathbb{I}}, \quad
   \text{Euclidean: } \bar u^{\mathbb{I}}=(u^{\mathbb{J}})^TC_{-,E}\epsilon^{\mathbb{J}\mathbb{I}}
\label{eq:SM on polarization}
\fe
where the subscripts L, E denote Lorentzian and Euclidean and the $\pm$ in the charge conjugate matrices reflect the conventional choices in 4D Clifford algebra \eqref{eq:C4}. The corresponding constraint on the boundary profile, for details see App. \ref{sec:SO(1,4)details}
\ie
  \text{dS: } 
  \bbarpsi^{i,\mathbb{I}}_{\zero,-} = \left(\bpsi^{i,\mathbb{J}}_{\zero,-}\right)^T C_{+,L} \epsilon^{\mathbb{J}\mathbb{I}}, \quad
  \text{EAdS: }
  \bbarpsi^{i,\mathbb{I}}_{\zero,+} = \left(\bpsi^{i,\mathbb{J}}_{\zero,-}\right)^T C_{-,E}\epsilon^{\mathbb{JI}}.
\fe

Now consider the 3-pt amplitude of two gravitino and a photon or graviton. This will be the seed for the WFC. If the bulk is given by a two derivative gravitational theory, then the amplitude is uniquely fixed to,
\ie
M({J\psi\psi})=M(JJJ)\left(l^{\mathbb{I}\mathbb{J}} \bar u^\mathbb{I}u^\mathbb{J}\right),\quad 
M(T\psi\psi)= M(JJJ)\left(l^{\mathbb{I}\mathbb{J}}\bar{u}^\mathbb{I}\slashed{\epsilon}_1 u^\mathbb{J}\right)\,,
\label{eq:JpsipsiTpsipsi amplitude undetermined lIJ}
\fe
where $l^{\mathbb{I}\mathbb{J}}$ is an SO(2) invariant tensor, $\delta
^{\mathbb{I}\mathbb{J}}$ or $\epsilon^{\mathbb{I}\mathbb{J}}$, and $M(JJJ)$ is the 3-pt Yang-Mills amplitude. Since the fermions are now real, the amplitude should now satisfy fermi-statistics, namely
\begin{equation}
    M(J\psi_2\psi_3)=-M(J\psi_3\psi_2),\quad M(T\psi_2\psi_3)=-M(T\psi_3\psi_2)\,.
\end{equation}
Now since the $M(JJJ)$ is totally anti-symmetric, the parenthesis in eq.(\ref{eq:JpsipsiTpsipsi amplitude undetermined lIJ}) should be \emph{even} under permuting $2\leftrightarrow3$. Incorporating the reality conditions in \eqref{eq:SM on polarization}, together with the algebraic properties of $C$ and gamma matrices \eqref{eq:C4}, we find the following identities:
\ie
 \bar u^\mathbb{I}_2u^\mathbb{J}_3
&=\epsilon^{\mathbb{I'I}}\epsilon^{\mathbb{J'J}}\bar u^\mathbb{J'}_3u^\mathbb{I'}_2, 
\quad \bar{u}_2^\mathbb{I}\slashed{\epsilon}_1 u^\mathbb{J}_3
=
\epsilon^{\mathbb{I'I}}\epsilon^{\mathbb{J'J}}\bar{u}_3^\mathbb{J'}\slashed{\epsilon}_1 u^\mathbb{I'}_2 \text{  (Lorentz)}\\ 
 \bar u^\mathbb{I}_2u^\mathbb{J}_3
&=\epsilon^{\mathbb{I'I}}\epsilon^{\mathbb{J'J}}\bar u^\mathbb{J'}_3u^\mathbb{I'}_2,
\quad
\bar{u}_2^\mathbb{I}\slashed{\epsilon}_1 u^{\mathbb{J}}_3=-\epsilon^{\mathbb{I'I}}\epsilon^{\mathbb{J'J}}\bar{u}_3^\mathbb{J'}\slashed{\epsilon}_1 u^{\mathbb{I'}}_2\text{  (Euclidean)}.
\fe
As one can see, the last term becomes $2\leftrightarrow 3$ symmetric if one contracts with $\epsilon^{\mathbb{IJ}}$. For the others, one simply contracts with $\delta^{\mathbb{IJ}}$. Thus the tensor $l^{\mathbb{I} \mathbb{J}}$ is determined to be, 
\ie
 M(T\psi\psi)=
\begin{cases}
    \text{L: }\delta^{\mathbb{I}\mathbb{J}}\; M(JJJ) \left(\bar{u}^\mathbb{I}\slashed{\epsilon} u^\mathbb{J}\right) \\
    \text{E: } \epsilon^{\mathbb{I}\mathbb{J}}\; M(JJJ) \left(\bar{u}^\mathbb{I}\slashed{\epsilon} u^\mathbb{J}\right)
\end{cases}
, \quad M({J\psi\psi})=\delta^{\mathbb{I}\mathbb{J}}\;  M(JJJ) \bar u^\mathbb{I}u^\mathbb{J}\;\text{(L and E)}\,.\label{eq:amplitudeswithlij}
\fe
The reality of the overall normalization is fixed by requiring the amplitude to be CPT even.

\subsection{Unitarity violation of four gravitino amplitude}
With the 3-pt amplitude completely determined, we can now conduct the ``four-particle" test for the four gravitino amplitude. This is most easily studied in the spinor-helicity formalism. See appendix \ref{sec:SO(1,4)details}, for conversion formulae. For the graviton-gravitino-gravitino amplitude we have,
\begin{equation}
 M_{\text{L}}\left(T^{+} \psi^{+} \psi^{{-}}\right)^\mathbb{IJ}=-\epsilon^{\mathbb{IJ}} \frac{[1 2]^5}{[2 3]^2[31]} ,\quad M_{\text{E}}\left(T^{+} \psi^{+} \psi^{{-}}\right)^\mathbb{IJ}=\delta^{\mathbb{IJ}} \frac{[1 2]^5}{[2 3]^2[31]}.
\end{equation}
where we have used the SM conditions in the chiral basis \eqref{eq:SMonchiral}. For the gravi-photon-gravitino-gravitino amplitude, we have,
\ie
 M_{\text{L}/\text{E}}\left(J^{{+}} \psi^{{-}} \psi^{{-}}\right)^{\mathbb{IJ}} =  
\epsilon^{\mathbb{IJ}}
\frac{ {}_{}\langle 2 3\rangle_{}^4 }{\langle1 2\rangle_{} \langle1 3\rangle_{}}.
\fe
 We now build the 4-pt amplitude $ M\left(\psi^{+} \psi^{-} \psi^{-} \psi^{+}\right)^{\mathbb{I}_1\mathbb{I}_2\mathbb{I}_3\mathbb{I}_4}$. Let's glue the 3-pt amplitudes from the $S$-channel.  The $S$- and $U$- channel residue can be related by $2\leftrightarrow 3$ exchange and one has,
\ie
M_{\text{L}}\left(\psi^{+} \psi^{-} \psi^{-} \psi^{+}\right)^{\mathbb{I}_1\mathbb{I}_2\mathbb{I}_3\mathbb{I}_4}
&=
\frac{[1 4]^3\langle2 3\rangle^3}{T}\left(\frac{\epsilon^{\mathbb{I}_1\mathbb{I}_2}\epsilon^{\mathbb{I}_3\mathbb{I}_4}}{S}+\frac{\epsilon^{\mathbb{I}_1\mathbb{I}_3}\epsilon^{\mathbb{I}_2\mathbb{I}_4}}{U}\right) \\
M_{\text{E}}\left(\psi^{+} \psi^{-} \psi^{-} \psi^{+}\right)^{\mathbb{I}_1\mathbb{I}_2\mathbb{I}_3\mathbb{I}_4}
&=
\frac{[1 4]^3\langle2 3\rangle^3}{T}\left(\frac{\delta^{\mathbb{I}_1\mathbb{I}_2}\delta^{\mathbb{I}_3\mathbb{I}_4}}{S}+\frac{\delta^{\mathbb{I}_1\mathbb{I}_3}\delta^{\mathbb{I}_2\mathbb{I}_4}}{U}\right).\label{eq:4psigravitonexchange}
\fe
Note the a massless $T$-channel pole is present even though we were considering $S$- and $U$-channel factorizations. This is a common feature for massless higher spins, i.e. the gluing of higher spin amplitudes in one channel reveals factorization poles in an other channel. Let us now consider the $T$ channel factorization,
\ie &\lim_{T\to0}M_{\text{L}}\left(\psi^{+} \psi^{-} \psi^{-} \psi^{+}\right)^{\mathbb{I}_1\mathbb{I}_2\mathbb{I}_3\mathbb{I}_4}
=\frac{[1 4]^3\langle2 3\rangle^3}{ST}
(\epsilon^{\mathbb{I}_1\mathbb{I}_2}\epsilon^{\mathbb{I}_3\mathbb{I}_4}-\epsilon^{\mathbb{I}_1\mathbb{I}_3}\epsilon^{\mathbb{I}_2\mathbb{I}_4})=-\frac{[1 4]^3\langle2 3\rangle^3}{ST}(\epsilon^{\mathbb{I}_1\mathbb{I}_4}\epsilon^{\mathbb{I}_2\mathbb{I}_3})
\\  &\lim_{T\to0}M_{\text{E}}\left(\psi^{+} \psi^{-} \psi^{-} \psi^{+}\right)^{\mathbb{I}_1\mathbb{I}_2\mathbb{I}_3\mathbb{I}_4}
=\frac{[1 4]^3\langle2 3\rangle^3}{ST}
(\delta^{\mathbb{I}_1\mathbb{I}_2}\delta^{\mathbb{I}_3\mathbb{I}_4}-\delta^{\mathbb{I}_1\mathbb{I}_3}\delta^{\mathbb{I}_2\mathbb{I}_4})=
\frac{[1 4]^3\langle2 3\rangle^3}{ST}(\delta^{\mathbb{I}_1\mathbb{I}_4}\delta^{\mathbb{I}_2\mathbb{I}_3})\,,
\label{eq:4psiphotonexchange}
\fe
where in the last line, we've rewritten the residue in a form of product of 3-pt amplitudes in the $T$-channel. We can see that there is a relative \emph{minus sign} in the Lorentzian case! Thus we see that the total energy pole of conserved spin-$\frac{3}{2}$ WFC has a non-unitary amplitude in dS space. From the Lagrangian perspective, this incompatibility manifests through the relative minus sign between the kinetic terms of the gravi-photon and graviton \cite{Pilch:1984aw}.

\subsection{Possible resolutions}
We've seen in the previous discussion, that the consistency between the flat-space limit and de Sitter isometry leads to the unitarity violation for the flat-space amplitude. Note that the discussion started with one derivative coupling for the 3-pt amplitude between two gravitinos. Had we started with three derivative couplings, the final $\langle T\bar\psi\psi\rangle$ would be dual to that of conformal supergravity, which in itself, is non-unitary. Since the inconsistency appears on the total-energy pole, one might avoid the problem by introducing new features that modify the degree of total energy pole, such that the leading pole no longer corresponds to supergravity.

First, one might consider adding  higher-derivative contact terms, which in turn induces a  higher-order $E_T$ pole. The residue for the leading order $E_T$ pole now simply becomes a polynomial amplitude. However, such an EFT-type solution is really a mirage. In flat space the EFT expansion is an expansion in  $p/\Lambda$, where $\Lambda$ is the cutoff. At low energies this is a convergent series. In (AdS)dS, now each extra power of $\Lambda$ comes with an extra $E_T$ and thus higher order in $1/\Lambda$ is more dominant than lower order when $E_T\rightarrow 0$, i.e. the EFT expansion is no longer well defined in the limit. This is equivalent to saying that in this limit, the WFC is proportional to the high energy limit of the amplitude, as stressed in \cite{Arkani-Hamed:2018kmz, DuasoPueyo:2025lmq}. Thus even if one were to consider turning on higher derivative terms, in the $E_T\rightarrow 0$ limit the flat-space amplitude takes its high energy limit form. Importantly, irrespective of the nature of the UV completion, which only determines how soft the WFC behaves as $E_T\rightarrow 0$, the low energy massless poles along with its negative residues are still present! Indeed the massless pole of high-energy string amplitudes is proportional to its field theoretic limit. 

The discussion above suggests that we need to increase the total energy pole via the exchange diagram. We can consider adding derivatives to the vertices of the exchange diagram, or introduce new higher spin exchanges. Adding higher derivatives to the 3-pt coupling simply leads us to conformal supergravity as discussed previously. 
Thus we turn to the introduction of additional higher-spin fields. Let us first analyze the self-consistency of introducing such fields at the amplitude level. 

The simplest possibility is to minimally couple a massless bosonic spin-$s$ field $\varphi_s$ ($s \ge 3$) two the gravitinos. The corresponding three-point amplitude is
\begin{equation}
M_L(\varphi_{s}^-\psi^{+}\psi^{-}) = \langle23\rangle^{-s}\langle13\rangle^{s+3}\langle12\rangle^{s-3}\,.
\end{equation}
where we have suppressed the symplectic indices and the structure constant. Gluing the three-point into four-point in the $S$-channel leads to:
\ie
\lim_{S\to0}M_{\text{L}}\left(\psi^{+} \psi^{-} \psi^{+} \psi^{-}\right)
&=
\frac{T^{s-3}}{S} [1 4]^3\langle2 3\rangle^3\,.
\fe
Note that for $s\geq 3$ the residue does not contain poles in other channel. Such exchange introduces a total-energy pole of order $p=2s-1$ and should do the job. However due to this new 3-pt amplitude, we should also consider the two $\varphi_s$ two $\psi$ amplitude. In the $S$-channel, we find, 
\ie
    \lim_{S\to0}M_L(\varphi_s^{+}\psi^{+} \varphi_s^{-}\psi^{-})=\frac{T^{-s}}{S}\langle 34 \rangle^3[12]^{2s}\langle 32 \rangle^{2s-3},
\fe
One can see the presence of a higher-order pole when $s\geq3$, which thus violates locality.



At this point, one could introduce a new fermionic massless higher-spin field,  $\chi_{s'}$ with spin $s' \geq 3s {-}\tfrac{1}{2}$, with mix-coupling between the gravitino and the bosonic spin $s$ field, 
\ie
M_L(\varphi_s^{+}\psi^{+}\chi_{s'}^{-})=\langle 1 2\rangle^{-s-3/2-s'} \langle 2 3\rangle^{s-3/2+s'} \langle 1 3\rangle^{-s+3/2+s'}. 
\fe
The exchange amplitude of this field then becomes the leading $E_T$-pole residue (at order $p=2s'{-}2s{-}4$) and now the two $\varphi_s$ two $\psi$ amplitude becomes local, 
\ie
    \lim_{S\to0}M_L(\varphi_s^{+}\psi^{+} \varphi_s^{-}\psi^{-})=\frac{T^{s'-3s -3/2}}{S}\langle 34 \rangle^3[12]^{2s}\langle 32 \rangle^{2s-3}\,.
\fe
However, this again generates another inconsistent interaction in mixed four-fermion amplitude  with the same total energy pole order:
\ie
 \lim_{S\to0}M_L(\chi_{s'}^{-} \psi^+ \chi_{s'}^{+}\psi^-)=\frac{T^{-s'-s-3/2}}{S}\langle 14 \rangle^3[32]^{2s'}\langle 12 \rangle^{2s'-3},
\fe
which is clearly non-local. Now, we're back to a similar problem that we began with. To wash it out, one would need to introduce another bosonic higher-spin field $\varphi_{s''}$ with $s'' \geq 2s'-5$ and the process repeats indefinitely.

Note that, the above procedure will lead to an infinite number of bosonic and fermionic higher-spin fields. Interestingly, this spectrum is reminiscent of Vasiliev-like construction \cite{Vasiliev:2003ev}, for supersymmetric dS theories \cite{Sezgin:2012ag,Hertog:2017ymy}. We will comment on this in the conclusion.

\section{Conclusions}

Starting from a flat-space S-matrix—already constrained by its own stringent consistency conditions—we systematically impose conformal Ward identities (CWIs) together with partial-energy pole constraints to construct the three- and four-point functions. Remarkably, these three ingredients suffice: additional expected constraints are automatically satisfied by the resulting solutions. For instance, the bootstrap outputs pass Manifest Locality Tests~\cite{Jazayeri:2021fvk, Bonifacio:2022vwa} and obey the appropriate Ward–Takahashi identities. Even more intriguingly, our analysis suggests that the flat-space amplitude may not need to be supplied as input at all. Once the leading total-energy pole is parameterized with the correct momentum dimensions, the CWIs themselves uniquely force it to coincide with the flat-space amplitude. This suggests that there is a deeper interplay between the constraints.

Unlike its flat-space counter part where the bootstrap guarantees a solution, here to solve CWI we are required to introduce ansatze for subleading poles that does not guarantee a solution. This is a reflection of the fact that not all consistent interacting theories in flat space are admissible in (EA)dS. As an application, we consider the WFC of conserved spin-3/2 operators. We find that the WT identity at 3-pts suggests that it is charged under the U(1), indicating that minimal coupling, which is absent in flat space, must emerge in (EA)dS. Furthermore, compatibility of reality conditions of flat-space amplitude and dS isometry leads to inconsistent factorization for the total energy pole residue of the 4 spin-3/2 WFC. Thus ruling out a conserved spin-3/2 operator with bulk Einstein gravity.

There are several immediate avenues for future investigation:

\begin{itemize}
    \item Supersymmetric  WFCs: As mentioned in the introduction, there is no prohibition of supersymmetry in dS. Any exact conformal theory in flat-space can be consistently put in dS. It will be interesting to consider the supersymmetric WFCs using massive on-shell superspace. An optimal candidate would be N=4 SYM theory. 
    
    \item The cutting rules display an intriguing feature: on the right-hand side they introduce a spurious singularity—the internal-energy pole—that is absent on the left-hand side. The requirement that this pole cancel imposes nontrivial constraints on the derivative expansion of lower-point WFCs, a condition known as manifest locality. Strikingly, our bootstrap procedure never needed to invoke this constraint explicitly; the conformal Ward identities alone proved sufficiently restrictive. It would be interesting to understand the origin of this redundancy. One possibility is that flat-space locality, already encoded in the S-matrix used as the seed, fully captures the necessary locality structure, leaving no additional independent constraints to be enforced. It will be interesting to clarify this.

    \item There has been considerable emphasis on the appearance of the flat-space scattering amplitude as the leading contribution in the vanishing total-energy limit. In fact, massive interactions in the bulk will also lead to massless amplitudes in the $E_T\rightarrow 0$ limit. This naturally raises the question of what constitutes the leading term in the $E_T\rightarrow 0$ expansion of higher spin WFCs, given the fact that there are no consistent massless higher-spin amplitudes. 
    
    For non-conserved higher-spin operators, the corresponding boundary states are typically double-trace operators—dual to multi-particle configurations \cite{Witten:2001ua}—or stringy excitations. In such cases, we expect that the leading contribution in the small-$E_T$
 expansion is governed by an appropriate collinear higher-point or high-energy limit of string amplitudes, respectively. On the other hand, less is known with regards to Vasiliev-like higher-spin theories, whose boundary operator are conserved higher-spin currents. It will be interesting to study the Fourier transform of its boundary correlation function, for which some are known, and study its leading $E_T$ term. We leave this for future study. 
\end{itemize}

\section*{Acknowledgments}
We thank  Mattia Arundine, Daniel Baumann, Harry Goodhew, Austin Joyce, Hikaru Kawai, Jiajie Mei and Suvrat Raju  for enlightening discussions. We especially thank Harry Goodhew for detailed comments on the draft. Y-t H  thanks Riken iTHEMS and the Yukawa Institute for
Theoretical Physics at Kyoto University. Discussions during
“Progress of Theoretical Bootstrap” were useful in completing this work.
Y-t H Z-X H and Y L are supported by the Taiwan National Science and Technology Council grant 112-2628-M-002-003-MY3 and 114-2923-M-002-011-MY5. W.-M. C. is supported by the NSTC through Grant NSTC 112-2112-M-110-013-MY3 and NSTC 114-2811-M-110-009.

\appendix

\section{Conventions and notation}
\label{sec: Conventions and notation}
\paragraph{Gamma Matrices} The explicit form of the $4\times 4$ gamma matrices we use in this paper are,
\begin{equation}
    \gamma_0 =\begin{pmatrix}
         ~-i& ~~0~~~  \\
         ~~0& ~~i~~~
    \end{pmatrix},\quad~~~
    \gamma_i=\begin{pmatrix}
         ~0&~~\sigma_i~~  \\
         ~\sigma_i&~~ 0 ~~
    \end{pmatrix}, \quad
    \gamma_5=\begin{pmatrix}
         ~0&~~-i~~  \\
         ~i&~~ 0 ~~
    \end{pmatrix}.
    \label{eq:gammamatrices}
\end{equation}
 The explicit form of the $2\times 2$ Pauli matrices are,
\begin{equation}
    \sigma_1=\left( \begin{array}{cc} 0 & 1 \\
1 & 0 \end{array} \right),\quad\sigma_2=\left( \begin{array}{cc} 0 & -i \\
i & 0 \end{array} \right),\quad\sigma_3=\left( \begin{array}{cc} 1 & 0 \\
0 & -1 \end{array} \right).
\end{equation}
where they satisfy the familiar equation for Pauli matrices\,,
\ie
(\sigma_i)^a_{~\,b} (\sigma_j)^b_{~\,c} =\delta^a_c \delta_{ij}+i\epsilon_{ijk} (\sigma^k)^a_{~\,c}\,.
\fe
The $SU(2)$ spinor indices can be raised and lowered by $\epsilon_{ab}$ and $\epsilon^{ab}$ as 
\ie
\epsilon^{cb} T^{ a}_{\,~ b }= T^{ a c}\,,~~~~\epsilon_{ca} T^{a}_{\,~b}= T_{cb}\,,
\fe
and moreover, we define
$
\epsilon_{12}=-\epsilon^{12}=1\,.
$
Then following the notation in \cite{Baumann:2020dch}, the 3D spatial momentum could be expressed in the spinor basis as
\ie
\vk_i (\sigma^i)_{ab}\equiv \vk_{ab}=\frac{1}{2}\left(\lambda_a\bar \lambda_b+\lambda_b\bar \lambda_a\right)=\lambda_{(a}\bar \lambda_{b)}\,.
\fe
We also define the inner product of two spinors\,,
\ie
\braketw{i}{j}\equiv \epsilon_{ab}\lambda^a_i  \lambda^b_j=\lambda^a_i  \lambda_{j,a}\,.
\fe
Now the on-shell condition could be written as
\ie
E^2=-\frac{1}{2} \vk_{ab}\vk^{ab}=\frac{\braketw{\lambda}{\bar\lambda}^2}{4}\,,
\fe
and a consistent choice is
\ie
E\equiv -\frac{\braketw{\lambda}{\bar \lambda}}{2}\,.
\fe
We could write down transverse polarization vectors in the helicity basis as,
\ie
\label{eq: SPH polarization}
(\epsilon^{(-)})^{ab}=\frac{\lambda^{a}\lambda^{b}}{\braketw{ \lambda}{\bar \lambda}}\,\,,~~~({\epsilon}^{(+)})^{ab}=\frac{\bar \lambda^{a}\bar \lambda^{b}}{\braketw{\lambda}{\bar \lambda}},
\fe
which satisfy
\begin{equation}
(\epsilon^{(+)})_{ab}(\epsilon^{(-)})^{ab}=1,\quad (\epsilon^{(\pm)})_{ab}(\epsilon^{(\pm)})^{ab}=0 ,\quad k_{ab}(\epsilon^{(\pm)})^{ab}=0.    
\end{equation}
And the 4D spinor helicity form of $\bar u$ and $u$ could be obtained firstly by expressing the boundary profiles in terms of $ \bar\lambda_a, \lambda_a$\cite{Flatspace},
\ie
    \bbarchi^{(+)}&=\Big(0,(\bar{\lambda})^a\Big),\quad\bbarchi^{(-)}=\Big(0,(\lambda)^a\Big),\\
	\bchi^{(-)}&=\left(\begin{array}{c}
	(\lambda)_a
	\\ 0
	\end{array}\right),\quad \bchi^{(+)}= \left(\begin{array}{c}
	(\bar \lambda)_a
	\\ 0
	\end{array}\right).
\fe 
Then we could use the relation between the polarizations and profiles to get the corresponding spinor helicity forms of $\bar u$ and $u$ \cite{Flatspace},
\ie
\label{eq:spin1/2polarization}    
\bar{u}^{(+)}&=\Big(i(\bar{\lambda})^a,(\bar{\lambda})^a\Big),\quad\bar{u}^{(-)}=\Big(-i(\lambda)^a,(\lambda)^a\Big),\\
	u^{(-)}&=\left(\begin{array}{c}
	(\lambda)_a
	\\ -i (\lambda)_a
	\end{array}\right),\quad u^{(+)}= \left(\begin{array}{c}
	(\bar \lambda)_a
	\\ i (\bar \lambda)_a
	\end{array}\right).
\fe
As a consistency check, we can find that they're also the eigenbases of $\gamma_5$, with $\gamma_5 u^{(\pm)} = \pm u^{(\pm)}, \, \bar u^{(\pm)} \gamma_5 = \pm \bar u^{(\pm)}$. And satisfy Dirac equation by construction.
\paragraph{Momentum Dependence and Energy Variables}
Throughout this paper, the momentum dependence of the operators (particles) in the WFCs (amplitudes) follows their position in the bracket from left to right unless otherwise stated. For example,
\begin{equation}
\langle\mathcal{O}\mathcal{O}\mathcal{O}\rangle=\langle\mathcal{O}_1\mathcal{O}_2\mathcal{O}_3\rangle,
\end{equation}
and similarly for the amplitudes.
Energy variables with multiple lower indices are defined as sum of the individual energies. For example,
\begin{equation}
    E_{13u}\equiv E_1+E_3+E_u.
\end{equation}

\section{Residual Gauge Symmetry: From Bulk to Boundary}
\label{sec:residualsym}
In this appendix, we discuss how the residual bulk gauge symmetry becomes the gauge transformation of boundary profiles in flat and curved backgrounds.

First we consider the U(1) gauge field which has the standard transformation rule,
\ie
\label{eq:temporalgaugebulkgauge for vector field}
\delta A_\mu(\vec x, \eta) = \partial_\mu \alpha(\vec x, \eta).
\fe
Note that the above holds regardless of the background, since the gauge parameter $\alpha(\vec x, \eta)$ is a \emph{scalar} thus the covariant derivative reduces to the partial derivative. Next we consider $A_\mu(\vec{x},\eta)$ in the temporal gauge $A_0(\vec{x},\eta)=0$. After gauge-fixing, there remains a residual gauge symmetry that leaves the condition unchanged,
\begin{equation}
   \delta A_0(\vec x, \eta) = 0 = \partial_0 \alpha_\partial(\vec x). 
\end{equation}
Therefore, taking the $\eta=0$ limit in \eqref{eq:temporalgaugebulkgauge for vector field} one will obtain the following gauge transformation of the boundary profile as a result of bulk residual gauge symmetry,
\ie
\delta A_{i,\zero}(\vec x) = \partial_i \alpha_\zero(\vec x).
\fe
Next we move on to the graviton, \emph{where the difference between flat and curved background starts showing up due to the spinning gauge parameter}. As a convention, we set the perturbation of the metric as $g_{\mu\nu} = g^{(0)}_{\mu\nu} +\kappa h_{\mu\nu}$.\footnote{It is straightforward to see that $g^{(0)}_{\mu\nu}=\eta_{\mu\nu}$ in flat space, and $g^{(0)}_{\mu\nu}=\frac{1}{H^2 \eta^2} \eta_{\mu\nu}$ in dS space. It is easy to extend the discussion to (E)AdS space, so we demonstrate the discussion in dS space from now on. When the metric go to boundary $\eta\rightarrow \epsilon$, the boundary metric is given by $g_{ij,\zero}^{(0)}=\frac{\delta_{ij}}{H^2\epsilon^2}$. We could absorb the additional dimensionless factor into the normalization of the boundary action, and send $g_{ij,\zero}^{(0)}$ to $\delta_{ij}$ in the  boundary profile discussion.} The gauge transformation on $h_{\mu\nu}$ is $\delta h_{\mu\nu}(\vec{x},\eta)=2\nabla_{(\mu}\xi_{\nu)}(\vec{x},\eta)$, where $\nabla_\mu$ can be expanded order by order in $\kappa$. We will from now on consider the transformation in the $\kappa^0$ order. For later convenience, we lay out explicitly $\delta h_{ij}$ in both spaces (suppressing the 4D coordinate dependence),
\ie
(\delta h_{ij})^{(0)}
= \begin{cases} 
2 \partial_{(i} \xi_{j)}, & \text{flat} \\
2 \partial_{(i} \xi_{j)} - \frac{2}{\eta} \delta_{ij} \xi_0, & \text{dS},
\end{cases}
\label{eq:spin-2ijvariation}
\fe
where we find that the curvature effect shows up in the dS as an additional scalar parameter. Next, we adopt the temporal gauge condition $h_{0\mu}(\vec{x},\eta)=0$, investigate the constraints of the parameters of the residual gauge symmetry and see how do they reflect on the boundary profiles. Starting from $\delta h_{00}$, we find,
\ie
(\delta h_{00})^{(0)}=0=\begin{cases}
2\partial_0 \xi_0 \;\rightarrow \;\xi_0(\eta,\vec x) = \sigma_\zero(\vec x) & \text{flat} \\
2\partial_0 \xi_0 + \frac{2}{\eta} \xi_0 \;\rightarrow \;\xi_0(\eta,\vec x) = \frac{\epsilon}{\eta} \sigma_\zero(\vec x),& \text{dS}
\end{cases}
\fe
where we have introduced a late-time cut-off $\epsilon$ in the dS to ensure the condition can be extended to the $\eta=\epsilon$ boundary. For the $\delta h_{i0}$ components, we then find constraints between $\xi_i(\vec{x},\eta)$ and $\sigma_{\partial}(\vec{x})$,
\ie
\label{eq:spin-2ih0variation}
(\delta h_{i0})^{(0)}=0=\begin{cases} 
\partial_i \sigma_\zero+ \partial_0 \xi_i
    \;\rightarrow\; 
    \xi_i(\vec x,\eta) 
    =
    \xi_{i,\zero}(\vec x) + \eta \partial_i \sigma_\zero(\vec x) &\text{flat}  \\
\partial_i \sigma_\zero+ \partial_0 \xi_i+ \frac{2}{\eta} \xi_i
    \;\rightarrow \;
    \xi_i(\vec x,\eta) 
    =
    \frac{\epsilon^2}{\eta^2}\xi_{i,\zero}(\vec x) - \frac{\eta}{3} \partial_i \sigma_\zero(\vec x). &\text{dS}
\end{cases}
\fe
However, we have an additional constraint from the Bunch-Davies boundary condition, which requires that $\lim_{\eta\to-\infty} h_{ij}<\infty$. The residual gauge transformation $\delta h_{ij}=\nabla_{(i} \xi_{j)}$ should preserve this property. It is straightforward to see that in flat space, this constraint is satisfied only when $\sigma_\zero=0$. On the contrary, in the dS, the power-law $\eta$ dependence ensures this condition and thus \emph{there are two parameters left}. Therefore, pushing \eqref{eq:spin-2ijvariation} to the boundary with the gauge-preserving $\xi_i$ in the flat space simply yields the boundary diffeomorphism.

In dS, taking $\eta\rightarrow \epsilon$ and keeping the leading order in $\epsilon$, we can see that at linearized order in $\kappa$, and considering the scalar classical solution $\lim_{t\to0}\phi|= (\frac{\eta}{\epsilon})^{3-\Delta_{\mathcal{O}_\phi}} \phi_\zero$, the gauge transformation on the boundary profile is \footnote{Here, we set $H=1$ for simplicity.}
\ie
(\delta h'_{ij,\zero})^{(0)} = 2 \partial_{(i} \xi'_{j,\zero)} - 2 \delta_{ij} \sigma'_\zero, \,
(\delta \phi_{\zero})^{(0)} =  \left[\sigma'_\zero(\vec x) (3-\Delta_{\mathcal{O}_\phi}) \phi_{\zero} + \xi'_{i,\zero}(\vec x) \partial_i \phi_{\zero} \right]
\fe
after redefining the boundary profile $h_{ij,\zero}'=\epsilon^2 h_{ij,\zero}, \xi'_{i,\zero}=\epsilon^2 \xi_{i,\zero},\sigma'_\zero=\epsilon \sigma_\zero$. We can see that the additional gauge parameter $\sigma_\zero(\vec x)$ is a physical degree of freedom, and the residual gauge symmetry on the boundary is boundary diffeomorphism plus Weyl transformation.

Let us at last consider the spin-3/2 (gravitino) case. We will consider $\mathcal{N}=1$ in the flat space and $\mathcal{N}=2$ in the EAdS.   (Below we combine the two symplectic Majorana fermions into a Dirac one for compactness). The transformation of the bulk gravitino at the linearized order is,\footnote{The $\hat\nabla_i$ in EAdS space is defined as $\hat{\nabla}_i = \nabla_i - \frac{\Gamma_i}{2},\Gamma_i=e^i_a \gamma^a=\frac{1}{t}\gamma_i+\mathcal{O}(\kappa)$, which is the covariant derivative in EAdS. The linear term is required to make the commutator equal to the background curvature tensor. In flat space, we take $\hat\nabla_i = \nabla_i$.}
\ie
(\delta \bpsi_{i})^{(0)} = (\hat\nabla_i \bm{\epsilon})^{(0)} = \begin{cases}
\partial_i \bm{\epsilon}  & \text{flat} \\
\partial_i \bm{\epsilon} -\frac{\gamma_i\gamma_{0,E}}{2\eta}\bm{\epsilon} -\frac{\gamma_i}{2\eta} \bm{\epsilon} & \text{EAdS}.
\end{cases}
\fe
In particular, after projecting to the minus eigenstate of $\gamma_0$ we have,
\ie
(\delta \bpsi_{i,-})^{(0)} = \begin{cases}
\partial_i \bm{\epsilon}_{-}  & \text{flat} \\
\partial_i \bm{\epsilon}_{-} -\frac{\gamma_i}{\eta} \bm{\epsilon}_{+} & \text{EAdS}.
\end{cases}
\label{eq:deltapsiibulk}
\fe
As before, now we would like to see the constraints on the parameters implied by maintaining the temporal gauge condition $\delta\bm{\psi}_0=0$. After that, we can then substitute the constraints and push the bulk variation of the spatial component \eqref{eq:deltapsiibulk} onto the boundary to obtain the transformation of the boundary profiles.

Requiring the temporal gauge unchanged leads to the following constraints on the residual gauge parameters,\footnote{The BD boundary condition on the physical degrees of freedom $\lim_{t\to-\infty}\delta h_{ij}<\infty, \lim_{t\to-\infty}\delta \bpsi_{i,-}<\infty$ implies nothing here.}
\ie
\label{eq:spin-32variation}
(\delta \bpsi_{0})^{(0)}=0=\begin{cases}
    \partial_0 \bm\epsilon^{(0)} \rightarrow \bm{\epsilon}^{(0)} = \bm{\epsilon}_\zero(\vec x) & \text{flat} \\
    \partial_0 \bm\epsilon^{(0)} -\frac{\gamma_0}{2\eta} \bm\epsilon \rightarrow \bm\epsilon_{\pm}^{(0)}=  \bm\epsilon_{\pm,\zero}(\vec x) (\frac{\eta}{\epsilon})^{\pm \frac{1}{2}}& \text{EAdS}.
\end{cases} 
\fe
Thus we see as in the graviton case, only in curved space there is an additional gauge parameter in the transformation of the boundary degrees of freedom. However, there is an important consistency of the local SUSY algebra: \emph{Two successive local SUSY transformations give the diffeomorphism.} Therefore, maintaining the temporal gauge of the graviton might lead to further constraint on the SUSY parameter. To see, we note that the above algebra implies the diffeomorphism parameter can be represented by two SUSY parameters as,
\begin{equation}
[\delta_{\epsilon_1},\delta_{\epsilon_2}]=\delta_{\xi}\quad\to\quad\xi_\mu=\bar\epsilon_1 \gamma_\mu \epsilon_2.
\end{equation}
Now, as analyzed previously, in both spaces the boundary $\xi_i$ are spatial functions (and similarly for $\xi_0$ in the curved space).\footnote{The previous dS analysis can be extended to EAdS without any change of the conclusion.} However, $\xi_0$ in the flat space is required to be zero in order to be compatible with the Bunch-Davies vacuum. This then leads to an additional constraint on the residual SUSY parameters in the flat space,
\begin{equation}
\xi_0=\bar{\bm{\epsilon}}_1 \gamma_0 \bm{\epsilon}_2 =\bm{\epsilon}_{1,-}^T C_- \bm{\epsilon}_{2,+}-\bm{\epsilon}_{1,+}^T C_- \bm{\epsilon}_{2,-}=0,
\end{equation}
where $\bm{\epsilon}_n\equiv\bm{\epsilon}_\partial(\vec{x}_n)$, the Lorentzian charge conjugate matrix $C_-$ satisfies \eqref{eq:C4} and we have used the Majorana condition on the parameter $\bar{\bm{\epsilon}}=\bm{\epsilon}^TC_-$. Therefore, we find that \emph{in the flat space one needs to further choose either $\bm{\epsilon}_+=0$ or $\bm{\epsilon}_-=0$.} In view of the boundary profile being set in the minus eigenstate of $\gamma_0$, non-triviality of the boundary transformation occurs only when $\bm{\epsilon}_+=0$.

At last, the bulk SUSY variation of the graviton reads $
\delta h_{ij} = \bar{\bm{\epsilon}} \gamma_{(i} \bpsi_{j)} = \bar{\bm{\epsilon}}_+ \gamma_{(i} \bpsi_{j)-} + \bar{\bm{\epsilon}}_- \gamma_{(i} \bpsi_{j)+}
$. Consider its boundary limit, the $\bar{\bm{\epsilon}}_- \gamma_{(i} \bpsi_{j)+}$ is absent in the flat space due to the previous reason $\bm{\epsilon}_+=0$. However, it is actually also absent in the EAdS despite the fact that there is a nonzero $\bm{\epsilon}_+$. The reason lies in the $\eta\to\epsilon$ behavior of $\bm{\psi}_{+}^i$ scales as $\bm{\psi}^i_+\sim \eta^2\bm{\psi}^i_-$, and thus is suppressed near the boundary. In conclusion, the residual gauge SUSY implies the following boundary symmetries in both spaces as,
\begin{equation}
\label{eq:3DN=1superconformal}
    \begin{cases}
(\delta \bpsi_{i,-,\zero})^{(0)} = \partial_i \bm{\epsilon}_{-,\zero}, \;
(\delta h_{ij,\zero})^{(0)} = \bar{\bm{\epsilon}}_+ \gamma_{(i} \bpsi_{j),-,\zero}  & \text{flat} \\
(\delta \bpsi_{i,-,\zero})^{(0)} = \partial_i \bm{\epsilon}_{-,\zero} - {\gamma_i} \bm{\epsilon}'_{+,\zero}, \;
(\delta h_{ij,\zero})^{(0)} = \bar{\bm{\epsilon}}_+ \gamma_{(i} \bpsi_{j),-,\zero} & \text{EAdS}.
\end{cases}
\end{equation}
where we have redefined $\bm{\epsilon}_{+,\zero}' = \epsilon^{-1} \bm{\epsilon}_{+,\zero}$. We can identify the symmetry associated with $\bm{\epsilon}_-$ as the boundary SUSY, and the additional gauge parameter $\bm{\epsilon}_+$ as the super-Weyl transformation. Following a similar discussion, we can also see the graviphoton transformation under the super-Weyl transformation as $\delta A_{i,\zero}=\frac{1}{2}i\bar{\bm{\epsilon}}_-\bpsi_{i,-,\zero}$.

In conclusion, we observe that in the flat space, the residual gauge symmetry of the bulk diffeomorphism and SUSY simply implies the corresponding boundary transformations, whereas in the curved space, addtional boundary symmetry (Weyl and super-Weyl) appears.
A similar discussion for Lorentzian AdS, including the treatment of the graviphoton in $\mathcal{N}=2$ SUSY, can be found in \cite{Papadimitriou:2017kzw}.


\paragraph{The (gamma) trace WT identity: from (Super)Weyl transformation}
Based on the previous discussion, we'll show we could understand the (gamma) trace part of our bootstrap result as arising from the (super)Weyl transformation, which is a bulk residual gauge symmetry. For the trace WT identity on the tensor field, we can extend the derivation found in the appendix of \cite{Baumann:2020dch}, leading to results that match those obtained from bootstrap in \eqref{eq:tchichitr} and \eqref{eq:tpsipsitraceidentity}.

We now turn to examine the gamma trace WT identity by studying how the boundary profile transforms under the super-Weyl transformation. Let us focus on the following terms in the wave function, 
\begin{equation}
\langle\bar\psi^k\psi^l\rangle\bar\psi_{\partial,k}\psi_{\partial,l}+\langle T^{ij}\bar\psi^k\psi^l\rangle\delta_{a(i}h^a_{j)}\bar\psi_{\partial,k}\psi_{\partial,l}.\label{eq:tpsipsigammatrWFC}
\end{equation}
Next, to derive the WT identity for the gamma trace identity of $\psi$,
\begin{equation}
  \langle T^{ij} \bar\psi^k\psi^l\rangle\sigma_l=\langle\bar\psi_{2}^{(j}\psi^{\underline{l}}\rangle\sigma^{i)},  
\end{equation}
we consider transformations on $\psi_{\partial,l}$ in \eqref{eq:tpsipsigammatrWFC}, using \eqref{eq:3DN=1superconformal} \emph{and keep the two terms to order } $\mathcal{O}(h^1)$,
\ie
&\langle\bar\psi^k\psi^l\rangle\bar\psi_{\partial,k}(\delta\psi_{\partial,l})+\langle T^{ij}\bar\psi^k\psi^l\rangle\delta_{a(i}h^a_{j)}\bar\psi_{\partial,k}(\delta\psi_{\partial,l})=0
\\
&\Rightarrow-\langle\bar\psi^k\psi^l\rangle\bar\psi_{\partial,k}\left(\sigma_a h^a_l\right)+\langle T^{ij}\bar\psi^k\psi^l\rangle\delta_{a(i}h^a_{j)}\bar\psi_{\partial,k}\left(\sigma_b\bar{e}^b_l\right)=0,
\fe
next strip off all the boundary profiles we can then relate the gamma trace part of $\psi$ to $\langle\bar\psi\psi\rangle$.
\begin{equation}
    \sigma_k\langle T^{ij} \bar\psi^k\psi^l\rangle=\sigma^{(i}\langle\bar\psi_{-3}^{j)}\psi^l\rangle.
\end{equation}
Next we apply \eqref{eq:3DN=1superconformal} and the gravyphoton transformation we mentioned below it to the variation of the boundary profiles $\epsilon_{\partial,i},\psi_{\partial,i}$, use the similar method one and focus on $\bm\epsilon_+$ terms one can then verify the following gamma trace WT identity of $\langle J^i\bar\psi^j\psi^k\rangle$,
\begin{equation}
 \langle J^i\bar\psi^j\psi^k\rangle\sigma_k =-\langle J_{1}^iJ^j\rangle.   
\end{equation}

\section{The Leading Total Energy Pole Structure}
\label{app:LeadingTotalEnergyPoleStructure}

In this appendix we review the structure of the leading total energy pole for all types of interactions arising from bulk contact and one-particle exchange processes. Our discussion essentially follows the analysis of AdS correlators in \cite{Raju:2012zr,Goodhew:2020hob}.

\paragraph{Leading $E_T$ Pole from Asymptotic Expansion of the Integrand}

In \cite{Raju:2012zr,Goodhew:2020hob}, the leading total energy pole structure is obtained from the integral in the far-past limit of the WFCs. For example, for the three-point pure scalar contact interaction $\langle T_1 \phi_2 \phi_3 \rangle$, the integrand of the WFCs in the far-past limit is
\ie
\lim_{\eta\to-\infty} \eta^{-2} (\epsilon_1 \cdot p_2)^2 \cdot \mathcal{K}_{\Delta=3}(\eta,p_1)\mathcal{K}_{\Delta_2}(\eta,p_2)\mathcal{K}_{\Delta_3}(\eta,p_3)=(\epsilon_1 \cdot p_2)^2  E_1 E_2^{\Delta_2-2} E^{\Delta_3-2}_3 \cdot \eta e^{i E_T \eta} \label{eq: farpastlimit}
\fe
where the scalar propagator is given by
\ie
&\text{Scalar}:\quad\ck^s_{\nu}=\left(\frac{z}{z_*}\right)^{\frac{3}{2}}\frac{K^{(2)}_{\nu}(Ez)}{K^{(2)}_{\nu}(Ez_*)},\quad \nu=\Delta-\frac{3}{2}
\fe
with $K_\nu$ the modified Bessel $K$-function. The leading total energy pole is then obtained from (we omit all powers of $z_*$, which can be absorbed into the definition of the boundary profile)
\ie
\int^0_{-\infty} dz (\epsilon_1 \cdot p_2)^2 E_1 E_2^{\Delta_2-2} E^{\Delta_3-2}_3 \cdot z e^{i E_T z} = -M_{T\phi\phi} \cdot \frac{ E_1 E_2^{\Delta_2-2} E^{\Delta_3-2}_3}{E_T^2} 
\fe
We see that the leading total energy pole is of order $2$. Similarly, we can obtain the total energy pole order for other cases by analyzing the integrand in the far-past limit. 

This naturally raises the following questions: Why can the leading total energy pole be extracted solely from this regime? Do contributions to the leading total energy pole from the region near the boundary actually vanish?

\paragraph{Leading $E_T$ Pole from Asymptotic Expansion of the Integrand}

To address these questions, let us start from the EAdS contact interaction. The integrand is built out of Bessel $K$-functions and are difficult to analyze in closed form over the full integration range. However, if we write down the full asymptotic expansion of the integrand in the far-past limit, we will see that it is given by a rational function multiplied by an exponential, which is easy to integrate term by term. This also allows us to identify precisely which part of the series contributes to the leading total energy pole.

To proceed, we first write down the large-$z$ expansion of the Bessel $K$-function \cite{abramowitz+stegun},
\small{
\ie
    \lim_{z\to\infty}K_{\nu}(z)
    &
    \equiv
    \sum_{n=1}^{\infty} k_n(z)\\
    &
    =
    \sqrt{\frac{\pi}{2z}}e^{-z}\left(1+\frac{4\nu^2-1}{8z}+\frac{(4\nu^2-1)(4\nu^2-9)}{2!(8z)^2}+\frac{(4\nu^2-1)(4\nu^2-9)(4\nu^2-25)}{3!(8z)^3}+\cdots\right)
    \label{eq:besselKseries},
\fe}
which is valid only for large $z$. It is therefore not legitimate to use this expansion to integrate over the entire $z$-range. Instead, we rely on the inequalities that control the convergence of the Bessel $K$-series \cite{abramowitz+stegun}. For $n\geq\nu-\frac{1}{2}$ we have upper and lower bounds on $K_\nu$,
\ie
    \left\vert K_\nu(z)-K_\nu^n(z)\right\vert<\left\vert K_\nu^{n+1}(z) \right\vert,\quad\rightarrow\quad K^n_\nu(z)-\left\vert K_\nu^{n+1}(z)\right\vert<K_\nu(z)<K^n_\nu(z)+\left\vert K_\nu^{n+1}(z)\right\vert,\label{eq:besselKinequality}
\fe
where $K_{\nu}^n(z)\equiv \sum_{r=1}^{n} k_r(z)$ denotes the $n$-th order truncation of the series \eqref{eq:besselKseries}. We now use these upper and lower bounds to bound the full integrand, taking $n=\lceil \nu- 1/2 \rceil = \lceil \Delta-2 \rceil$ for the Bessel $K$ in each propagator. 

Without loss of generality, we focus again on the $\langle T \phi \phi \rangle$ case. The integrand is then bounded by
\ie
&E_1 E_2^{\Delta_2-2} E^{\Delta_3-2}_3  \cdot z^{1/2} (\epsilon_1 \cdot p_2)^2 \cdot (K^{n_1}_{3/2}(E_1 z)-\left\vert K_{3/2}^{n_1}(E_1 z)\right\vert)  (K^{n_2}_{\Delta_2-2}(E_2 z)-\left\vert K_{\Delta_2-2}^{n_2}(E_2 z)\right\vert)\\
&\cdot (K^{n_3}_{\Delta_3-2}(E_3 z)-\left\vert K_{\Delta_3-2}^{n_3}(E_3 z)\right\vert)\\
&< \eta^{-2} (\epsilon_1 \cdot p_2)^2 \cdot \mathcal{K}_{\Delta=3}(\eta,p_1)\mathcal{K}_{\Delta_2}(\eta,p_2)\mathcal{K}_{\Delta_3}(\eta,p_3) \\
&< E_1 E_2^{\Delta_2-2} E^{\Delta_3-2}_3  \cdot z^{1/2} (\epsilon_1 \cdot p_2)^2 \cdot (K^{n_1}_{3/2}(E_1 z)+\left\vert K_{3/2}^{n_1}(E_1 z)\right\vert)  (K^{n_2}_{\Delta_2-2}(E_2 z)+\left\vert K_{\Delta_2-2}^{n_2}(E_2 z)\right\vert)\\
&\cdot (K^{n_3}_{\Delta_3-2}(E_3 z)+\left\vert K_{\Delta_3-2}^{n_3}(E_3 z)\right\vert) \label{eq:bound}
\fe
with $n_1=1$, $n_2=\lceil \Delta_2-2 \rceil$, and $n_3=\lceil \Delta_3-2 \rceil$. Since the truncated series is a rational function times an exponential, each term in the upper and lower bounds can be integrated as \eqref{rational exp integral} 
which implies that the leading $E_T$ singularity comes from the term with the \emph{largest} $L$ (power of $z$) in the integrand. Applying this total energy pole kinematics to the inequality and taking the leading contribution, we see that the upper and lower bounds on the integral of \eqref{eq:bound} converge to the common value
\begin{equation}
    \int_0^\infty dz (\epsilon_1 \cdot p_2)^2  E_1 E_2^{\Delta_2-2} E^{\Delta_3-2}_3 \cdot \eta e^{i E_T \eta}
\end{equation}
where the integrand is precisely the far-past limit given in \eqref{eq: farpastlimit}. Therefore, we confirm that the leading total energy pole is indeed captured by integrating the far-past form of the integrand. A similar analysis applies in (EA)dS. 

Based on this result, we now analyze the leading total energy pole order for each case considered in the main text, and in particular extend the pole-order counting formula of \cite{Raju:2012zr,Goodhew:2020hob} to fermionic WFCs.

\paragraph{Contact Interaction}  

Let us begin by introducing the propagators in dS space (the corresponding EAdS discussion is analogous),
\ie
&\text{Scalar}:\quad\ck^s_{\nu}=\left(\frac{\eta}{\eta_*}\right)^{\frac{3}{2}}\frac{H^{(2)}_{\nu}(-E\eta)}{H^{(2)}_{\nu}(-E\eta_*)},\quad \nu=\Delta-\frac{3}{2}
\\
&\text{ Spin-$l$}:\quad\ck_l=\Hat{\Pi}\left(\frac{\eta}{\eta_*}\right)^{-l}\ck^{s}_{l-\frac{1}{2}}\left(E,\eta\right)
\\
&\text{Spin-$\frac{1}{2}$}:\quad\ck_{\frac{1}{2}}=\left(\frac{\eta}{\eta_*}\right)^{\frac{3}{2}}e^{iE\left(\eta-\eta_*\right)}\left(1-\slashed{\hat{p}}\right)
\\
&\text{Spin-$\frac{3}{2}$}:\quad\ck_{\frac{3}{2}}=\Hat{\Pi} \left(\frac{\eta }{\eta ^{*}}\right) \left(\left(\frac{1+iE\eta }{1+iE\eta ^{*}}\right)\left(\frac{\eta }{\eta ^{*}}\right)^{-\frac{3}{2}} \ +\left(\frac{1}{1+iE\eta ^{*}}\right)\frac{( E\eta )^{-1/2}}{( E\eta _{*})^{-3/2}}\slashed{\hat{p}}\right) e^{iE\eta }
\label{eq:dSbulkbdypropgs}
\fe
where $H^{(2)}_n(z)$ is the Hankel function of the second kind, $\eta^*$ is the late-time coordinate, $\hat{\Pi}$ collectively denotes the transverse traceless projectors and we have suppressed all indices. Note that the spin-$1/2$ propagator corresponds to massless bulk fields, while the spin-$3/2$ propagator corresponds to the conserved case. 

Using the far-past behaviour of the Hankel function,
\ie
 \lim_{z\to\infty}H_\nu^{(2)}(z)\sim\sqrt{\frac{1}{z}}e^{-i\left(z-\frac{\nu\pi}{2}\right)},
\fe
one finds that the far-past limits of bosonic and fermionic bulk--boundary propagators behave as
\begin{equation}
 \text{Bosonic}: \lim_{\eta\to-\infty}\mathcal{K}\sim E^{\Delta-2}\eta^{\#}e^{i E\eta},\quad\text{Fermionic}: \lim_{\eta\to-\infty}\mathcal{K}\sim E^{\Delta-\frac{3}{2}}\eta^{\#}e^{i E\eta}  \label{eq:bulkbdypast}
\end{equation}
where, since we are focusing on the dependence on boundary kinematics, we do not write explicitly the power $\#$ of $\eta$, and again suppress all tensor structures. 

Combining \eqref{eq:bulkbdypast} with the dS contact Feynman rules, the relation between the boundary profile and the amplitude polarization in the total energy pole limit \cite{Flatspace}, and the power-counting in time following \cite{Raju:2012zr,Goodhew:2020hob}\footnote{The powers of time arise from the metric and the propagators. By enumerating all possible contraction patterns, one finds that the total power of time is always equal to the total number of derivatives.}, we see that, under the far-past approximation to the integrand,
\begin{equation} 
    \lim_{E_T\to0}\left\langle\mathcal{O}\mathcal{O}\mathcal{O}\right\rangle=\frac{\left(\prod_{a=1}^3E_a^{k_a}\right)}{E_T^p}\times M, \quad p=\left[ M\right] 
\label{eq:contactleadingresidue}
\end{equation}
where $M$ denotes the flat-space amplitude, $k_a = \Delta_a -2 \text{ (bosonic)}, k_a =  \Delta_a -3/2 \text{ (fermionic)}$ are the powers of the energy prefactors appearing in \eqref{eq:bulkbdypast}, $p$ is the leading pole order, $[M]$ counts the purely kinematical (momentum/energy/polarization) mass dimension of $M$, for example $[M_{J\chi\chi}]=1.$

\paragraph{Exchange Interaction}

Having seen that the leading total energy pole residues are proportional to the flat-space amplitude for contact processes, we now turn to one-particle exchange processes. Our method combines several approaches in the literature \cite{Raju:2012zr, arkanihamed2017cosmologicalpolytopeswavefunctionuniverse}. The goal remains the same: extract the leading pole residue by inspecting the far-past regime of the integral and then determine the leading pole order.

We begin by examining the general structure of scalar exchange in the $s$-channel,
\begin{equation}
    \int d\eta_1 d\eta_2 \;\left(\eta_1^{n_1}\eta_2^{n_2}\right)\tilde{L}_{n_1}(\eta_1;\vec{p}_1,\vec{p}_2)\cdot G(\eta_1,\eta_2;\vec{p}_s)\cdot\tilde{R}_{n_2}(\eta_2;\vec{p}_3,\vec{p}_4),
\end{equation}
where we do not specify the external particle content at this stage. The functions $\tilde{L}_{n_1},\tilde{R}_{n_2}$ encode the left/right bulk--boundary propagators and the vertices, while $G$ denotes the scalar bulk--bulk propagator. The prefactor $\eta_1^{n_1}\eta_2^{n_2}$ arises from the metric and propagators, and, as in the contact case, a single vertex containing time derivatives can contribute multiple possible values of $n_1,n_2$.

To analyze the far-past structure of the double time integral, it is useful to perform the change of variables
\begin{equation}  
    \eta_1\equiv\tau+\delta,\quad\eta_2\equiv\tau-\delta,
\end{equation}
and to impose the far-past limit by taking $\tau\to-\infty(1-i\epsilon)$ while keeping $\delta$ unbounded. With these definitions, the far-past limit of the integral takes the form
\footnote{
    Here we use the fact that, near the far past, the bulk--bulk propagator satisfies
    \begin{equation}
        G\sim \eta_1\eta_2G_{\text{Feyn}}.
    \end{equation}
    This can be shown either by directly taking the limit in the explicit dS expression for $G$, or by analyzing the EOM near the far past. The far-past limits of the bulk--boundary propagators were already discussed in \eqref{eq:bulkbdypast}. Furthermore, the maximal power of $\tau$ always arises from the change of variables in the factor $\eta_1^{n_1}\eta_2^{n_2}$, which underlies the counting of $L$.
} 
\ie
&\left(\prod_{a}E_a^{k_a}\right)\lim_{E_T\to0}\int_{-\infty}^{\Lambda} d\tau \;\tau^{L} e^{i E_T\tau}\int_{-\infty}^{\infty}d\delta \;e^{i(E_{12}-E_{34})\delta} \mathcal{L}\cdot G_{\text{Feyn}}(E_s,2\delta)\cdot\mathcal{R}
\\
&= \lim_{E_T\to0}\left(\frac{\prod_{a}E_a^{k_a}}{E_T^{L+1}}\int_{-\infty}^{\infty}d\delta\; e^{i(E_{12}-E_{34})\delta} \mathcal{L}\cdot G_{\text{Feyn}}(E_s,2\delta)\cdot\mathcal{R}\right)
\fe
where $L=[ M]+1$ and we focus on $L\geq 0$. Here, $\mathcal{L}$ and $\mathcal{R}$ are the flat-space analogues of $\tilde{L}$ and $\tilde{R}$, and are independent of $\delta$. The factor $\tau^L$ represents the leading power of $\tau$ in the integrand.

The $\delta$-integral implements the correct amplitude factorization in the $E_T\to0$ limit. This is seen by the Feynman parametrization of $G_{\text{Feyn}}$:
\begin{equation}
    \int_{-\infty}^{\infty}d\delta \;e^{i(E_{12}-E_{34})\delta}G_{\text{Feyn}}=\int_{-\infty}^{\infty}d\delta \;e^{i(E_{12}-E_{34})\delta}\int_{-\infty}^{\infty}\frac{d\omega}{2\pi i}\frac{e^{2i\omega\delta}}{\omega^2-E_s^2+i\epsilon}=\frac{1}{\frac{1}{4}\left(E_{34}-E_{12}\right)^2-E_s^2},
\end{equation}
where, in the last equality, we have performed the $\delta$-integral first to generate a delta function $\delta(E_{12}-E_{34}+2\omega)$, which localizes the $\omega$ integral. In the limit $E_T\to 0$, $(E_{34}-E_{12})^2\to 4E_{34}^2$, and the final denominator becomes indistinguishable from the $s$-channel Mandelstam invariant. Thus, our claim is established, with total energy pole order $p=L+1$. For integer-spin exchange \cite{Raju:2012zr}, the same strategy can be applied to obtain the leading total energy pole structure, leading to the same conclusion.

Combining this analysis with that of the contact interaction, we obtain the following general structure (covering both contact and exchange WFCs) for the leading total energy pole,
\ie
\lim_{E_T\to0}\left\langle\mathcal{O}\mathcal{O}\mathcal{O}\mathcal{O}\right\rangle=\frac{\left(\prod_{a}E_a^{k_a}\right)}{E_T^p}\times M,\quad p=[ M]+ (n -3)
\fe
where $k_a = \Delta_a -2 \text{ (bosonic)}, k_a =  \Delta_a -3/2 \text{ (fermionic)}$ and $n$ denotes the number of external legs. One can verify this leading structure by comparison with the explicit results in \cite{Goodhew:2020hob}. 
\section{Derivation of the Fermionic Exchanged Cutting Rules}
\label{app: Feynman Rule from boundary action}
In this appendix we are going to derive the cutting rules for massless spin-1/2 and conserved spin-3/2 exchanges in the (EA)dS. Since derivations in EAdS are conceptually the same as in the dS, for the spin-1/2 case we will work in EAdS while for the spin-3/2 case we will work in dS. 
\paragraph{$\Delta=3/2$ Spin-1/2}
Let us start by considering the massless spin-1/2 field $\bm{\chi}$ in the EAdS. Under a change of variable $\bchi(\vec p,z) := (\frac{z}{\epsilon})^{3/2} \bchi'(\vec p,z)$ the EOM of $\bm{\chi}'$ can be simplified to the following form,
\ie
\label{eq:chi'eom}
(\gamma_{0,E} \partial_z -i \gamma^i p_i) \bchi'(\vec p,z) &= \frac{1}{z}\frac{\delta L_{\text{int}}}{\delta \bar \bchi}, 
\fe
where $\gamma_{0,E}$ follows \eqref{eq:4Dgammafrom5D}. In particular, the $\slashed{\nabla}$ operator now acts on $\bm{\chi}'$ identical to the flat-space whereas the interaction term is dressed with a $1/z$ factor, reflecting the nature of the time-dependent background. Given the similarity of the EOMs between EAdS \eqref{eq:chi'eom} and the flat space, the bulk-to-boundary and bulk-to-bulk propagators of $\bm{\chi}'$ are thus the flat-space ones dressed with overall $z$ factors. Importantly, \emph{the discontinuity of the bulk-to-bulk propagator can still be expressed as a product of discontinuity of the bulk-to-boundary propagator as in the flat-space} \cite{Flatspace}. 

Now, what do these simplifications imply in the cutting rule? To see, we recall that the on-shell action is given by substituting the classical solution in to the boundary action \cite{FermionLagTwoPoint}, which reads,
\ie
S_{b,\text{cl}} =-\frac{i}{2}\int d^3 x\ \epsilon^{-3} \bar \bchi_b \bchi_b = -\frac{i\epsilon^{-3}}{2}\intp{} \bar\bchi'_b(-\vec p)\bchi'_b(\vec p).
\fe
One can absorb the overall cut-off $\epsilon^{-3}$ into the boundary profiles $\bm{\chi}_\zero$ and recover the flat-space boundary action. This then leads to the same Feynman rules as the flat space with only changes in time-dependent vertices (dressed with $(z/\epsilon)^{3/2}$). Given the fact that the discontinuity is an operation on the internal energy \eqref{eq:disc of boson G}, we then conclude that the cutting rule remains the same as the flat space.

\paragraph{$\Delta=5/2$ Spin-3/2}
Next, for the conserved spin-3/2 field, the situation is more subtle. Following the analysis in \cite{Corley:1998qg}, one can see that the EOM for the conserved spin-3/2 field with $\Delta=5/2$ does not reduce to the massless or massive spin-3/2 EOM in flat or (EA)dS space via a simple field redefinition. 
Nevertheless, we can still follow the approach outlined in our companion paper~\cite{Flatspace} and utilize the structure of the EOM to relate the conserved spin-$\tfrac{3}{2}$ field to the scalar propagator and to establish analogous cutting rules.
We will from now on thoroughly focus on the transverse gamma-traceless component of the classical solution, as the WFC \emph{should be independent of the internal gauge choice}. For the cutting rule of external longitudinal or gamma-trace components, they can be straightforwardly verified from the WT identities.

As an illustrative case, let us focus on the gravitational interaction in dS, noting that the result for Euclidean AdS (EAdS) is expected to be the same. The interaction term is
\[
L_{\text{int}} = g\, h_{\mu\nu}\, \bar{\psi}_\rho\, V^{\mu\nu\rho\sigma}\, \psi_\sigma ,
\]
where $g$ is the coupling and $V$ defines the gravitational interaction vertices. We further introduce
\[
H^{(n)}_{\rho \sigma} := V^{\mu\nu\rho\sigma}\, h^{(n)}_{\mu\nu} ,
\]
as the contraction of $V$ with the $n$th-order classical solution of the graviton. In this short-hand notation, the EOM of the $n$th order solution of  $\bm{\psi}_i^{\Hat{T}}$ then reads,
\footnote{
    In the EOM, we tune the effective mass term such that $\Delta=5/2$ is obtained from the near boundary behavior of the free Bunch-Davies solution and the two-point function yields the expected conserved two-point function.
    For comparison, we note that the expanded EAdS EOM in \cite{Corley:1998qg} takes the form 
    \ie
    (\eta\gamma_0\partial_0 + i \eta \vec{\gamma}\cdot \vec{p} + \frac{1}{2}\gamma_0 - 1)\bm{\psi}^{\hat{T},{(n_\psi+n_h+1)}}_i(\eta,\vec p)
    =
    \int \frac{d^3 q}{(2 \pi)^3} g H^{(n_\psi+n_h),j}_{i}(\eta,\partial_0,\vec q)\bm{\psi}^{\hat{T},{(n_\psi)}}_j(\eta,\vec p - \vec q),
    \fe 
    which differs from our EOM by a phase of momentum.
}
\ie
 (-\eta\gamma_0 \partial_\eta + i \eta \vecslashed{p} + \frac{1}{2}\gamma_0 - i)\bpsi^{\hat{T},{(n_p+n_h+1)}}_i(\eta,\vec p)=\int \frac{d^3 q}{(2 \pi)^3} g H^{(n_h),j}_{i}(\eta,\partial_\eta,\vec q)\bpsi^{\hat{T},{(n_p)}}_j(\eta,\vec p - \vec q).
\fe
Similar to what we did in the "alternative way" to derive the cutting rule in flat space \cite{Flatspace}, we can decompose $\bm{\psi}_i^{\Hat{T}}$ into $\gamma_0$ eigenstates and iteratively obtain the solution. The benefit of this method is that the $n$th order solution that shares the same $\gamma_0$ eigenvalue with the boundary profile behaves very alike a massive scalar, thus greatly simplifies the derivation of the discontinuity. To be more explicit, the free solution reads,
\begin{equation}
    \bpsi_{i}^{\hat{T},(0)}(\vec p,\eta)= \left(1-\frac{\vecslashed{p}}{\eta E^2}(\eta\partial_\eta - \frac{1}{2})\right)\left[\frac{\epsilon^3}{\eta^3}K_{\phi,\Delta=5/2}(\vec p,\eta)\right]\bpsi^{\hat{T}}_{i,\zero,-}(\vec p)
    \equiv  K^{\hat{T}}_{\bpsi}(\vec p,\eta)\bpsi^{\hat{T}}_{i,\zero,-}(\vec p),
\end{equation}
where we note that the above equation contains both the $\gamma_0$  eigenstates and $K_{\phi,\Delta=5/2}$ is the bulk-to-boundary propagator of the $\Delta=5/2$ scalar. The first-order solution is given by,
\ie
\label{ds: classical solution}
\bpsi^{\hat{T},(1)}_{i}(\vec p,\eta)
    &=
    \left[ 1 - \frac{\vecslashed{p}}{\eta\cdot E^2}(\eta\partial_\eta - \frac{1}{2})\right]
    \cdot
    g\int d\eta' \frac{\epsilon^4}{\eta^3 \eta^{'}} G_{\phi,\Delta=5/2}(\vec p, \eta, \eta')\cdot\left(\frac{1+i\gamma_0}{2}\right)\\
    &\quad\cdot
    \left[\frac{1}{\eta^{'}}(\eta'\partial_{\eta'} + \frac{3}{2})\frac{1}{\eta^{'}}
    +
    \frac{i\vecslashed{p}}{\eta'}
    \right]
    \int\frac{d^3 q}{(2\pi)^3} H^{(0),j}_{i}(\eta',\partial_{\eta'},\vec q)\bpsi^{\hat{T},(0)}_j(\eta',\vec p - \vec q) \\
    &\quad-
    \frac{i\vecslashed{p}}{\eta\cdot E^2}\left(\frac{1+i\gamma_0}{2}\right)\cdot
    g \int \frac{d^3 q}{(2\pi)^3}
     H^{(0),j}_{i}(\eta,\partial_\eta,\vec q)\bpsi^{\hat{T},(0)}_j(\eta,\vec p - \vec q),
\fe
where $G_{\phi,\Delta=5/2}$ is the $\Delta=5/2$ scalar bulk-to-bulk propagator. The first term (first two lines) above corresponds to the minus $\gamma_0$ eigenstate while the last term is the plus eigenstate. The solution of $\bar{\bm{\psi}}_i^{\Hat{T}}$ can be similarly obtained, where the free solution reads,
\begin{equation}
    \bbarpsi_{i}^{\hat{T},(0)}(\vec p,\eta) =\bbarpsi^{\hat{T}}_{i,\zero,+}(\vec p)\left[\frac{\epsilon^3}{\eta^3}K_{\phi,\Delta=5/2}(\vec p,\eta)\right]\left(1+(\overleftarrow{\partial}_\eta \eta- \frac{1}{2} )\frac{\vecslashed{p}}{\eta E^2}\right) 
    \equiv \bbarpsi^{\hat{T}}_{i,\zero,+}(\vec p) K^{\hat{T}}_{\bbarpsi}(\vec p,\eta)
\end{equation}
and the first-order solution reads,
\ie
\bbarpsi_{i}^{\hat{T},(1)} (\vec p_s ,\eta) 
    &= 
        g\int
        \frac{d^3 q}{(2\pi)^3} \int d\eta' \,
        \bbarpsi^{\hat{T},(0)}_{j}(\eta',\vec p - \vec q)
        H^{(0),j}_{i}(\eta',\partial_{\eta'},\vec q)
        \cdot
        \left(\frac{1-i\gamma_0}{2}\right) \\
    &\quad\cdot
        \left[\frac{1}{\eta^{'}}(\eta'\partial'_0 + \frac{3}{2})\frac{1}{\eta^{'}}
        -
        \frac{i\vecslashed{p}}{\eta'}
        \right]
        \cdot
    \frac{\epsilon^4}{\eta^3 \eta^{'}} G_{\phi,\Delta=5/2}(\vec p, \eta, \eta')\cdot
        \left[ 1 + (\overleftarrow{\partial_\eta}\eta - \frac{1}{2})\left(\frac{\vecslashed{p}}{E^2 \eta}\right)\right]
       \\
    &\quad+ 
        g \int\frac{d^3q}{(2\pi)^3} \bbarpsi_{j}^{\hat{T},(0)}(\eta,\vec p -\vec q) H^{(0),j}_{i}(\eta,\partial_0,\vec q)
        \cdot
        \left(\frac{1-i\gamma_0}{2}\right)\cdot \frac{i\vecslashed{p}}{E^2 \eta}.
\fe

One can then use these classical solutions to compute the WFCs. Using the notation defined in section \ref{subsec:cuttingrules}, the relavent 3 and 4-pt WFCs in our example of deriving the spin-3/2 cutting rule are,
\ie
\label{ds: Fermion Fey}
c_{3,\bbarpsi_1 T_2 \bpsi_3} 
    &=
    \epsilon^{-2}\int d\eta \ \bbarpsi^{\hat{T},(0)}_{j}(\eta,\vec p_1) \cdot ig H^{(0),ji}(\vec p_2,\eta) \cdot \bpsi^{\hat{T},(0)}_{i}(t,\vec p_3)\\
c_{4,T_1 \bbarpsi_2 T_3 \psi_4}
    &=
    \frac{\epsilon^{-2}}{2} 
    \int d\eta\ 
    \bbarpsi^{\hat{T},(0)}_{j}(\vec p_2,\eta) 
    \cdot igH^{(0),ji}(\vec p_1,\eta) \cdot  
    \bpsi^{\hat{T},(1)}_{i}(\vec p_s,\eta) \\
    &\quad+ 
    \bbarpsi^{\hat{T},(1)}_{j}(-\vec p_s,\eta) 
    \cdot igH^{(0),ji}(\vec p_3,\eta) \cdot
    \bpsi^{\hat{T},(0)}_{i}(\vec p_4,\eta).
\fe
Now, let us examine the discontinuity of the first-order solution. We first note that, since any even function in $E$ vanishes under the discontinuity operation, the discontinuity of the first-order solution is then given by (A similar expression holds for the conjugate field),
\ie
\label{ds: first order solution discontinuity}
&\disc_{E_s} \psi^{\hat{T},(1)}_{i,\zero}(\vec p_s, \eta)\\ 
&=
	g\int d\eta' \left( 1-\frac{\vecslashed{p}_s}{\eta\, E^2_s}(\eta\partial_\eta - \frac{1}{2}) \right) \left( \frac{\epsilon^3}{\eta^3} \frac{\epsilon}{\eta^{'}}
	\disc_{E_s} G_{\phi,\Delta=5/2}(\vec p_s, \eta, \eta') \right)\cdot\left(\frac{1+i\gamma_0}{2}\right)\\
    &\quad\cdot
    \left[\frac{1}{\eta^{'}}(\eta'\partial_{\eta'} + \frac{3}{2})\frac{1}{\eta^{'}}
    +
    \frac{i\vecslashed{p}}{\eta'}
    \right]
    \int\frac{d^3 q}{(2\pi)^3} H^{(0),j}_{i}(\eta',\partial_{\eta'},\vec q)\bpsi^{\hat{T},(0)}_j(\eta',\vec p - \vec q).
\fe
At last, use the discontinuity structure of $G_\phi$,
\ie
\disc_{E_s} G_{\phi,\Delta=5/2}(\vec p_s, \eta, \eta') = \frac{1}{E^2} \disc_{E_s} K_{\phi,\Delta=5/2}(\vec p_s, \eta)\, \disc_{E_s} K_{\phi,\Delta=5/2}(\vec p_s, \eta'),
\fe
and substitute it to the discontinuity of the 4-pt WFC,
\ie
\label{ds: gravitino disc 1 order}
&\disc_{E_s}c^{\hat T}_{4,T\bar\psi T \psi}
= 
\frac{1}{2} 
\int d\eta\, \bar \psi^{\hat{T},(0)}_{\zero, j}(\vec p_2, \eta)\, (ig H^{(0),j}_i(\vec p_1, \eta))\, \left( \disc_{E_s} \psi^{\hat{T},(1)}_{i, \zero}(\vec p_s, \eta) \right)\\
&\quad+ 
\left( \disc_{E_s} \bar \psi^{\hat{T},(1)}_{\zero, j}(-\vec p_s, \eta)\right) (ig H^{(0),j}_{i}(\vec p_3, \eta))\, \psi^{\hat{T},(0)}_{i, \zero}(\vec p_4, \eta),
\fe
we then finish the derivation of the cutting rule of exchanging the bulk conserved spin-3/2 particle,
\ie
    \begin{aligned}
        \disc_{E_s} c^{\hat T}_{4,s}
        &=
        \disc_{E_s} c^{\hat T}_{3, A, i} (\vec p_1, \vec p_2, E_s) 
        \cdot \left[\frac{1+i\gamma_0}{2} \cdot \frac{i\vecslashed{\hat p}_s}{2 E_s^2} \cdot \frac{1-i\gamma_0}{2}\right]^{AB}
        \disc_{E_s} c^{\hat T}_{3, B, i} (\vec p_3, \vec p_4, E_s).
    \end{aligned}
\fe
Moreover, by employing the longitudinal/gamma-trace WT identity, one can show that the internal trace and longitudinal components make no contribution to the discontinuity of the three-point WFCs, while the external trace and longitudinal components are trivially consistent with the cutting rule. In other words, one can express the cutting rule on the \emph{full} components as,
\begin{equation}
	\begin{aligned}
	&\disc_{E_s} c_{4,s}
	=
	\disc_{E_s} c_{3, A, i} (\vec p_1, \vec p_2, E_s) 
	\cdot \left[\frac{1+i\gamma_0}{2} \cdot \frac{i\hat{\Pi}^{ij}_s \vecslashed{\hat p}_{s}}{2 E_s^2} \cdot \frac{1-i\gamma_0}{2}\right]^{AB}
	\disc_{E_s} c_{3, B, j} (\vec p_3, \vec p_4, E_s).
	\end{aligned}
\end{equation}
Here, $\hat\Pi_{(3/2,3/2), ij, s}$ is the gamma-traceless, transverse projector defined in \eqref{eq:gammadecompose}. This result matches the flat space cutting rule, with internal energy factor scaling as $1/E_s^{2\Delta_s-3}$ to be consistent with dilatation symmetry. We thus confirm the validity of~\eqref{discCOT} for conserved $\Delta_s=5/2$ spin-3/2 exchange.\footnote{
    In \eqref{discCOT}, we've already plugged in the 4-component embedding for the 2-component spinors which we give in the flat space paper \cite{Flatspace}.}
\section{The Clifford Algebra of $SO(1,4)$}
\label{sec:SO(1,4)details}
Our goal in this appendix is to establish the 3D Dirichlet boundary conditions compatible with the $SO(1,4)$ SM condition for (EA)dS fermions. We will begin by inspecting how the same $SO(1,4)$ SM condition is realized differently in $\text{(EA)dS}_4$ due to the different charge conjugations defined therein. Next, we identify the canonical conjugate components of the fermions to avoid fixing them simultaneously. After all these, we can then establish the correct Dirichlet boundary conditions for (EA)dS that are also compatible with the SM condition.

Let us start from introducing the 4 and 5D charge conjugation operators, the gamma matrices, and the Dirac conjugation. These would enable us to do the 4D realization of the 5D SM condition. Some nice reviews on the Clifford algebra in general spacetime dimension and signatures can be found in \cite{Freedman:2012zz, Ortin:2015hya, Polchinski:1998rr}. 

\paragraph{$4D$ Realizations of $SO(1,4)$} The 5D charge conjugation operator obeys the following algebraic properties,
\begin{equation}
    C_5^T = -C_5, \quad C_5 \gamma_M C_5^{-1} = \gamma_M^T, \;(M=0,1,2,3,5)\label{eq:C5}
\end{equation}
where $\gamma_0^2=-1=-\gamma_5^2$ (the explicit form of $\gamma_M$ can be found in \eqref{eq:gammamatrices}). In our convention, we have $C_5=i\gamma_5\gamma_2$\footnote{The overall $i$ is chosen such that $C_5^2=1$ for convenience.} and $[C_5,\gamma_0]=\{C_5,\gamma_5\}=0$.  The 4D charge conjugation operator, on the other hand, obeys similar algebraic relation as \eqref{eq:C5} but with an additional freedom of choosing the overall sign in transposing the 4D gamma matrices,
\begin{equation}
C^T_{4\pm,\mathfrak{s}} = -C_{4\pm,\mathfrak{s}},\quad C_{4\pm,\mathfrak{s}} \gamma_{\mu,\mathfrak{s}} C^{-1}_{4\pm,\mathfrak{s}} =\pm \gamma_{\mu,\mathfrak{s}}^T,\; (\mu=0,1,2,3)\label{eq:C4}
\end{equation}
where $\mathfrak{s}=L,E$ denoting the Lorentzian / Euclidean signature of the 4D space. More explicitly, the 4D gamma matrices in both spaces can be represented by the 5D ones as the following,
\ie
\label{eq:4Dgammafrom5D}
&\gamma_{\mu,L}=\{\gamma_0,\gamma_1,\gamma_2,\gamma_3\} \quad \text{and} \quad \gamma_{\mu,E}=\{\gamma_5,\gamma_1,\gamma_2,\gamma_3\},
\\
&\gamma_{5,L}=(-i\gamma_0\gamma_1\gamma_2\gamma_3)=\gamma_5\quad \text{and} \quad \gamma_{5,E}=(-\gamma_{1}\gamma_{2}\gamma_{3}\gamma_{5})=i\gamma_0.
\fe
 Given the 5D relation \eqref{eq:C5}, the $+$ convention of the 4D charge conjugation operator in both spaces can then be identified as,\footnote{It is also easy to find that in our basis $C_{4-,L}=\gamma_2\;,C_{4-,E}=i\gamma_1\gamma_3$.}
 \begin{equation}
     C_5=C_{4+,\mathfrak{s}}.
 \end{equation}
 Moreover, one can also find out the relation between 5D and 4D Dirac fermions,\footnote{For the Dirac conjugate, we define $\bar\chi^{\mathbb{I}}=(\chi^{\mathbb{I}})^{\dagger}D_{\pm}, D_{\pm}^2=1, D_\pm \gamma_\mu D_\pm^{-1}=\pm\gamma_\mu^\dagger.$ In the main context, we use $D_5=D_{5-}=D_{4,L}=D_{4-,L}=i\gamma_0,\;D_{4,E}=D_{4+,E}=I_4.$}
\ie
\chi^{\mathbb{I}}_5&= \chi^{\mathbb{I}}_4\\
\bar\chi^{\mathbb{I}}_5&=(\chi^{\mathbb{I}}_5)^\dagger D_5=\bar\chi^{\mathbb{I}}_{4,\mathfrak{s}} D_{4,\mathfrak{s}} D_5  = \begin{cases}
    \bar\chi^{\mathbb{I}}_{4,L} \;(\text{dS})\\
    \bar\chi^{\mathbb{I}}_{4,E}\gamma_{5,E}\;(\text{EAdS}).\label{eq:5Dto4DDirac}
\end{cases}
\fe
Having equipped with all the ingredients, we can now do the 4D realizations of the 5D SM condition. First, in 5D, the SM condition relates the the fermion with its Dirac conjugate as,
\ie
\bar\chi^{\mathbb{I}}_5=\chi^{\mathbb{J},T}_5\epsilon^{\mathbb{J}\mathbb{I}}C_5, \label{eq: Boundary Majorana 5D}
\fe
where the symplectic indices $\mathbb{I},\mathbb{J}$ are even-dimensional. In this paper, we focus on the simplest one, $\mathbb{I}=1,2$.\footnote{We set $\epsilon_{12}=-\epsilon_{21}=1$.} Now, together with the identification between 5D and 4D fermions \eqref{eq:5Dto4DDirac}, we can then establish the 4D realizations of the $SO(1,4)$ SM condition \eqref{eq: Boundary Majorana 5D} in both spaces,
\ie
SO(1,4)\; \text{SM Condition in 4D:}\;\begin{cases}
   (\text{dS}) \;\bar\chi^{\mathbb{I}}_{4,L}=\chi^{\mathbb{J},T}_4\epsilon^{\mathbb{J}\mathbb{I}}C_{4+,L} \\
    (\text{EAdS}) \;\bar\chi^{\mathbb{I}}_{4,E}=\chi^{\mathbb{J},T}_4\epsilon^{\mathbb{J}\mathbb{I}}C_{4-,E}.\label{Maj 4D}
\end{cases}
\fe
From now on, we will omit the subscript $4$ on the fermion and the charge conjugate. We stress here that, \eqref{Maj 4D} applies both for spin-1/2 and 3/2 fields with the understanding on 3/2 that it applies to the 1/2 representation of it.
Let us now determine the correct Dirichlet boundary conditions compatible with \eqref{Maj 4D}. 
\paragraph{Spin-1/2 Boundary Conditions and Polarizations} First we recall that for fermions in spacetime with a spacelike boundary, it is natural to choose the boundary conditions to be $\gamma_{0,\mathfrak{s}}$ eigenstates. Second, the canonical conjugate is equivalent to the Dirac conjugate, which identifies $\{\bar\chi^\mathbb{I}_\mathfrak{s},\chi_\mathfrak{s}^\mathbb{I}\}$ as a canonical pair. Therefore, the canonical pair after projecting to $\gamma_{0,\mathfrak{s}}$ eigenstates is\footnote{$\chi^{\mathbb{I}}_{\pm,L}=\frac{i\pm\gamma_{0,L}}{2}\chi^{\mathbb{I}}_{L},\;\chi^{\mathbb{I}}_{\pm,E}= \frac{1\pm\gamma_{0,E}}{2}  \chi^{\mathbb{I}}_{E},\;\bar\chi^{\mathbb{I}}_{\pm,L}=\bar\chi^{\mathbb{I}}_{L} \frac{i\pm\gamma_{0,L}}{2},\; \;\bar\chi^{\mathbb{I}}_{\pm,E}=\bar\chi^{\mathbb{I}}_{E} \frac{1\pm\gamma_{0,E}}{2}$.},
\begin{equation}
    \left\{\bbarchi^\mathbb{I}_{\pm,\mathfrak{s}},\bchi_{\pm,\mathfrak{s}}^\mathbb{I}\right\},
    \label{eq:4DSMcanonicalpair}
\end{equation}
where we have used that $D_{4,\mathfrak{s}}$ commutes with $\gamma_{0,\mathfrak{s}}$ in both spaces in our convention. The above means that, \emph{fermions with the same symplectic indices and $\gamma_{0,\mathfrak{s}}$ eigenstates cannot be chosen as the boundary conditions.}  This is in turn consistent with the SM condition \eqref{Maj 4D}. Indeed, after projecting both sides of \eqref{Maj 4D} to $\gamma_{0,\mathfrak{s}}$ eigenstates, we find,
\ie
\begin{cases}
&\text{dS: } \;\bbarchi^{2}_{\pm,L}=\bchi^{1,T}_{\pm}C_{+,L}\text{,  }\;  \bbarchi^{1}_{\pm,L}=-\bchi^{2,T}_{\pm}C_{+,L} \\
&\text{EAdS: }\;\bbarchi^{2}_{\pm,E}=\bchi^{1,T}_{\mp}C_{-,E}\text{,  } \;\bbarchi^{1}_{\pm,E}=-\bchi^{2,T}_{\mp}C_{-,E}, \label{4D Maj in decomposition}
\end{cases}
\fe
which shows in both spaces the SM condition does not lead to simultaneous fixing of the canonical pair \eqref{eq:4DSMcanonicalpair}. Note that the EAdS result follows from $[C_{-,E},\gamma_{0,E}]=0,\;\gamma_{0,E}^T=-\gamma_{0,E}$.

For completeness, below we list down the only two compatible sets of the Dirichlet boundary conditions,
\ie
\begin{cases}
    \label{Dirichlet Condition SM}
    &\text{dS Dirichlet: } (\bar{\bm{\chi}} ^{1}_{+,\zero}, \bar {\bm{\chi}}^{2}_{-,\zero},\bm{\chi}^1_{-,\zero},\bm{\chi}^2_{+,\zero}),(\bar{\bm{\chi}}^{1}_{-,\zero}, \bar{\bm{\chi}}^{2}_{+,\zero},\bm{\chi}^1_{+,\zero},\bm{\chi}^2_{-,\zero}) \\
    &\text{EAdS Dirichlet: } (\bar{\bm{\chi}}^{1}_{+,\zero}, \bar{\bm{\chi}}^{2}_{+,\zero},\bm{\chi}^1_{-,\zero},\bm{\chi}^2_{-,\zero}),(\bar{\bm{\chi}}^{1}_{-,\zero}, \bar{\bm{\chi}}^{2}_{-,\zero},\bm{\chi}^1_{+,\zero},\bm{\chi}^2_{+,\zero}).
\end{cases}
\fe

The spin-1/2 on-shell polarizations satisfy the Dirac equation with positive energy, $\slashed{P} u =\bar u \slashed{P}=0$. Its relation to the boundary profiles that are $\gamma_{0,\mathfrak{s}}$ eigenstates have been established in \cite{Flatspace}, now with the additional symplectic indices. Below we list for completeness their explicit forms,
\ie
\begin{cases}
    &\text{dS: } u^{1}=(1-i\slashed{\hat{p}})\bm{\chi}^1_{-,\zero}, u^{2}=(1+i\slashed{\hat{p}})\bm{\chi}^2_{+,\zero}, 
    \bar u^1=\bar{\bm{\chi}}^1_{+,\zero}(1+i\slashed{\hat{p}}), \bar u^2=\bar{\bm{\chi}}^2_{-,\zero}(1-i\slashed{\hat{p}}) \\
    &\text{EAdS: } u^{1}=(1-i\slashed{\hat{p}})\bm{\chi}^1_{-,\zero}, u^{2}=(1-i\slashed{\hat{p}})\bm{\chi}^2_{-,\zero}, 
    \bar u^1=\bar{\bm{\chi}}^1_{+,\zero}(1+i\slashed{\hat{p}}), \bar u^2=\bar{\bm{\chi}}^2_{+,\zero}(1+i\slashed{\hat{p}}). \\
\end{cases}
\label{eq:SMpolarizationandprofiles}
\fe
Note that, together with the SM condition of the boundary profiles \eqref{Maj 4D}, the polarizations also satisfy identical relations,
\ie
\begin{cases}
   &\text{dS: } \bar u^{2}_{L}=u^{1,T}_{L}C_{+,L}\text{,  }  \bar u^{1}_{L}=-u_L^{2,T}C_{+,L} \\
   &\text{EAdS: } \bar u^{2}_{E}=u_E^{1,T}C_{-,E}\text{,  } \bar u^{1}_{E}=-u_E^{2,T}C_{-,E}.
\end{cases}
\label{eq:spin1/2SMpolarization}
\fe
\paragraph{Spin-3/2 Boundary Conditions and Polarizations} Similar to what we did for the discussion of spin-1/2, we decompose all the spin-3/2 fermion in the basis of $\gamma_{0,\mathfrak{s}}$ and try to impose the compatible Dirichlet boundary condition with the SM Condition. 
The canonical conjugate of the spin 3/2 fermion in the temporal gauge is defined by $\delta \mathcal{L}/\delta (\partial_0\psi_\mathfrak{s}^{\mathbb{I},i}) = \bar\psi^{\mathbb{I}}_{j,\mathfrak{s}} \gamma^{j0i}_{\mathfrak{s}}$. Since the 4D on-shell polarization is transverse gamma-traceless, therefore we will focus on this component. It is straightforward to show that, after decomposing $\psi_T^i$ as \eqref{eq:gammadecompose}, the canonical conjugates of $\psi_{\widehat{T}}^i,\;\psi_{\underline{T}}^i$ would also share the same transverse gamma trace(less) property. Therefore, the conjugate pair that we avoid to impose boundary condition on both sides is,
\begin{equation}
    \left\{\bbarpsi^{\mathbb{I},i}_{\Hat{T},\mathfrak{s}},\bpsi^{\mathbb{I},i}_{\Hat{T},\mathfrak{s}}\right\}.
\end{equation}
Below we provide for completeness all possible Dirichlet boundary conditions,
\ie
\label{All Dirichlet Condition SM 3/2}
\begin{cases}
    &\text{dS Dirichlet: } (\bar{\bm{\psi}}^{1,i}_{+,\zero}, \bar{\bm{\psi}}^{2,i}_{-,\zero},\bm{\psi}^{1,i}_{-,\zero},\bm{\psi}^{2,i}_{+,\zero}),(\bar{\bm{\psi}}^{1,i}_{-,\zero}, \bar{\bm{\psi}}^{2,i}_{+,\zero}, \bm{\psi}^{1,i}_{+,\zero},\bm{\psi}^{2,i}_{-,\zero}) \\
    &\text{EAdS Dirichlet: } (\bar{\bm{\psi}}^{1,i}_{+,\zero}, \bar{\bm{\psi}}^{2,i}_{+,\zero},\bm{\psi}^{1,i}_{-,\zero},\bm{\psi}^{2,i}_{-,\zero}),(\bar{\bm{\psi}}^{1,i}_{-,\zero}, \bar{\bm{\psi}}^{2,i}_{-,\zero}, \bm{\psi}^{1,i}_{+,\zero},\bm{\psi}^{2,i}_{+,\zero}).
\end{cases} 
\fe
In the main content, we choose the first parenthesis for both space as the boundary conditions. The explicit relation between the polarization and the chosen boundary conditions can then be established as,\footnote{Here we recall that the $\pm$ sign in front of each $i\slashed{\Hat{p}}$ is identical to the $\pm$ eigenvalue of the boundary profiles.}
\ie
\begin{cases}
    \text{dS: }& u^{1}\epsilon_{i}=(1-i\slashed{\hat{p}})\hat\Pi_{ij}\bm{\psi}^{1,j}_{-,\zero},\; 
   u^{2}\epsilon_{i}=(1+i\slashed{\hat{p}})\hat\Pi_{ij}\bm{\psi}^{2,j}_{+,\zero},\\ 
    &\bar u^1\epsilon_{i}=\bar{\bm{\psi}}^{1,j}_{+,\zero}\hat\Pi_{ji}(1+i\slashed{\hat{p}}), \;\bar u^2\epsilon_{i} =\bar{\bm{\psi}}^{2,j}_{-,\zero}\hat\Pi_{ji}(1-i\slashed{\hat{p}}) \\
    \text{EAdS: }& u^{1}\epsilon_{i}=(1-i\slashed{\hat{p}})\hat\Pi_{ij}\bm{\psi}^{1,j}_{-,\zero}, \;u^{2}\epsilon_{i}=(1-i\slashed{\hat{p}})\hat\Pi_{ji}\bm{\psi}^{2,j}_{-,\zero} \\
    &\bar u^1\epsilon_{i}=\bar{\bm{\psi}}^{1,j}_{+,\zero}\hat\Pi_{ji}(1+i\slashed{\hat{p}}),\; \bar u^2\epsilon_{i}=\bar{\bm{\psi}}^{2,j}_{+,\zero}\hat\Pi_{ji}(1+i\slashed{\hat{p}}). \\
\end{cases}
\label{Polarization 3/2}
\fe
where in the LHS of above equalities we suppress the summation over $\gamma_{5,\mathfrak{s}}$ chirality in the above equation.  Note that, the spin-3/2 on-shell polarizations also satisfy the SM condition similar to the spin-1/2 case,
\ie
\begin{cases}
   &\text{dS: } \bar u^{2}\epsilon_i=(u^{1}\epsilon_i)^TC_{+,L}\text{,  }\;  \bar u^{1}\epsilon_i=-(u^{2}\epsilon_i)^TC_{+,L} \\
   &\text{EAdS: } \bar u^{2}\epsilon_i=(u^{1}\epsilon_i)^TC_{-,E}\text{,  }\; \bar u^{1}\epsilon_i=-(u^{2}\epsilon_i)^TC_{-,E}.
\end{cases}
\label{eq:spin3/2SMpolarization}
\fe
\paragraph{Spinor Variables} Having established the relation between the boundary profiles and the polarizations, let us now work out how the SM conditions \eqref{eq:spin1/2SMpolarization} reflected on the spinor variables.
As a remark, we recall that the symplectic indices for spin-3/2 polarizations are written in the spin-1/2 parts. Thus we will focus on the SM spin-1/2 spinor variables. First, the spin-1/2 polarizations can be expressed in terms of the $\gamma_{5,\mathfrak{s}}$ eigenstates as
\begin{equation}
u^\mathbb{I}=u^{(+)}(|p]^\mathbb{I})+u^{(-)}
(|p\rangle^\mathbb{I}),
\end{equation}
where in our convention $|p]$ is positive helicity while $\ket{p}$ is negative and thus $(\pm)$ denotes the $\gamma_{5,\mathfrak{s}}$ eigenstates. Next, the SM conditions \eqref{eq:spin1/2SMpolarization} imply the following SM relations,
\ie
\bar u^{2} =\bar{u}^{(+)}([p|^1)+u^{(-)}
(\langle p|^1), \quad\bar u^{1} = -\bar{u}^{(+)}([p|^2)-u^{(-)}
(\langle p|^2),
\label{eq:SMonchiral}
\fe
where we have used $\{C_{+,L},\gamma_{5,L}\}=0,\;\gamma_{5,L}^T=-\gamma_{5,L}$ and $[C_{-,E},\gamma_{5,E}]=0,\;\gamma_{5,E}^T=\gamma_{5,E}$.
Note that the above means that $[p|^{\mathbb{I},a}=|p]_{\; b}^\mathbb{I}\epsilon^{ab},\; \langle p|^{\mathbb{I},a}=|p\rangle_{\; b}^\mathbb{I}\epsilon^{ab}$ in both spaces.

\newpage

\bibliography{newrefs}
\bibliographystyle{JHEP}

\end{document}